# A Library of Theoretical Ultraviolet Spectra of Massive, Hot Stars for Evolutionary Synthesis

Claus Leitherer

*Space Telescope Science Institute[1], 3700 San Martin Drive, Baltimore, MD 21218*

*leitherer@stsci.edu*

Paula A. Ortiz Otálvaro

*Instituto de Física, Universidad de Antioquia, Medellín, Colombia*

*pauortizo@gmail.com*

Fabio Bresolin

*Institute for Astronomy, University of Hawaii at Manoa, 2680 Woodlawn Drive,*

*Honolulu, HI 96822*

*bresolin@ifa.hawaii.edu*

Rolf-Peter Kudritzki

*Institute for Astronomy, University of Hawaii at Manoa, 2680 Woodlawn Drive,*

*Honolulu, HI 96822*

*kud@ifa.hawaii.edu*

Barbara Lo Faro

*Dipartimento di Astronomia, Università di Padova, Vicolo Osservatorio 2,*

*35122 Padova, Italy*

*barbara.lofaro@gmail.com*

Adalbert W. A. Pauldrach

*Universitäts-Sternwarte München, Scheinerstr. 1, 81679 München, Germany*

*uh10107@usm.uni-muenchen.de*

Max Pettini

*University of Cambridge, Institute of Astronomy, Madingley Rd., Cambridge CB3 0HA,*

*United Kingdom; and International Centre for Radio Astronomy Research, The*

*University of Western Australia, 35 Stirling Highway, Crawley WA 6009, Australia*

*pettini@ast.cam.ac.uk*

and

Samantha A. Rix

*Isaac Newton Group of Telescopes, Apartado de Correos 321, E-38700 Santa Cruz de La*

*Palma, Canary Islands, Spain*

*srix@ing.iac.es*

# Abstract

We computed a comprehensive set of theoretical ultraviolet spectra of hot, massive stars with the radiation-hydrodynamics code WM-Basic. This model atmosphere and spectral synthesis code is optimized for computing the strong P Cygni-type lines originating in the winds of hot stars, which are the strongest features in the ultraviolet spectral region. The computed set is suitable as a spectral library for inclusion in evolutionary synthesis models of star clusters and star-forming galaxies. The chosen stellar parameters cover the upper left Hertzsprung-Russell diagram at $L \gtrsim 10^{2.75} L_\odot$ and $T_{\text{eff}} \gtrsim 20{,}000$ K. The adopted elemental abundances are 0.05 $Z_\odot$, 0.2 $Z_\odot$, 0.4 $Z_\odot$, $Z_\odot$, and 2 $Z_\odot$. The spectra cover the wavelength range from 900 to 3000 Å and have a resolution of 0.4 Å. We compared the theoretical spectra to data of individual hot stars in the Galaxy and the Magellanic Clouds obtained with the International Ultraviolet Explorer (IUE) and Far Ultraviolet Spectroscopic Explorer (FUSE) satellites and found very good agreement. We built a library with the set of spectra and implemented it into the evolutionary synthesis code Starburst99 where it complements and extends the existing empirical library towards lower chemical abundances. Comparison of population synthesis models at solar and near-solar composition demonstrates consistency between synthetic spectra generated with either library. We discuss the potential of the new library for the interpretation of the rest-frame ultraviolet spectra of star-forming galaxies. Properties that can be addressed with the models include ages, initial mass function, and heavy-element abundance. The library can be obtained both individually or as part of the Starburst99 package.

## 1. Introduction

Many of the key diagnostic spectral lines of hot, massive stars are in the satellite ultraviolet (UV) wavelength region from 900 to 3000 Å. This spectral range contains strong telltale signatures of stellar winds, whose most prominent examples are S VI $\lambda$939, O VI $\lambda$1035, P V $\lambda$1123, C III $\lambda$1176, N V $\lambda$1241, O IV $\lambda$1342, O V $\lambda$1371, Si IV $\lambda$1398, C IV $\lambda$1550, and N IV $\lambda$1719 (Pellerin et al. 2002; Walborn et al. 1985, 2002). Weaker photospheric absorption blends of the most abundant ions, most notably Fe II/III/IV/V, introduce strong blanketing at all wavelengths. These wind and photospheric lines hold the key for understanding the fundamental parameters of individual stars and are widely used for determining heavy-element abundances ($Z$), mass-loss rates ($\dot{M}$), wind velocities $v_\infty$, and other physical properties (Puls 2008).

The same spectral lines are also important diagnostics for unresolved stellar populations containing massive, hot stars (Leitherer 2009) and can be used to infer fundamental properties such as ages or the initial mass function (IMF). However, as many different stellar types contribute to the integrated spectrum of an entire galaxy, the properties of the underlying stellar population cannot be inferred from a simple comparison with an individual stellar spectrum but must be determined from evolutionary synthesis modeling. Significant progress in this direction has been made by employing spectral synthesis codes such as Starburst99 (Leitherer et al. 1999; Vázquez & Leitherer 2005; Leitherer & Chen 2009 and references therein). Such codes assume a particular star-formation law and IMF and use stellar evolution models to follow the stellar population over time. By comparing the observed spectrum with ones synthesized using a range of different model parameters, one can then constrain the properties of the underlying young stellar population.

Robert, Leitherer, & Heckman (1993) compiled the original stellar UV library in Starburst99. This library was built from spectra of Galactic O stars observed with the IUE satellite. de Mello, Leitherer, & Heckman (2000) subsequently extended this library to later spectral types and lower masses by adding a set of Galactic B-star spectra from the IUE archive. In order to account for sub-solar heavy-element abundances, Leitherer et al. (2001) generated an O-star library using HST UV spectra of stars in the Large and Small Magellanic Clouds (LMC and SMC, respectively). Finally, we extended the wavelength range to the region shortward of Lyman-$\alpha$ by adding far-UV spectra of hot stars in the Galaxy (Pellerin et al. 2002) as well as in the LMC and SMC (Walborn et al. 2002), all obtained with FUSE.

Comparison of synthetic spectra generated with these libraries using the Starburst99 code to observations of local and distant galaxies generally leads to quite good agreement (e.g., Leitherer 2009). A serious limitation of these synthetic spectra is their restriction to near-solar metallicity. The leverage provided by the Galaxy and the Magellanic Clouds is only modest, and extending the libraries to include empirical spectra of significantly more metal-poor stars outside the Local Group would be prohibitively expensive in terms of telescope time. Furthermore, the S/N and spectral resolution of stellar spectra obtained with currently available UV spectrographs very often lags the quality of rest-frame UV spectra of high-redshift galaxies observed with optical detectors. Rix et al. (2004) chose an alternative approach by utilizing theoretical library spectra rather than empirical ones. In their study, they focused on modeling the strongest photospheric absorption blends near 1425 and 1978 Å, whose metallicity dependence allows an important consistency check with abundances derived from rest-frame optical emission-line spectroscopy. Rix et al. calculated the complete UV spectrum from 900 to 2100 Å but considered the resulting wind lines as too unreliable for comparison with observations. This is rather disappointing, as the wind lines are the strongest spectral features and very often the only stellar lines that can be reliably detected in a spectrum of a star cluster or a galaxy.

Since the work of Rix et al. (2004) was completed, significant progress in the modeling of hot-star winds has been made. Most importantly, inclusion of wind structure and X-ray ionization in the wind provides a much more realistic treatment of the radiative transfer and the associated hydrodynamics in the WM-Basic model atmosphere code (Pauldrach, Hoffmann, & Lennon 2001). This and other progress allowed us to use WM-Basic for a much more realistic modeling of the wind lines in hot stars. Here we discuss the outcome of this effort. In Section 2 we describe the input physics used for the generation of the stellar spectra. An overview of the covered parameter space is in Section 3. Details of computations, including the derived stellar-wind parameters are in Section 4. In Section 5 we compare the theoretical spectra with observations of hot stars in the Galaxy, the LMC, and the SMC. The generation of the library and its implementation in Starburst99 is discussed in Section 6. We test the consistency of the synthetic population spectra generated with the new and the prior, empirical library in Section 7. A parameter study of the spectra of single and mixed populations is performed in Section 8. Finally, we present our conclusions in Section 9.

## 2. Model Atmospheres

Modeling the atmospheres of hot stars poses tremendous challenges due to severe departure from local thermodynamic equilibrium (LTE) because of the intense radiation, low densities, and the presence of supersonic stellar winds initiated by the transfer of momentum from the stellar radiation field to the atmospheric plasma. Fortunately, rapid progress during recent years has led to an astounding degree of sophistication in the latest generation of models (e.g., Puls 2008, and references therein). The main challenges are in the areas of (i) non-LTE, (ii) atomic data, (iii) line-blanketing, and (iv) radiatively driven winds. Several groups have independently developed model atmospheres for OB stars, with a different emphasis on one or more of these aspects. Plane-parallel models, such as ATLAS (Kurucz 2005) or TLUSTY (Lanz & Hubeny 2003, 2007), cannot account for spectral lines forming in the wind and are obviously not suitable for our purpose. The major spherical hot-star model atmospheres are PoW-R (Hamann & Gräfener 2004), PHOENIX (Hauschildt & Baron 1999), CMFGEN (Hillier & Miller 1998), WM-Basic (Pauldrach et al. 2001), and FASTWIND (Puls et al. 2005). PoW-R and PHOENIX are not optimized for the modeling of OB-stars but rather Wolf-Rayet (W-R) and late-type stars, respectively. FASTWIND's main application is the computation of optical and near-infrared H and He lines. On the other hand, CMFGEN and WM-Basic are both widely used for modeling the UV spectra of hot stars and we can build on prior experience. CMFGEN uses an analytical approximation for the wind structure, which must be assumed a priori, whereas WM-Basic solves the radiative transfer and the hydrodynamics self-consistently *to derive* the density structure of the wind.

A cost/benefit analysis of CMFGEN and WM-Basic, including computational resources, led us to choose WM-Basic[2]. WM-Basic is a PC application that calculates unified (photospheric + wind), spherically extended, fully blanketed, non-LTE model atmospheres and can also take into account the wind hydrodynamics self-consistently. A typical run on a Dell Optiplex 755 desktop with an Intel dual-core 3.2 GHz processor running Vista Ultimate SP1 takes less than 10 minutes. Such short run times allowed us to generate thousands of models over the course of the project. A complete model atmosphere calculation consists of three main blocks: (i) the solution of the hydrodynamics; (ii) the solution of the non-LTE model (calculation of the radiation field and the occupation numbers); (iii) the computation of the synthetic spectrum. The three cycles are interdependent and must therefore be solved iteratively. In the first step, the hydrodynamics is solved for a set of effective temperature ($T_{\text{eff}}$), surface gravity (log $g$), stellar radius ($R$), and abundances ($Z$), together with prespecified line force multiplier parameters (LFP) to describe the radiative line acceleration. The continuum force is approximated by the Thomson force, and a constant temperature structure ($T(r) = T_{\text{eff}}$) is assumed in this step. In a second step the hydrodynamics is solved by iterating the complete continuum force (which includes the opacities of all ions up to and including the Fe group elements) and the temperature structure (both are calculated using a spherical grey model), as well as the density and velocity structure. In a final outer iteration cycle, these structures can again be iterated together with the line force obtained from the spherical non-LTE model.

---

[2] The code is freely available and can be downloaded at http://www.usm.uni-muenchen.de/people/adi/Programs/Programs.html.

The main part of the code consists of the solution of the non-LTE model. The radiation field (Eddington flux and mean intensity), final temperature structure, occupation numbers, opacities, and emissivities are computed using detailed atomic models for all important ions. For the solution of the radiative transfer equation, the influence of UV and EUV line blocking is properly taken into account in addition to the standard continuum opacities and source functions. Moreover, the shock source functions produced by radiative cooling zones are also included. Line blanketing, which is a direct consequence of line blocking, is considered for the calculation of the final non-LTE temperature structure via luminosity conservation and the balance of microscopic heating and cooling rates. The last step of the complete cycle consists of the computation of the synthetic spectrum.

We simplified the iteration cycle in this work by applying further constraints, analogous to Rix et al. (2004). The LFPs were calculated outside the main WM-Basic cycle from the theoretical wind momentum versus luminosity relation and its dependence on metallicity (Kudritzki & Puls 2000). Radiative momentum is converted into mechanical momentum with little efficiency variation among hot stars. Therefore a tight relation between $L$ and ($\dot{M}\, v_\infty$) exists. This empirical relation is also followed closely by the hydrodynamically self-consistently computed models. One can thus take advantage of this relation to determine wind parameters and LFPs using previously established scaling relations. This will be discussed in more detail in Section 4. The final products of this hybrid procedure are the synthetic spectra and ionizing fluxes, as well as the hydrodynamic parameters of the wind, i.e., $\dot{M}$ and $v_\infty$.

WM-Basic does not account for the geometrical effects of wind inhomogeneities. There are various direct and indirect indications that hot star winds are not smooth but clumpy, i.e., that there are small-scale density inhomogeneities which redistribute the matter into over-dense clumps and an almost void inter-clump medium (Puls, Markova, & Scuderi 2008). Theoretically, such inhomogeneities have already been expected from the first hydrodynamic wind simulations (Owocki, Castor, & Rybicki 1988) because of the presence of a strong instability inherent to radiative line-driving. This can lead to the development of strong reverse shocks, separating over-dense clumps from fast, low-density wind material. Such density contrasts do not change the strongest UV lines which arise from the dominant ionizing state (e.g., Si IV $\lambda$1398 or C IV $\lambda$1550); however, lines related to trace ions (O VI $\lambda$1035, O V $\lambda$1371) may well be affected by such a wind structure. As a clarification we note that while WM-Basic does not consider the geometrical effects of clumps, it does account for the associated X-ray emission, as we will detail in the following paragraph.

Rix et al. (2004) used an earlier version of WM-Basic to produce a model grid for comparison with observed weak photospheric lines in galaxy spectra. At that time, WM-Basic used a simplified treatment to account for X-ray emission in the wind which often is non-negligible for the modeling of the stellar-wind lines. The sources of the OB star X-ray emission are shocks propagating through the stellar wind (Lucy & White 1980, MacFarlane & Cassinelli 1989), where the shocks result from the previously discussed strong hydrodynamic instability. A simple model assumes a random distribution of shocks in the wind, where the hot shocked gas is collisionally ionized and excited and emits spontaneously into and through an ambient "cool" stellar wind with a kinetic temperature of the order of $T_{eff}$. Such a model provides constraints on shock temperatures, filling factors, and emission measures. Feldmeier et al. (1997) further developed and refined this model, allowing for post-shock cooling zones of radiative and adiabatic shocks. Their theoretical framework has been implemented in WM-Basic and was used for our modeling.

## 3. Parameter Space and H-R Diagram Coverage

We defined a grid of ($L$, $T_{eff}$) covering the extreme luminous, hot part of the Hertzsprung-Russell (H-R) diagram. Our goal was to fill the parameter space relevant for stars contributing to the UV luminosity of a young stellar population. Such stars have zero-age main sequence (ZAMS) masses $\gtrsim$ 5 $M_\odot$ (corresponding to $L \gtrsim 10^{2.75}$ $L_\odot$) and $T_{eff} \gtrsim$ 20,000 K. Stars with lower $L$ and/or $T_{eff}$ do not have significant winds and/or do not have strong wind lines in the UV (Robert et al. 1993; de Mello et al. 2000; Leitherer et al. 2001). Even if strong winds were present, these stars would make a negligible contribution to the UV luminosity of a population unless they were observed in a single stellar burst devoid of massive stars. We were guided by the set of stellar evolutionary tracks with high mass loss released by the Geneva group (Schaller et al. 1992; Schaerer et al. 1993a, 1993b; Charbonnel et al. 1993; Meynet et al. 1994) to choose relevant ($L$, $T_{eff}$) grid points. We followed and interpolated between tracks for solar chemical abundances from the ZAMS until the W-R phase was reached. The derived basic parameters are summarized in Table 1. The table gives the model identifier (col. 1), the current mass (col. 2), log $L$ (col. 3), $T_{\rm eff}$ (col. 4), and $R$ (col. 5). The surface escape velocity $v_{\rm esc}$ in col. 6 has been corrected for continuum radiation pressure by Thompson scattering but not for rotational acceleration. The entries in cols. 7 – 10 give the scaling factors that were applied to the ZAMS surface abundances of the evolution models. The Geneva tracks with solar chemical composition assume $N({\rm H}) = 0.68$, $N({\rm He}) = 0.30$, $N({\rm C}) = 4.863 \times 10^{-3}$, $N({\rm N}) = 1.237 \times 10^{-3}$, and $N({\rm O}) = 1.0537 \times 10^{-2}$ by mass on the ZAMS and list any element variations as they appear in the course of evolution. We adopted their values. A blank in Table 1 indicates the original ZAMS abundances. The factors are multiplicative and denote an enhancement of He, N, O and a decrease of C. For several models the CNO variations are quite substantial, and one would expect this to have dramatic consequences for the wind parameters. However, the effects are quite moderate because of the identity of the wind driving lines. At temperatures below ~25,000 K, most of the driving lines come from Fe-group elements. Only if $T_{\rm eff} \approx 40,000$ K, lines from CNO elements take over (Mokiem et al. 2007). Since any element enhancement of N and O is to a large degree compensated by a decrease of C, the net effect of CNO variations on $\dot{M}$ is quite small, as pointed out by Vink & de Koter (2002). This is rather fortunate because the exact process by which CNO products make their way to the stellar surface is thought to depend on stellar properties such as rotational velocity which may itself depend on $Z$. Extrapolation to very low metallicity environments would otherwise introduce rather large uncertainties in the wind properties.

Finally, col. 11 of Table 1 gives the spectral types corresponding to the position in the H-R diagram. These spectral types were derived using the observational scale of Martins, Schaerer, & Hillier (2005a) for O stars and the B-star scale of Conti, Crowther, & Leitherer (2008).

We supplemented the grid defined by the solar metallicity tracks with ten additional points to the left of the ZAMS. These models have no observational counterparts at solar $Z$ but are useful to bracket the predicted H-R diagram at low metallicity, where stars are hotter and more luminous. These additional models were obtained by extrapolating selected evolutionary tracks to higher $T_{\text{eff}}$. They were added to Table 1 at their appropriate location and are flagged with italics.

In Figure 1 we show the distribution of the grid in the H-R diagram. There is a total of 86 data points in this figure, including the ten models to the left of the ZAMS. Ignoring the latter, one can easily recognize the location of the ZAMS defined by the leftmost open symbol at any given luminosity. The figure shows excellent coverage of the relevant parameter space. Note that stars with $L < 10^4$ $L_\odot$ make little contribution to the UV luminosity, and the least luminous model with a ZAMS mass of 5 $M_\odot$ was primarily included for exploratory reasons. Stars with modified surface abundances are highlighted in the figure. These stars are around the entry phase to becoming W-R stars. Since some of the strong UV P Cygni lines correspond to the elements with modified abundances, the effects on the resulting UV spectra can be substantial. However, the affected stars tend to be rare and short-lived so that the impact on the population spectrum is negligible. Of bigger concern is the omission of W-R stars in our models. We opted for excluding them in this work for two reasons. (i) WM-Basic is not optimized for modeling W-R stars and a separate set of model atmospheres would be needed. (ii) More importantly, stellar evolution models are quite uncertain for late stellar phases. While the overall connections between O and W-R stars are understood, the details of W-R evolution are far from final, and predictions for the line spectrum emitted by a W-R population would be rather speculative. Fortunately, W-R stars in the local universe do not contribute significantly to the UV line spectrum, except for the He II λ1640 emission line, which should therefore be viewed with care.

The stellar parameters of the models having non-solar $Z$ are identical to those for solar $Z$, except for the abundances themselves. Four additional sets with 0.05 $Z_\odot$, 0.2 $Z_\odot$, 0.4 $Z_\odot$, and 2 $Z_\odot$ were generated. The chosen heavy element abundances simply reflect the choice provided by the stellar evolution models. The evolution models with non-solar metallicities are scaled versions of the solar $Z$ models with scaling factors of 0.05, 0.2, 0.4, 2.0 and fixed element ratios.

## 4. Generation of the Model Grid

The computed spectra cover the wavelength range from 900 to 3000 Å at a fixed resolution of ~0.4 Å. This resolution is well-suited for comparison with observations of both local star-forming galaxies and distant star-forming galaxies. From an astrophysical perspective, a velocity resolution of ~100 km s$^{-1}$ at 1400 Å is sufficiently large to resolve all major wind lines, which have widths in excess of 1000 km s$^{-1}$. A related input parameter is the rotation velocity, for which we adopted 100 km s$^{-1}$ on the main-sequence, and 30 km s$^{-1}$ for all stars off the main-sequence and having $T_{\rm eff} < 30{,}000$ K. These values were used at all metallicities.

WM-Basic requires the explicit input of the abundances of elements 1 through 30. We adopted the solar abundances by Asplund, Grevesse, & Sauval (2005) as our reference for $Z_\odot$. These abundances reflect the revised, lower abundances for the Sun. They are essentially identical to those listed by Asplund et al. (2009). It is important to realize that the heavy-element abundances used in the stellar evolution models are not consistent with the abundances of Asplund et al. However, the stellar evolution models are only used to generate a realistic ($L$, $T_{\text{eff}}$) grid. Therefore this inconsistency is not a concern. While the revised solar abundances have little effect on the evolutionary tracks of massive stars, they do affect the computed UV spectra via changed opacities and wind properties.

We followed the approach of Rix et al. (2004) and pre-specified a set of LFPs $k$, $\alpha$, and $\delta$ using previously established scaling relations. LFPs were originally introduced by Castor, Abbott, & Klein (1975) and Abbott (1982) to extrapolate from the line force of a single line to a realistic ensemble of millions of lines. $k$ parameterizes the line opacities in units of the Thompson scattering opacity. $\alpha$ is the ratio of the line force from optically thick lines to the total line force and corresponds to the exponent of the line-strength distribution function. $\delta$ accounts for the change of the line force due to ionization variations in the wind. Kudritzki & Puls (2000) provided a calibration for the relation between the LFPs and the basic stellar parameters via the wind-momentum versus luminosity relation. We adopted their relation, together with updates and modifications by Kudritzki (2002) and Markova et al. (2004). In Table 2 we list the stellar-wind parameters for solar abundances obtained in this way. The table gives the LFPs (cols. 2 – 4), $\dot{M}$ (col. 5), and $v_\infty$ (col. 6) for all 86 models. As a reminder, the lack of clumping in the WM-Basic models may lead to inaccurate mass-loss rates. The wind parameters show the well known strong positive correlations between $\dot{M}$ and $L$ as well as between $v_\infty$ and $v_{esc}$. Table 3 through Table 6 gives the corresponding data for models with 2 $Z_\odot$, 0.2 $Z_\odot$, 0.4 $Z_\odot$, and 0.05 $Z_\odot$, respectively. Both $\dot{M}$ and $v_\infty$ decreases with decreasing $Z$.

We calculated atmospheric structures and synthetic spectra with WM-Basic for each of the 86 models. In Figure 2 we show as an example the computed spectrum for model 37. The corresponding star has $T_{eff}$ = 37,800 K, log $g$ = 3.7, $Z = Z_\odot$, which is equivalent to spectral type O6 III. Prominent spectral lines are identified. Most of these lines are formed in the stellar wind, as can be seen from the blueshifted absorption components with velocities exceeding 2000 km s$^{-1}$. S VI $\lambda$939, O VI $\lambda$1035, P V $\lambda$1123, N V $\lambda$1241, and C IV $\lambda$1550 are the strongest features that are uniformly present in O stars. Si IV $\lambda$1398 is weak in this particular model but can become a strong line in supergiants whose denser winds lead to recombination from $Si^{4+}$ (which is the dominant ionization stage) to $Si^{3+}$ (Walborn & Panek 1985; Drew 1989). Other features present in the spectrum in Figure 2 are C III $\lambda$1176, O IV $\lambda$1342, O V $\lambda$1371, S V $\lambda$1502, He II $\lambda$1640, and N IV $\lambda$1719. While these lines are quite conspicuous in the spectrum of an individual O star, some of these lines are often not detectable in the spectrum of a typical stellar population whose numerous B stars do not show these lines and dilute the O star contribution. In addition to the identified strong spectral lines, blends of photospheric features make the localization of the true continuum challenging. The spectrum shown here has *not* been manually rectified but has been obtained using the theoretical prediction for the continuum location from WM-Basic. We draw the reader's attention to the wavelength region longward of 1800 Å, which is devoid of both stellar-wind and photospheric lines. This region has little diagnostic value for constraining an O-star population.

The O-star spectrum in Figure 2 can be contrasted with the B-star spectrum in Figure 3 which shows model 56 with $T_{\text{eff}} = 15{,}300$ K, $\log g = 1.9$, $Z = Z_\odot$, corresponding to a B3 Ia supergiant. The high-ionization lines present in the O-star spectrum are much weaker or completely absent. The strongest lines are, among others, CIII $\lambda 1176$, C II $\lambda 1335$, Si IV $\lambda 1398$, and Fe III $\lambda 1893$. This star has a terminal wind velocity of $v_\infty = 300$ km s$^{-1}$, resulting in comparatively small blueshifts of the absorption components. The line-blanketing by photospheric absorption lines is significant, in particular at the shortest wavelengths where the *rectified* flux is situated close to the zero level.

The stellar-wind lines are sensitive to the properties of shocks in the outflow. The shocks occur in the denser layers of the winds and grow from small-scale instabilities (Lucy 1982; Owocki et al. 1988; Feldmeier, Puls, & Pauldrach 1997). The standard model incorporated in WM-Basic assumes randomly distributed shocks where the hot shocked gas is collisionally ionized and emits X-ray photons due to spontaneous decay, radiative recombination, and bremsstrahlung. The ambient cool stellar wind can re-absorb part of the emission due to K- and L-shell processes if the corresponding optical depths are large. The X-ray emission ($L_x$) predicted by this simple model scales with $\dot{M}$, and since $\dot{M}$ scales with $L_{\text{Bol}}$, a correlation between $L_x$ and $L_{\text{Bol}}$ is expected (Owocki & Cohen 1999). This correlation is well established observationally (Chlebowski & Garmany 1991; Sana et al. 2006). In the case of O stars a tight relation of $L_x \approx 10^{-7} L_{\text{Bol}}$ with little dispersion is found. B stars have less homogeneous X-ray properties and have on average lower $L_x$ for a fixed $L_{\text{Bol}}$, presumably a result of their much thinner winds. Cassinelli &

Cohen (1994) determined $L_x \approx 10^{-8.5} L_{Bol}$ for a small sample of normal B stars. These empirical relations suggest no additional $Z$ dependence of the X-ray luminosity other than that introduced by the $Z$ dependence of the stellar bolometric luminosity. From first principles, one would expect the natural scaling of the X-ray luminosity to be with the wind density parameter ($\dot{M}/v_\infty$) and not with stellar parameters such as $L_{Bol}$. If so, the metallicity dependence of $\dot{M}$ would lead to a stronger than observed $Z$ dependence of $L_x$. However, if a radial power-law scaling of the filling factor is introduced, then the observed $L_x$ vs. $L_{Bol}$ relation can be understood theoretically (Owocki & Cohen 1999). The scatter in the observed $L_x/L_{Bol}$ relation may indicate the existence of an additional parameter affecting $L_x$. Kudritzki et al. (1996) found a correlation of $L_x$ with the filling factor, which itself correlates with the density parameter ($\dot{M}/v_\infty$). This then turns into the relation $L_x \propto L_{Bol}^{1.34} (\dot{M}/v_\infty)^{-0.38}$. Adopting a metallicity dependence of ($\dot{M}/v_\infty$) $\propto Z^{0.7}$ (Mokiem et al. (2007) leads to a weak scaling relation of $L_x \propto Z^{0.25}$. Because of the weakness of this scaling and the absence of observational verification, we will rely on the simplest empirical relation between $L_x$ and $L_{Bol}$ and assume no additional $Z$ dependence is introduced at $Z \neq Z_\odot$.

An example of the sensitivity of some spectral lines to the X-ray properties of the wind is shown in Figure 4 where we varied $L_x/L_{Bol}$ from $10^{-7.5}$ to $10^{-6.2}$ for model 31. $L_x$ affects different spectral lines in a different manner. As a rule of thumb, lines related to trace ions (O VI, C III) are affected the most. C IV λ1550, which is a prime diagnostic line, displays little variation with $L_x$. While other stellar models behave somewhat differently than model 31, the overall result is a moderate sensitivity to $L_x/L_{Bol}$ across the H-R diagram. Guided by observations, we adopted $L_x = 10^{-6.85} L_{Bol}$ and $L_x = 10^{-8.5} L_{Bol}$ for O and B stars, respectively. We used these relations for all models at all metallicities.

The second shock-related parameter entering the models is $v_{turb}/v_\infty$, the shock velocity relative to the wind terminal velocity. This ratio determines the temperature in the post-shock region. Observational constraints on this parameter come from the blue absorption component of the UV P Cygni lines which suggest a non-monotonic velocity law. This is interpreted as evidence for large turbulent motions related to shocks. The corresponding shock velocities are typically 10% of $v_\infty$ (Groenewegen, Lamers, & Pauldrach 1989; Prinja, Barlow, & Howarth 1990). The result of varying $v_{turb}/v_\infty$ is shown in Figure 5 where we have plotted the spectrum for model 31 using three values of $v_{turb}/v_\infty$ = 0.08, 0.1, and 0.2. The results are similar to those in the previous figure: the principal diagnostic lines of N V λ1241 and C IV λ1550 show no significant change, whereas lines related to trace ions (e.g., Si IV λ1398) vary with shock velocity and post-shock temperature. We used $v_{turb}/v_\infty = 0.1$ for all models, as suggested by observations.

The third parameter entering our modeling is the run of the shock temperature, which is parameterized in terms of a power-law exponent γ. γ couples the jump velocity of the shock to the outflow velocity (see Pauldrach et al. 2001). The parameter study for γ is in Figure 6, which is again consistent with the prior results. In the present work we adopted γ = 0.5, as found by Pauldrach et al. from detailed line fitting for several well observed O stars. A fourth parameter $m$, specified as the ratio of the local wind outflow velocity to the sound velocity, describes the inner boundary of the shocks. We found the wind lines to be quite insensitive to $m$ and do not show the test spectra. $m = 10$ was used for all models.

## 5. Comparison with IUE and FUSE Observations

The synthetic spectra computed with WM-Basic have been extensively tested and compared with observations (e.g., Pauldrach 2003). Excellent agreement has been found once the uncertainties of the stellar parameters are taken into account. Although redoing such tests for our newly generated model grid may sometimes only reinforce previous results, we nevertheless performed a set of comparisons with available data for Galactic, LMC, and SMC stars covering the relevant wavelength region.

For a test of the models at wavelengths longer than 1150 Å we chose available data collected with the IUE satellite. The O- and B-star atlases of Walborn et al. (1985, 1995) are accompanied by the fully reduced data sets that are publicly available. We retrieved the full set of 186 OB spectra. The data are already continuum normalized and have a spectral resolution of 0.25 Å. In order to match the data to the models, we rebinned the data to a spectral resolution of 0.4 Å. We used the spectral types assigned to our model spectra (see Table 1) to identify three observational spectra with closely matching spectral types. In most, but not all, cases the match was exact. We then averaged the three observed spectra to generate a mean observational template with good S/N for comparison with the models. This was done for all 86 models at solar chemical composition and for a few sub-solar models with LMC/SMC counterparts in the atlas. The latter test, however, is not very meaningful because of the dearth of such spectra in the atlas and their types, which are affected by observational bias focusing on the most peculiar stars. In the following we restrict our discussion to Galactic stars in representative evolutionary phases covering the extremes of the H-R diagram.

We begin with the model triplet 21, 22, and 24. These three models are an evolutionary sequence of a very luminous star with an initial mass of 80 M$_\odot$ that is evolving from the main sequence as spectral type O3 I to O5 I and then becoming a B1.5 Ia star. The stellar parameters can be found in Table 1. The comparison with the observations is in Figure 7, Figure 8, and Figure 9 for models 21, 22, and 24, respectively. A summary of the parameters of the observed spectra is in Table 7, which lists the spectral type of each model (col. 1), the HD numbers of the stars (cols. 2 – 4), and the average log $L$ and $T_{\text{eff}}$ for each of the three named stars (cols. 5 and 6). Model 21 and the O3 I spectrum agree quite well (Figure 7). It should be kept in mind for this and the following comparisons that this is not an optimized fit. Had we adjusted the parameters of the model to match those of the observations, the agreement would much improve. Of course, this is not our goal because we wish to generate a fixed grid of library stars for use in population synthesis models. The numerous narrow absorption lines in the observations (e.g., at 1526 Å) are of interstellar origin and obviously not present in the models. The same applies to Lyman-α. The observed spectrum at the shortest wavelengths in Figure 7 is dominated by noise and should not be compared to the model. *O V λ1371 is the only strong line always showing significant disagreement.* This line is weaker in the observations and even an adjustment of the stellar parameters would not result in agreement without leading to disagreement for other lines. Bouret, Lanz, & Hillier (2005) were able to significantly improve the fit to the O V λ1371 line by introducing clumping in their wind models. A clumpy medium shifts the ionization equilibrium of trace ions towards lower ionization stages due to enhanced recombination. Since WM-Basic does not account for this effect, O V λ1371 must therefore be considered uncertain. In addition, in some models the strength of one or the other wind line is an overestimate, implying a higher wind optical depth in the models than in the observed spectrum. These overestimates may again arise from using unclumped wind models which tend to overestimate $\dot{M}$ and the ionization state of the wind. In contrast, $v_\infty$ always appears to be well matched by the models, and so are the shapes of the saturated wind lines, which depend on the velocity law of the wind.

The comparison between model 22 and the O5 I spectrum is in Figure 8. As in the previous figure, the agreement is quite good. The star has about the same luminosity as in the previous case but is 4,000 K cooler. As a result of the lower $T_{\text{eff}}$, the O V $\lambda$1371 line in the model is much weaker and in better agreement with the observations. Note Si IV $\lambda$1398, which is stronger than in model 21 because of the lower $T_{\text{eff}}$.

The final evolutionary point for the 80 $M_\odot$ star considered here corresponds to spectral type B1.5 Ia. The comparison between model 24 and the data is in Figure 9. The spectrum is markedly cooler and displays lower ionization stages. The lower wind velocities connected with the lower surface escape velocities result in narrow lines, which are in good agreement with the observations. Overall, the comparison gives confidence in the models and suggests that they successfully reproduce the UV spectra from the hottest O- to early B stars with very high luminosity.

The next model triplet 35, 37, and 39 follows a star with initial mass 50 $M_\odot$ from the early main-sequence (O3 V) through the early post-main-sequence (O6 III) to the evolved state as B1 Ia. The sequence is most relevant for the purposes of a library because a 50 $M_\odot$ star is often considered representative of the massive star population in a giant H II region. The outcome of the comparison between models and observations is reproduced in Figure 10, Figure 11, and Figure 12, which echo the results and conclusions of the previous evolutionary sequence. O V λ1371 in model 35 is still a slight mismatch to the observations but agrees better than in the case of the O3 I star in Figure 7. We note that the generally strongest and most easily observed lines of N V λ1241 and C IV λ1550 agree best with the observations.

The final comparison between the models and the IUE data shown here is for a relatively low-mass star with an initial mass of 30 $M_\odot$. The evolutionary sequence has spectral types O6.5 V (model 51), O8.5 III (model 53), and B0.5 I (model 55). Figure 13, Figure 14, and Figure 15 show the modeled and observed spectra. The generally lower luminosity and temperature lead to increased line blanketing, which is correctly accounted for in the models. All major wind lines are in good agreement.

Testing our models at wavelengths below 1150 Å can be done using spectra obtained with the FUSE satellite. The data presented by Pellerin et al. (2002) for Galactic OB stars and by Walborn et al. (2002) for LMC and SMC stars are available electronically and were used by us for this purpose. Since the data were fully reduced and continuum normalized, no further processing except for rebinning to a spectral resolution of 0.4 Å was done.

Severe contamination by the interstellar Lyman and Werner bands of $H_2$ renders the FUSE spectra of Galactic stars essentially useless for a model comparison. Taresch et al. (1997) combined WM-Basic with spectral synthesis models for the molecular and atomic/ionic interstellar lines to disentangle stellar and interstellar blends. This approach, however, is beyond the scope of the present paper, and the FUSE data for the Galactic stars will not be further discussed here. LMC and SMC stars have lower extinction and weaker $H_2$ bands so that their FUSE spectra can provide some guidance for the models. Yet, even for these stars, $H_2$ contamination is significant and needs to be taken into account. The work of Walborn et al. (2002) includes 28 OB stars in each of the LMC and the SMC. The corresponding H-R coverage is rather sparse and suitable observational counterparts could be found for only ~45% of the models. In the following we will discuss representative cases for different stellar parameters in both the LMC and the SMC. We will compare the LMC and SMC spectra to models with $Z = 0.4\ Z_\odot$ and $Z = 0.2\ Z_\odot$, respectively, but note that on average massive SMC stars tend to be somewhat more metal-rich than our $0.2\ Z_\odot$ models.

In Figure 16 we compare model 18 with $Z = 0.4\ Z_\odot$ to the LMC star Sk−67D211. This is one of the hottest models with an equivalent spectral type of O2 I, which corresponds reasonably well to the O2 III (f*) classification of Sk−67D211. Almost all the absorption lines in the observations that disagree with the model are interstellar $H_2$ bands, interstellar Lyman lines, or interstellar low-ionization metal lines. Walborn et al. (2002; their Figure 17) provide identifications of all interstellar features in the FUSE wavelength region. The dominant spectral line is the P Cygni profile of O VI λ1035, which is well reproduced by this model. The other strong wind-line is the S VI λ939 doublet, whose longer-wavelength component agrees quite well with the data. The other component at 933 Å is strongly blended with the Lyman lines close to the series limit and is not useful for a comparison.

The next comparison at LMC chemical composition was made for a late O star (see Figure 17). We chose model 53 and the LMC star Sk−67D101 whose spectral type is O8 II((f)). In addition to the previously mentioned lines, the C III feature at 1176 Å appears at this temperature. Model and observations are in reasonable agreement. The final comparison for the LMC is between model 6 and Sk−66D41 in Figure 18. Sk−67D41 (spectral type B0.5 Ia) is the closest match for model 6 in the FUSE sample. While the $T_{eff}$ for its spectral type agrees well with that of model 6, its luminosity is lower by a factor of ~4. This may be the reason for mismatch around 1120 Å where the spectral lines of Si III λ1113, P V λ1123, and Si IV λ1125 are located. Most other spectral lines, in particular the important C III λ1176 are in good agreement.

As we did for the LMC, we will discuss comparisons between models and observations for the SMC using a very hot, an intermediate temperature, and a cool early-type star. We begin with model 2, whose observational counterpart is the SMC star NGC 346-3 with a spectral type of O2 III(f*). The comparison in Figure 19 suggests good agreement. The prominent wind doublets of S VI λ939 and O VI λ1035 agree well when taking into account interstellar contamination in the observations. Note in particular Lyman-β at the blue edge of the O VI line. As we mentioned before, the heavy-element abundance of the $Z = 0.2\ Z_\odot$ series is lower than observed in massive SMC stars. As a result, all spectral features in the model spectrum are somewhat weaker than in the observations.

The comparison between the spectrum of model 62 and of the SMC star AV 47 is reproduced in Figure 20. AV 47 has a spectral type of O8 III (f)) and stellar parameters that are a good match to the model. O VI λ1035 is strongly blended with interstellar Lyman-β and C II λ1036 but otherwise the agreement is reasonable. C III λ1176 shows excellent agreement. The final comparison is for the B0.5 Iaw star AV 488 and model 6 in Figure 21. The arguments given during the previous discussion of the LMC B supergiant Sk−66D41 apply here as well.

The comparison between the set of model spectra and observations of Galactic as well as LMC and SMC stars suggests no major significant *systematic* disagreement for the major diagnostic lines, except for OV λ1371, which tends to be stronger in the models than in the observations. Random deviations of individual lines can generally be attributed to mismatches of stellar parameters and could be improved or even eliminated if desired. We conclude that the calculated set of spectra reliably reproduces the spectra of luminous OB stars and is suitable as a library for spectral synthesis.

## 6. Implementation in Starburst99

We combined the 86 WM-Basic spectra for the five chemical compositions to generate ten electronic files containing the fully blanketed spectra as well as the continuum fluxes. This follows the approach we took for the generation of the optical high-resolution library in Starburst99 (González Delgado et al. 2005; Martins et al. 2005b). Each spectrum is uniquely tied to a position in the H-R diagram by its log $L$, $T_{\text{eff}}$, and log $g$. The assignment of a particular spectrum is done via a nearest neighbor search in stellar parameters. While a full two-dimensional interpolation might seem preferable, the gain in precision is miniscule, but the increase in computing time can be an issue on less powerful computers. We experimented with several schemes in our algorithm: linear versus logarithmic, different weighting factor in x- and y direction, and search in log $L$ versus log $g$. In the end we opted for a nearest neighbor search in log $T_{\text{eff}}$ and log $g$.

Since the WM-Basic library is restricted to $L \gtrsim 10^{2.75}\ L_\odot$ and $T_{\rm eff} \gtrsim 20{,}000$ K (see Figure 1), a switch to a different, existing library in Starburst99 must be performed at the boundaries. Here we transition to the standard Kurucz library whose continuum fluxes are in excellent agreement with the WM-Basic fluxes at the boundary. However, the Kurucz library as implemented in Starburst99 has a spectral resolution of only 10 Å in the UV. Therefore most of the spectral-line information is lost once the transition from WM-Basic to Kurucz has occurred. For a typical single stellar population, the loss of spectral resolution becomes noticeable in Starburst99 at an age of about 50 Myr after the onset of star formation. On the other hand, an equilibrium population, i.e., a population forming stars continuously and having equal numbers of stellar birth and death, will always be dominated in the UV by the stellar types represented by the WM-Basic library. This case is relevant for the interpretation of the spectra of star-forming galaxies, whose integrated UV spectra are usually modeled under the assumption of continuous star formation.

The five metallicity values of the WM-Basic library match those of the Geneva evolution models in Starburst99. Therefore we linked the WM-Basic spectra for each $Z$ value to the corresponding set of Geneva tracks. The Padova tracks in Starburst99 have chemical compositions of 0.02 $Z_\odot$, 0.2 $Z_\odot$, 0.4 $Z_\odot$, $Z_\odot$, and 2.5 $Z_\odot$, which is somewhat different from the Geneva values at the lowest and highest metallicities but still acceptably close. Therefore we followed the same approach for the Padova tracks as we did for the Geneva ones. We note that this mirrors the method we used for the implementation of the optical high-resolution library of Martins et al. (2005b). The UV and the optical libraries in Starburst99 are fully compatible and allow a consistent spectral synthesis from 900 to 7000 Å at resolutions of 0.4 Å and 0.3 Å in the UV and optical, respectively.

## 7. Comparison of Synthetic Spectra from Empirical and Theoretical Libraries

As a first test of the new WM-Basic library, we compared synthetic population spectra generated by Starburst99 both with the new, theoretical library and with the existing, empirical UV libraries. The empirical libraries in Starburst99 are based on observations obtained with three telescopes: The UV library of hot stars with solar chemical composition longward of Lyman-α uses IUE spectra (Robert et al. 1993; de Mello et al. 2000). The wavelength coverage is 1200 – 1800 Å at a spectral resolution of 0.75 Å. The library at sub-solar Z is comprised of a mix of LMC and SMC stars. The spectra were taken with HST's FOS and STIS spectrographs, cover the range 1200 – 1600 Å, and have a spectral resolution of 0.75 Å (Leitherer et al. 2001). Spectra in the 1000 – 1200 Å regime were obtained with the FUSE satellite to build the far-UV library in Starburst99 by Robert et al. (2003). There are two separate libraries, one for solar and another one for sub-solar chemical composition. As with the HST library, the sub-solar FUSE library is a hybrid generated with LMC and SMC stars. The FUSE libraries have a spectral resolution of 0.12 Å.

We generated a model series for a standard single stellar population, i.e., an instantaneous burst of star formation following a Salpeter IMF with mass limits of 1 and 100 $M_\odot$ and following the Geneva evolutionary tracks at high mass loss. Since our goal is to identify similarities and differences between the two sets of libraries, we will restrict our discussion to instantaneous burst models, rather than models with continuous star formation whose spectra exhibit more pronounced degeneracies and are less useful for our immediate purpose.

In Figure 22 we show the comparison with the IUE library at $Z_\odot$. The spectral resolution of the WM-Basic library was decreased by a factor of two to match that of the IUE library. Plotted are spectra covering ages from 1 to 20 Myr at nine time steps. 20 Myr is the approximate evolutionary time-scale of a 10 $M_\odot$ star. Both spectral libraries are complete at (and beyond) this age. Evidently the overall agreement between the spectra generated with the two sets of libraries is excellent for all ages shown in this figure. Even though the strengths of some unsaturated wind lines in individual stellar spectra may not fully agree with the observations (see Section 5), these stellar spectra contribute little to the integrated light of a large stellar population. Owing to the stellar lifetimes and the weighting by the IMF, massive luminous stars with strong winds and saturated lines are far more important contributors to the population spectrum.

The age evolution of the population is most clearly reflected in the weakening of the N V $\lambda$1241 and C IV $\lambda$1550 P Cygni profiles with time, which results from the decreasing O-star contribution. Si IV $\lambda$1398 exhibits the well-known luminosity effect and is strongest when massive O supergiants appear between 3 and 5 Myr. N IV $\lambda$1719 behaves similarly. Predictably, O V $\lambda$1371 is present in the 1 Myr old model generated with WM-Basic but absent in the empirical spectrum. This is the only significant discrepancy. The O V line disappears after ~2 Myr when the hottest, massive stars have evolved to somewhat cooler temperatures. Numerous narrow absorption lines in the empirical spectra have interstellar origin. Obviously such contamination is absent in the theoretical spectra, thereby greatly facilitating automatic spectral fitting procedures that utilize the full spectrum. We draw attention to Si IV whose interstellar components contribute significantly to the total profile and introduce a major uncertainty in the interpretation of this line. Lyman-α has strong damping wings in the empirical data due to Galactic H I absorption. This affects the blue edge of the N V P Cygni profile. Again, the library generated with WM-Basic does not have this issue. Finally, we point out a shortcoming in the normalization of the empirical spectra around 1450 Å. Severe blanketing make the continuum determination in this wavelength region a challenge and causes the incorrect continuum location in the IUE data.

The comparison between the WM-Basic and the empirical spectra with $Z_\odot$ in the wavelength region 1000 – 1200 Å is in Figure 23. The FUSE spectra used in the empirical library were degraded in spectral resolution by a factor of four to match the resolution of the WM-Basic library. Since the empirical spectra are heavily contaminated by interstellar absorption lines, few conclusions can be drawn from this figure. The strongest and most discrepant lines, such as the one at 1050 Å, are due to the $H_2$ Werner bands. O VI λ1035 is the most prominent spectral feature in both the theoretical and empirical spectra. The agreement between models and observations is quite good for all ages considered. At the youngest ages, a strong P Cygni profile appears in the models. The empirical spectra lack the emission component, which is largely the result of the underlying interstellar absorption. The feature at ~1037 Å is mostly $H_2$, with some contribution from C II λ1037. C III λ1176 is the only strong spectral line not contaminated by interstellar lines. The empirical and theoretical spectra are in good agreement for this line.

Starburst99 uses a library composed of LMC and SMC stars to synthesize empirical spectra of metal-poor populations. In Figure 24 we compare WM-Basic spectra with $Z = 0.4\ Z_\odot$ to these empirical spectra. The WM-Basic spectra were degraded in resolution by a factor of 2. Since there are few B stars in the empirical library, the comparison at ages older than ~15 Myr should be taken with a grain of salt. As for the solar metallicity case, the agreement between the theoretical and empirical spectra is rather encouraging. The strongest wind features N V $\lambda$1241 and C IV $\lambda$1550 agree except at a very early age when the empirical C IV has stronger absorption at low velocities. O V $\lambda$1371 is too strong in the models as well. We caution that model deficiencies may not be the only reason for the disagreement. Massive, hot stars spend the initial 10 – 20% of their life-time in optically obscured ultra-compact H II regions where they are not accessible to observations in the optical (Wood & Churchwell 1989). This is reflected in the composition of the empirical library which is deficient in very hot stars close to the ZAMS. In contrast, the theoretical library includes such stars and will produce their spectral signatures in the population spectrum. One might therefore argue that the first 1 – 2 Myr in the evolution of a single stellar population are not observable and the comparison in Figure 24 (and the other corresponding figures) is not relevant. Nevertheless, there are reasons to expect the O V line to be unreliable in the models because of our neglect of wind clumping (see Sections 2 and 5).

Our final comparison series is shown in Figure 25, where the wavelength region from 1000 to 1200 Å is considered. The theoretical spectra are the same as those in Figure 24, except for the resolution, which was left unchanged at 0.4 Å. The empirical spectra are again based on FUSE (Walborn et al. 2002) and were degraded in resolution by a factor of four. The FUSE library is incomplete for spectral type late-O and later, as is noticeable in the absence of spectral lines after ~7 Myr (see Figure 25). The figure is in agreement with the results of the earlier figures. The O VI λ1035 line is reasonably well reproduced, and so is the C III λ1176 line.

The comparison of synthetic population spectra generated with the new WM-Basic library on the one hand and with the existing library on the other suggests rather good agreement at solar and mildly sub-solar metallicity. The only significant exception is the O V λ1371 line, which is stronger in the models at ages younger than ~2 Myr. While no such comparisons can be done at other metallicities, the agreement found with the current empirical libraries gives confidence in our extrapolation to more extreme chemical compositions.

## 8. Parameter Study of Single and Mixed Populations

In this section we identify and discuss the most relevant properties of the stellar-population spectra calculated with the new theoretical library. We focus on the behavior with chemical abundances and ignore parameters like the IMF or the star-formation history since they were addressed before (e.g. Leitherer, Robert, & Heckman 1995; Leitherer et al. 2001), and the new library does not significantly affect the previously derived conclusions. Smith, Norris, & Crowther (2002) released a set of SEDs for hot, massive populations generated with WM-Basic and CMFGEN for O- and W-R stars, respectively. Except for the W-R contribution and the lower spectral resolution of ~10 Å in the non-ionizing satellite-UV, the spectral library of Smith et al. produces synthetic spectra which are fully consistent with those discussed here. In particular, the predicted UV luminosities are identical.

We generated sets of synthetic spectra for the five available chemical compositions. Two standard stellar populations were considered: a single stellar population with ages between 0 and 20 Myr and a population with constant star formation with ages between 0 and 50 Myr. The latter population is in quasi-equilibrium at the upper age. A Salpeter IMF between 1 and 100 $M_\odot$ was adopted. In Figure 26 and Figure 27 we show the computed spectra at wavelengths below and above 1950 Å, respectively. The split in two wavelength regions was chosen purely for the purpose of optimizing the display. In each figure we have plotted spectra at representative time steps for $Z =$ 0.05 $Z_\odot$ and 2 $Z_\odot$. These two metallicities bracket the range of chemical abundances considered. Below 1950 Å (Figure 26), the previously discussed strong wind lines of O VI $\lambda$1035, N V $\lambda$1241, Si IV $\lambda$1398, C IV $\lambda$1550, and N IV $\lambda$1719 dominate the spectra at an early age and fade with the disappearance of the O stars and their powerful winds. Si IV, C IV, and N IV have a very strong $Z$ dependence and become rather inconspicuous at $Z = 0.05$ $Z_\odot$. In contrast, O VI and N V are still strong at the lowest $Z$ and could be detectable even in an observed galaxy spectrum of moderate quality at that metallicity.

The wavelength region longward of 1950 Å (Figure 27) is devoid of strong wind lines even at the highest metallicity. The only significant spectral feature is C III $\lambda$2298 which appears when intermediate O stars contribute. O stars have comparatively few spectral lines in this wavelength region. At later ages, when B stars become important, photospheric and wind features originating from singly and doubly ionized stages will appear.

The spectra generated for the case of continuous star formation are reproduced in Figure 28. As we did for the instantaneous models, we are plotting the lowest and highest metallicities to bracket the complete $Z$ range. The model age is 50 Myr, at which time an equilibrium has been reached. The main spectral features are similar to those seen in the instantaneous burst models with ages of a few Myr. The spectra in Figure 28 are appropriate for observations of large stellar assemblies, such as star-forming galaxies, for which causality arguments suggest quasi-continuous star formation (Rix et al. 2004). The four strongest features predicted for the UV spectra of galaxies with active star formation are O VI $\lambda$1035, N V $\lambda$1241, Si IV $\lambda$1398, and C IV $\lambda$1550. The O VI line is less useful when observed at cosmological redshifts because it is compromised by the Lyman-$\alpha$ forest. Zoomed reproductions of these four stellar-wind lines at all metallicities are in Figure 29 through Figure 32. C IV (Figure 29) is the strongest feature. The absorption component is strongly sensitive to $Z$, whereas the emission shows less variation with metallicity. The blue absorption edge is blended with Si II $\lambda$1527 and $\lambda$1533 which originate in B stars. Since Si II $\lambda$1527 is a resonance transition, it usually has a strong interstellar component as well. The behavior of N V (Figure 30) is qualitatively similar to that of C IV. When compared to observations, the presence of strong interstellar Lyman-$\alpha$ needs to be taken into account. A strong damped Lyman-$\alpha$ wing may affect the blue edge of the absorption component. O VI (Figure 31) shows a much weaker dependence on Z. Except for the $Z = 0.05$ $Z_\odot$ model, the profiles are relatively similar. The behavior of the O VI line results from the counteracting effects of chemical composition and stellar temperature on the $O^{5+}$ column density. The dominant ionization stage in O star winds is $O^{3+}$, and the mean ionization fraction of $O^{5+}$ increases monotonically with stellar temperature. Therefore, to first order, $O^{5+}$ column density becomes independent of the stellar mass loss because the ionization fraction and the total oxygen column density have the opposite dependence on $\dot{M}$ (via the $\dot{M}$ vs. $Z$ relation). We caution that Lyman-β can be a strong interstellar line, which would affect the blue absorption wing of O VI. Si IV (Figure 32) is strongly sensitive to metallicity but the line is predicted to be rather weak for any chemical composition. Unless data with excellent S/N are available, this line will be difficult to detect in galaxy spectra. Si IV is strong only during a relatively short period when luminous hot supergiants appear. When observed in a singular burst at the right epoch, e.g., at ages around 3 – 4 Myr, the Si IV line may in fact become a rather conspicuous feature. This is the case in the spectra of individual star clusters. The interstellar contribution to Si IV further compromises the virtue of this line as a stellar population tracer. The relative strength of the interstellar relative to the stellar part of Si IV is greater than in the other spectral features discussed here because of the lower ionization energy required for $Si^{3+}$ (33 eV) compared with $C^{3+}$ (49 eV) or $N^{4+}$ (77 eV). These values work in favor of an interstellar contribution in Si IV and a stellar contribution in C IV and N V.

Smith et al. (2002) and Sternberg, Hoffmann, & Pauldrach (2003) published grids of ionizing fluxes for hot, massive stars using WM-Basic models, however without the inclusion of X-rays. Since X-rays only affect the spectral energy distributions in the extreme UV shortward of the He II edge in a direct way (and indirectly modify some spectral lines in the optical), our results for the hydrogen Lyman continuum are for the most part identical to those of Smith et al. and Sternberg et al. Therefore will not discuss the ionizing fluxes predicted by our models. It is, however, of interest to discuss the properties of the Lyman-break since our high-resolution spectra provide insight into the effects of line-blanketing around 912 Å. Knowing the intrinsic Lyman discontinuity is essential for estimating the escape fraction of H-ionizing photons from star-forming galaxies, a parameter which in turn is crucial for quantitative assessments of the photon budget during the epoch of reionization (McCandliss et al. 2009). Spectra for the wavelength region around the Lyman-break are shown in Figure 33. The adopted parameters are the same as before for the continuous star formation models. While the ionizing fluxes just below the Lyman edge vary by ~0.2 dex over the full $Z$ range, the Lyman discontinuity is essentially independent of $Z$ because the effects of blanketing are very similar above and below the Lyman break. Independent of $Z$, the Lyman discontinuity is 0.4 – 0.5 dex for the chosen stellar population.

## 9. Conclusions

Model atmospheres of hot, massive stars have reached a state of sophistication to allow the computation of fully synthetic high-resolution spectra in the space-UV. This wavelength range contains some of the major diagnostics of hot stars and young stellar populations. Theoretical models can complement empirical spectra in stellar libraries, in particular in the UV where observations are a limited resource and often become unfeasible. Rix et al. (2004) pioneered this approach by synthesizing the photospheric absorption lines seen in the spectra of star-forming galaxies with WM-Basic models. The method allowed Rix et al. to determine metal abundances in two galaxies and has since been applied in several other cases (e.g., Halliday et al. 2008). At that time, the models had to be limited to stellar photospheric lines because of the yet unexplored effects of shocks on the emergent wind line profiles. Our new work extends the study of Rix et al. to include strong and weak stellar-wind features in the UV from 900 to 3000 Å by taking into account shock heating. This is a major advance since stellar-wind features are the strongest lines in the UV and can by observed in galaxy spectra having much lower S/N than that required for the detection of faint photospheric features. Since the shocks form in the wind above the stellar photosphere, the photospheric absorption lines are little affected, and our results for the photospheric lines are quite similar to those of Rix et al.

The strongest wind lines are O VI $\lambda$1035, C III $\lambda$1176, N V $\lambda$1241, Si IV $\lambda$1398, C IV $\lambda$1550, and N IV $\lambda$1719. Several of these lines are readily observed in the restframe-UV spectra of local star-forming galaxies and of distant star-forming galaxies obtained from space or from the ground, respectively. We show that the modeled lines agree rather well with observations of individual stars when such observations are available. Owing to their strength and wavelength, N V $\lambda$1241, Si IV $\lambda$1398, and C IV $\lambda$1550 are the key diagnostic lines in synthetic spectra of standard stellar populations. These lines correlate with age and metallicity and display a behavior that can be readily understood in terms of the properties of radiatively driven hot-star winds. This makes them excellent diagnostics for stellar population studies at low and high redshift. Their large equivalent widths of several Å and line widths of ~$10^3$ km s$^{-1}$ permit comparisons even with data of relatively low S/N and spectral resolution.

While the new models can be universally applied to a wide range of data, users should be aware of some caveats. Although the models populate the upper H-R diagram quite densely, resource limitations require us to truncate the coverage at $L \approx 10^{2.75}\ L_\odot$ and $T_{\text{eff}} \approx 20{,}000$ K. Therefore the models cover the UV spectra of single stellar populations only up to ages of ~20 Myr. Equilibrium populations, as applicable to galaxy spectra, are not affected by this limitation as stars of ages less than 20 Myr will always dominate the UV in this case. Another caveat refers to the exclusion of W-R stars, which are not accounted for because WM-Basic is not optimized for modeling their winds. Test calculations show that W-R stars do not contribute significantly to the UV continuum and are responsible only for the He II λ1640 line unless the IMF is unrealistically biased towards extremely massive stars. Therefore the omission of W-R stars is not a major concern unless there is specific interest in the He II λ1640 line.

Our models are assumed to have a homogenous density distribution. This assumption is not fully met. As a result, spectral lines associated with trace ions calculated in our models do sometimes not agree with the observations. The most discrepant example (and the only line strong enough to be of practical concern) is O V λ1371, which is too strong in the models. The underlying physical cause is understood and efforts are underway to remedy the situation in the next generation of models.

The current suite of models covers metal abundances of 0.05 $Z_\odot$, 0.2 $Z_\odot$, 0.4 $Z_\odot$, $Z_\odot$, and 2 $Z_\odot$. The choice of these values is dictated by the heavy-element abundances of the available stellar evolutionary tracks. These metallicities cover the range observed in galaxies in the local and distant universe. However, there is interest in pushing the limits to lower values in anticipation of the discovery of very metal-poor primordial galaxies. Evolutionary models for stars forming in these environments are currently under development (S. Ekström, private communication). Once these models become available, we will study the UV spectra resulting from such low metal abundance. We also plan to address individual element variations, such as an enhanced α/Fe ratio, on the UV line profiles. Such variations are expected in extreme starburst environments and have been suggested as the interpretation of the behavior of the Si IV and C IV lines in some star-forming galaxies. Once self-consistent synthetic model spectra are at hand, such claims will stand on a much more solid footing.

The new spectral library is part of the Starburst99 package, which can be downloaded at http://www.stsci.edu/science/starburst/PCStarburst99.html. Currently, the library is fully integrated into the PC version of Starburst99. The full package, including the installer is at http://www.stsci.edu/science/starburst/Starburst99v600.zip. By downloading and installing the package users can extend the parameter study performed in the present work. In particular, variations of the IMF have not been addressed here but may be of interest for some applications.

*Acknowledgments.* Paula Ortiz wishes to thank STScI and the Space Astronomy Summer Program 2009 for their hospitality and support. A. Pauldrach's research was supported by the Deutsche Forschungsgemeinschaft under grant Pa 477/9-1.

# Figures

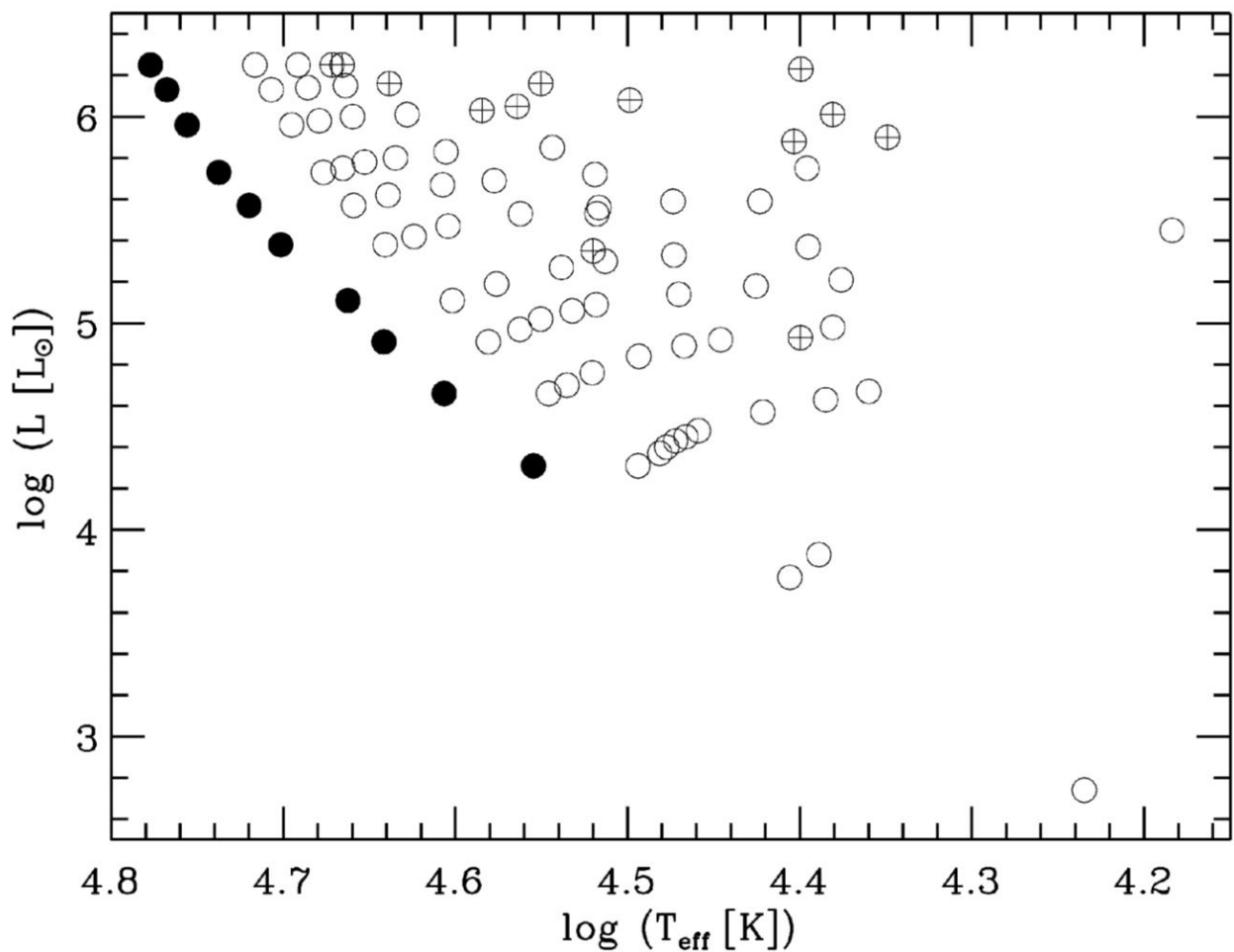

Figure 1. — Location of models in the H-R diagram. ○: models with $L$ and $T_{eff}$ taken from stellar evolution models; ●: additional models accounting for the ZAMS shift at lower Z; ⊕: models with modified He, C, N, O abundances from stellar evolutionary tracks.

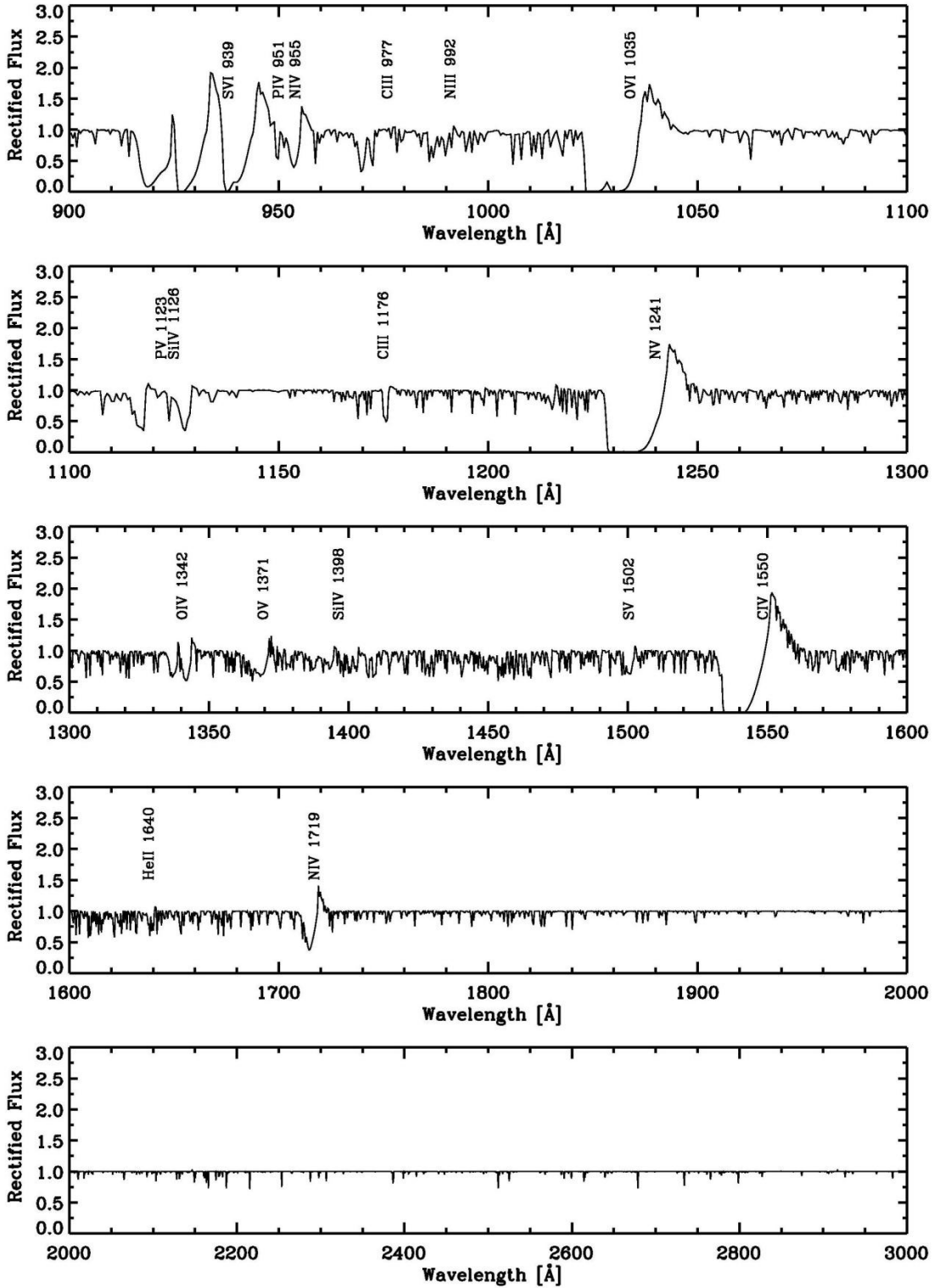

Figure 2. — Calculated UV spectrum for model 37 ($T_{\text{eff}} = 37{,}800$ K; $\log g = 3.7$; $Z = Z_\odot$). The corresponding spectral type is O6 III. The most important spectral lines are identified.

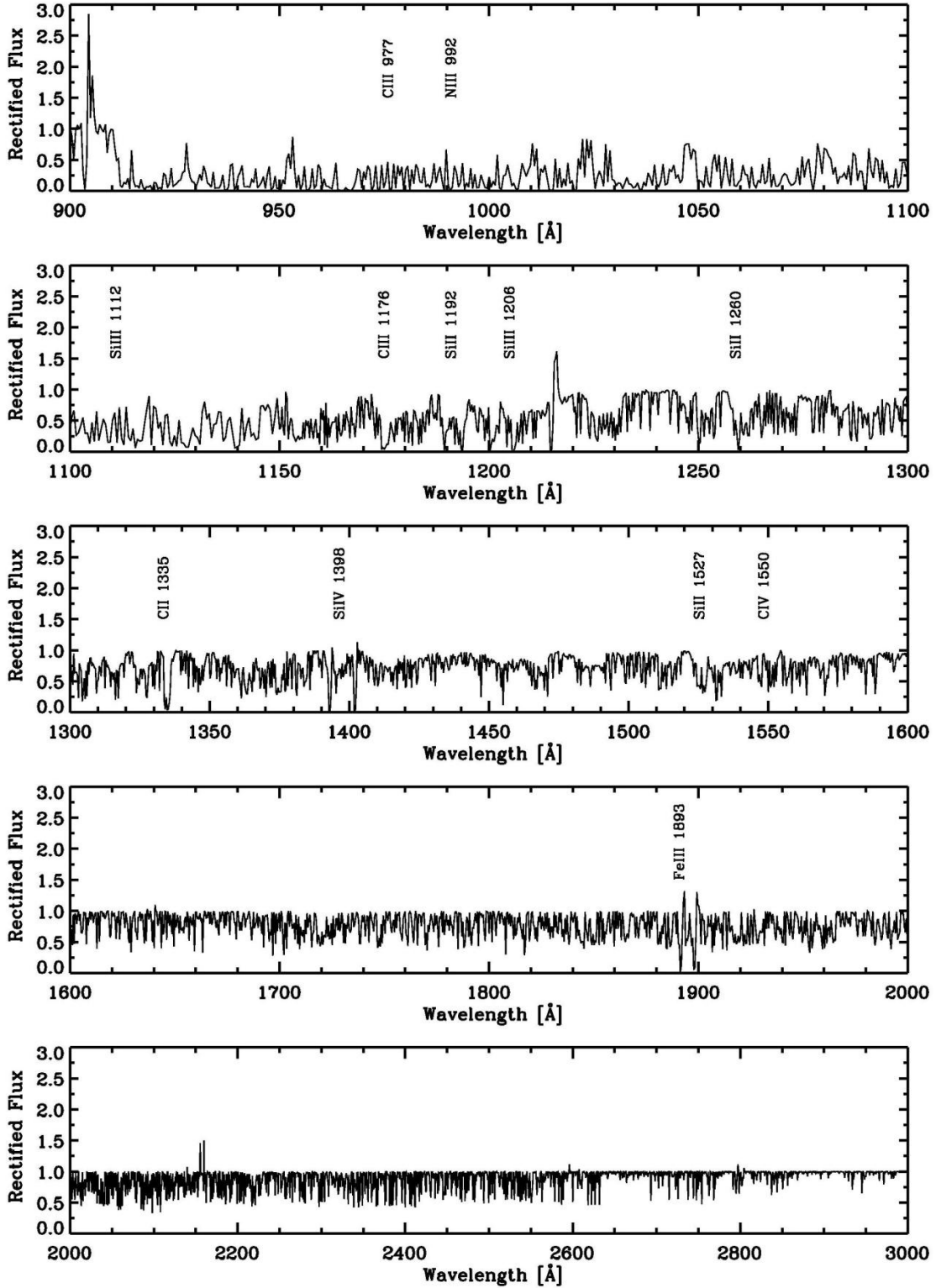

Figure 3. — Same as Figure 2 but for model 56 ($T_{\rm eff}$ = 15,300 K; log $g$ = 1.9; $Z = Z_\odot$). The corresponding spectral type is B3 Ia.

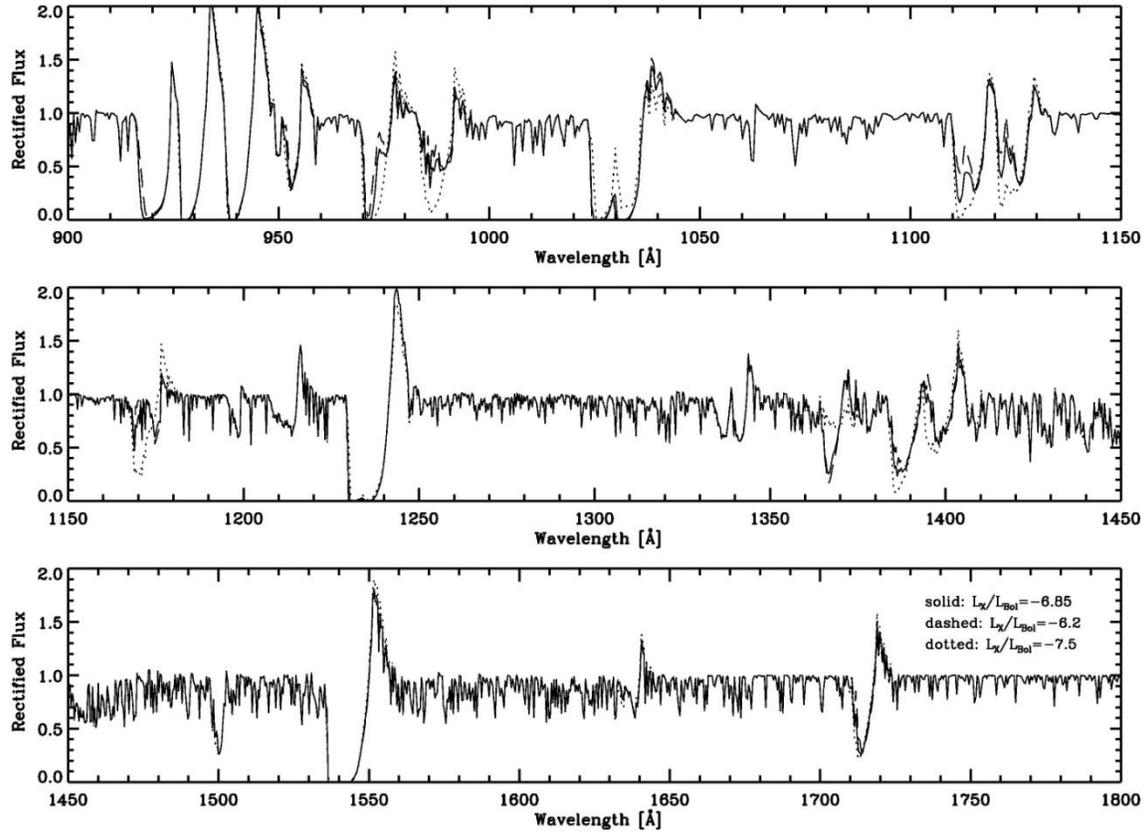

Figure 4. — Influence of $L_x/L_{Bol}$ on the calculated UV spectrum. Solid spectrum: $\log L_x/L_{Bol} = -6.85$; dashed: $-6.2$; dotted: $-7.2$. Shown is model 31 ($T_{eff} = 35{,}000$ K; $\log g = 3.4$; $Z = Z_\odot$).

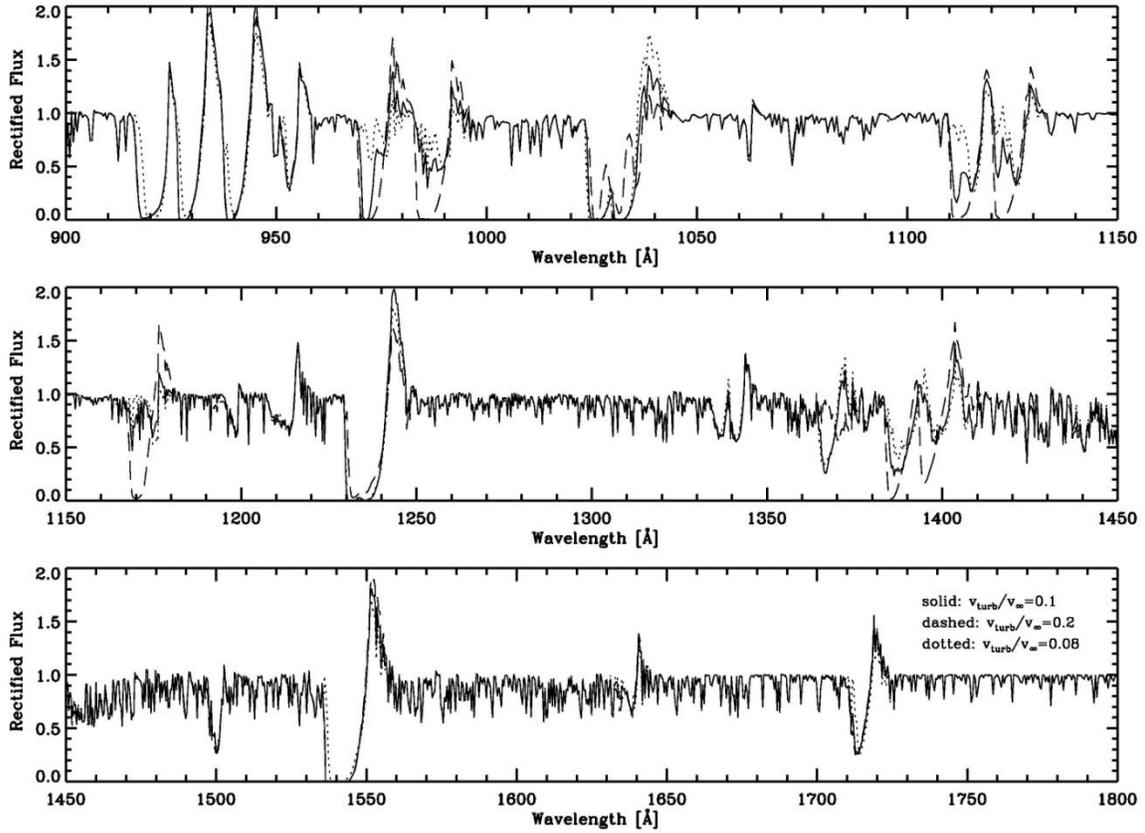

Figure 5. — Same as Figure 4, but for $v_{\text{turb}}/v_\infty$. Solid spectrum: $v_{\text{turb}}/v_\infty = 0.1$; dashed: 0.2; dotted: 0.08.

Figure 6. — Same as Figure 4 but for γ. Solid spectrum: $\gamma = 0.5$; dashed: 1.0; dotted: 0.2.

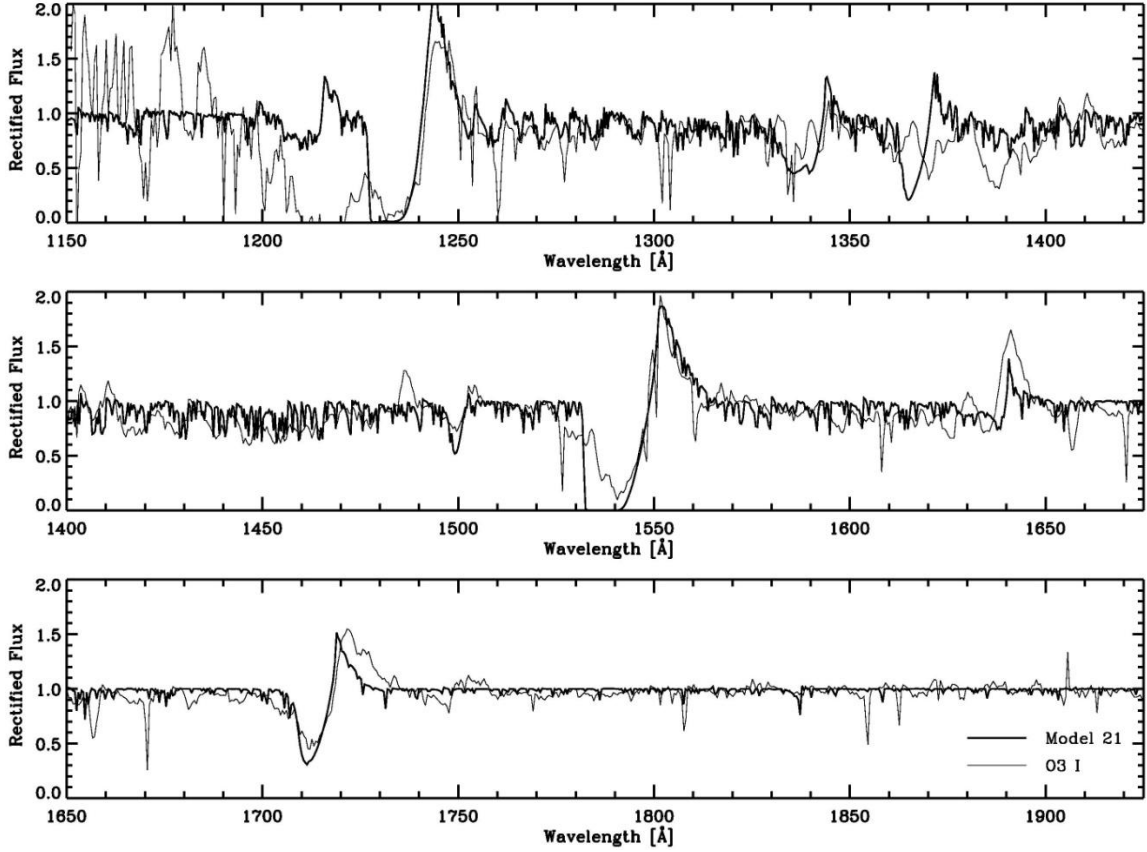

Figure 7. — Comparison of the spectrum of model 21 ($T_{\text{eff}} = 42{,}500$ K; $\log L = 6.01$; $Z = Z_{\odot}$) with the mean O3 I spectrum of three stars observed with IUE. See Table 7 for details of the comparison stars. The strong Lyman-α absorption ($\lambda = 1216$ Å) in the observed spectrum is interstellar; it is therefore not seen in the model spectrum.

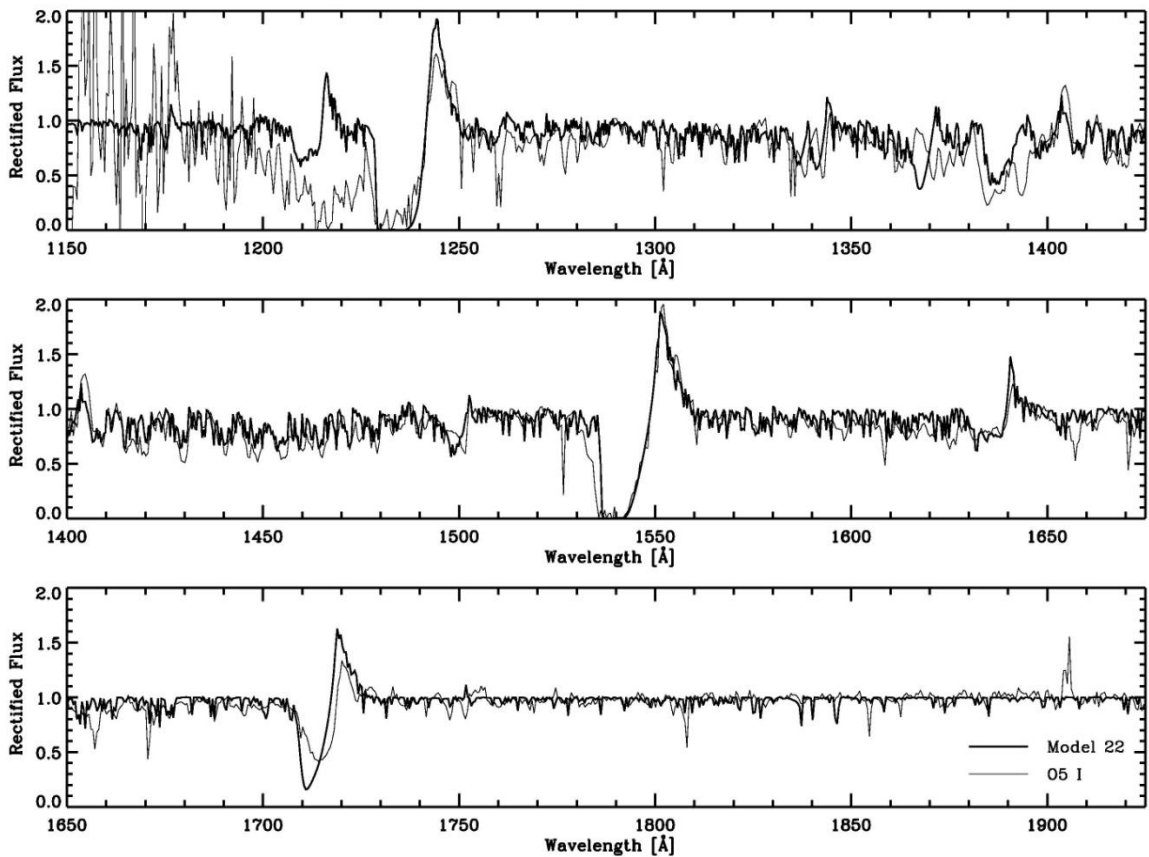

Figure 8. — Same as Figure 7, but for model 22 ($T_{\rm eff}$ = 38,400 K; log $L$ = 6.03; $Z = Z_\odot$) and a mean O5 I comparison spectrum.

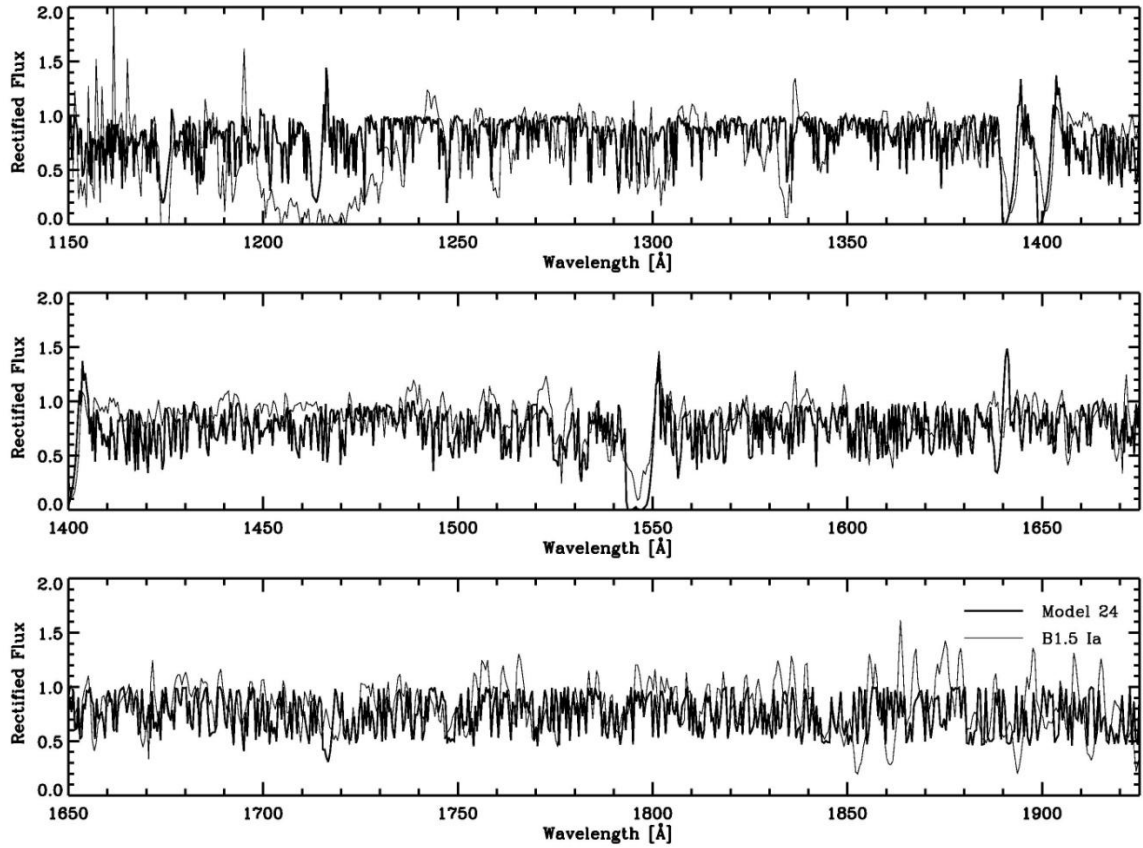

Figure 9. — Same as Figure 7, but for model 24 ($T_{\text{eff}} = 24{,}000$ K; $\log L = 6.01$; $Z = Z_\odot$) and a mean B1.5 Ia comparison spectrum.

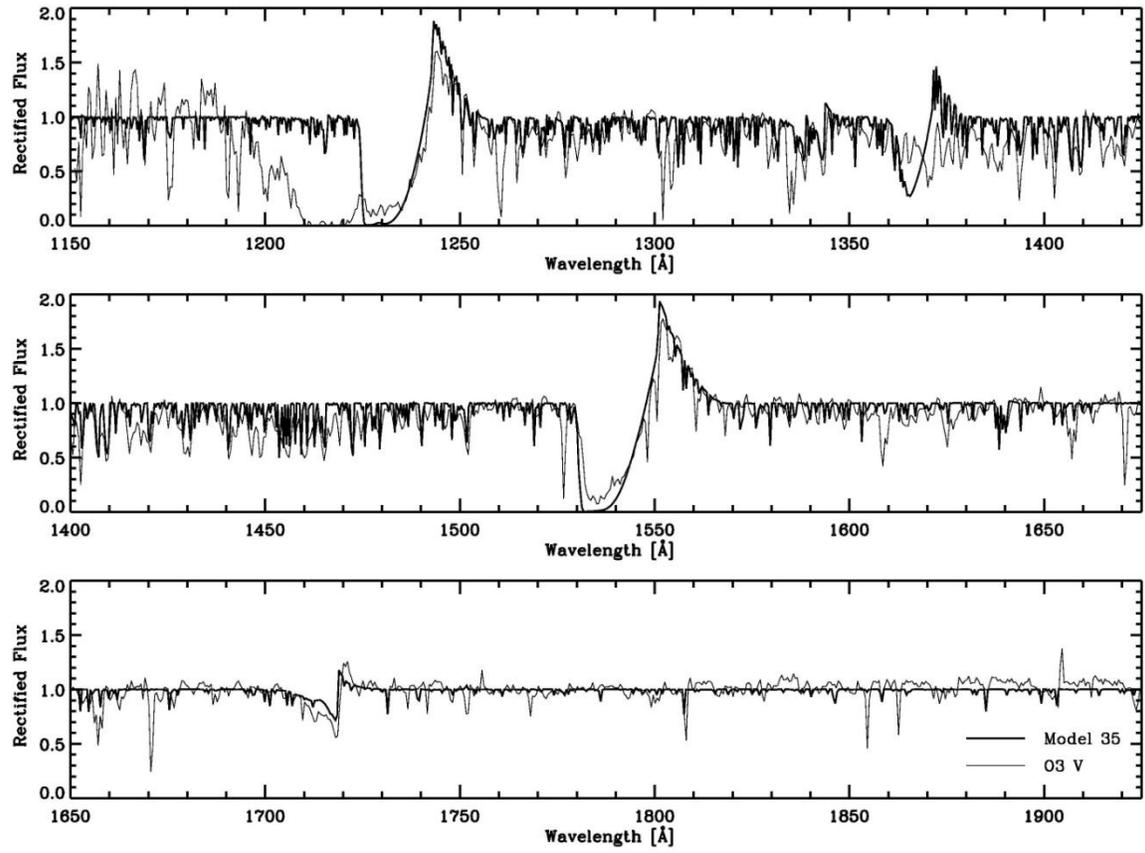

Figure 10. — Same as Figure 7, but for model 35 ($T_{\rm eff}$ = 43,600 K; log $L$ = 5.62; $Z = Z_\odot$) and a mean O3 V comparison spectrum.

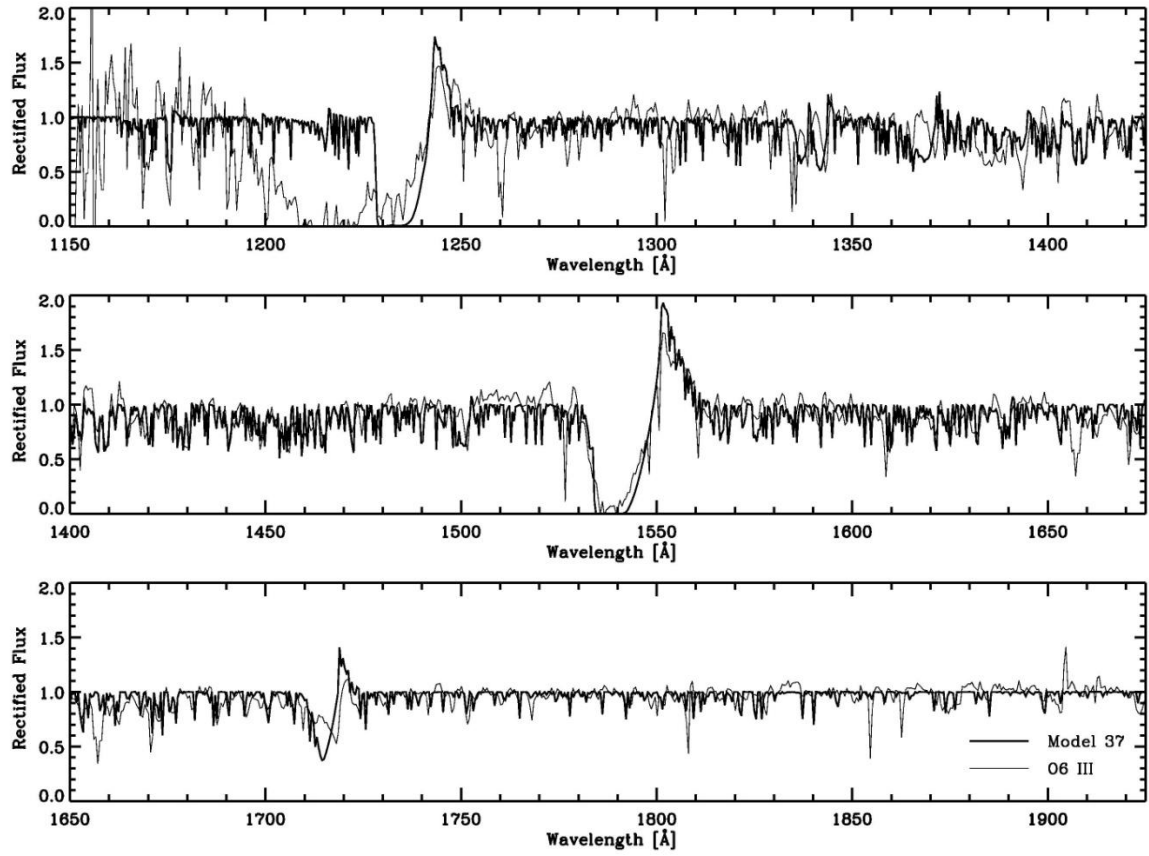

Figure 11. — Same as Figure 7, but for model 37 ($T_{\text{eff}} = 37{,}800$ K; $\log L = 5.69$; $Z = Z_\odot$) and a mean O6 III comparison spectrum.

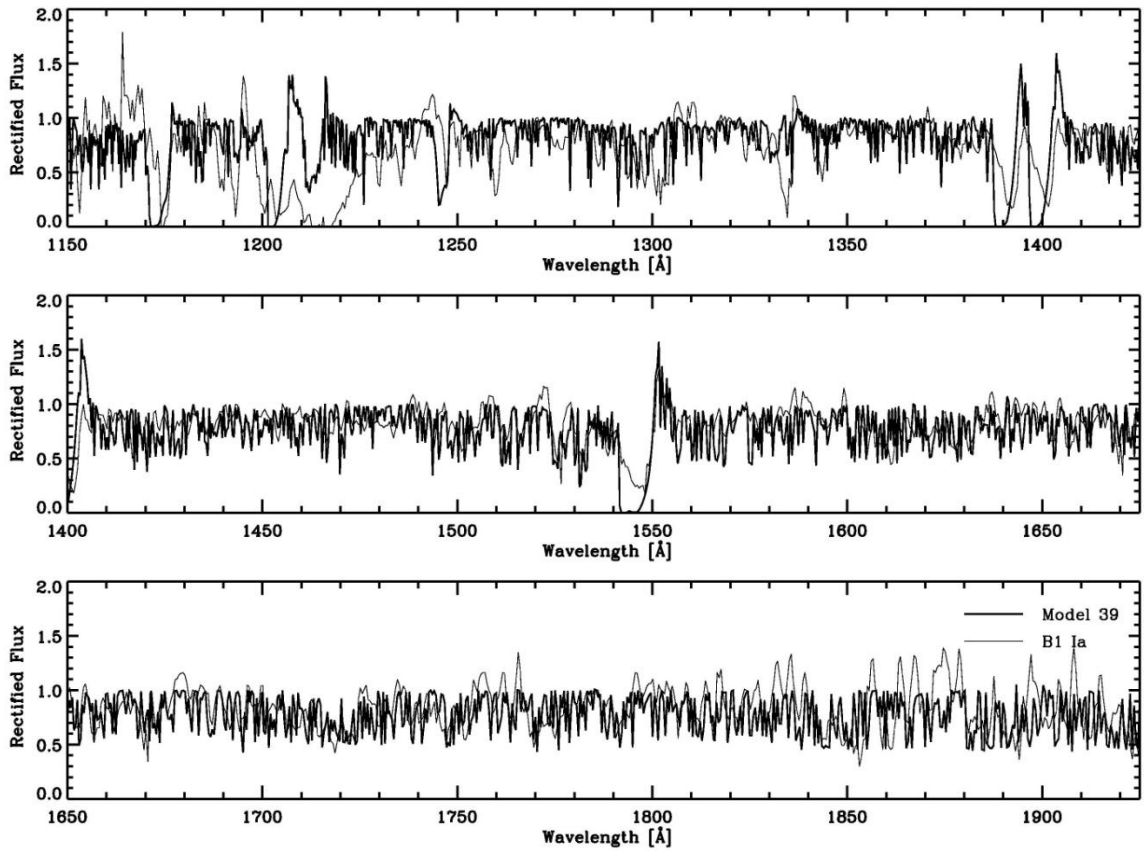

Figure 12. — Same as Figure 7, but for model 39 ($T_{\text{eff}}$ = 24,900 K; log $L$ = 5.75; $Z = Z_\odot$) and a mean B1 Ia comparison spectrum.

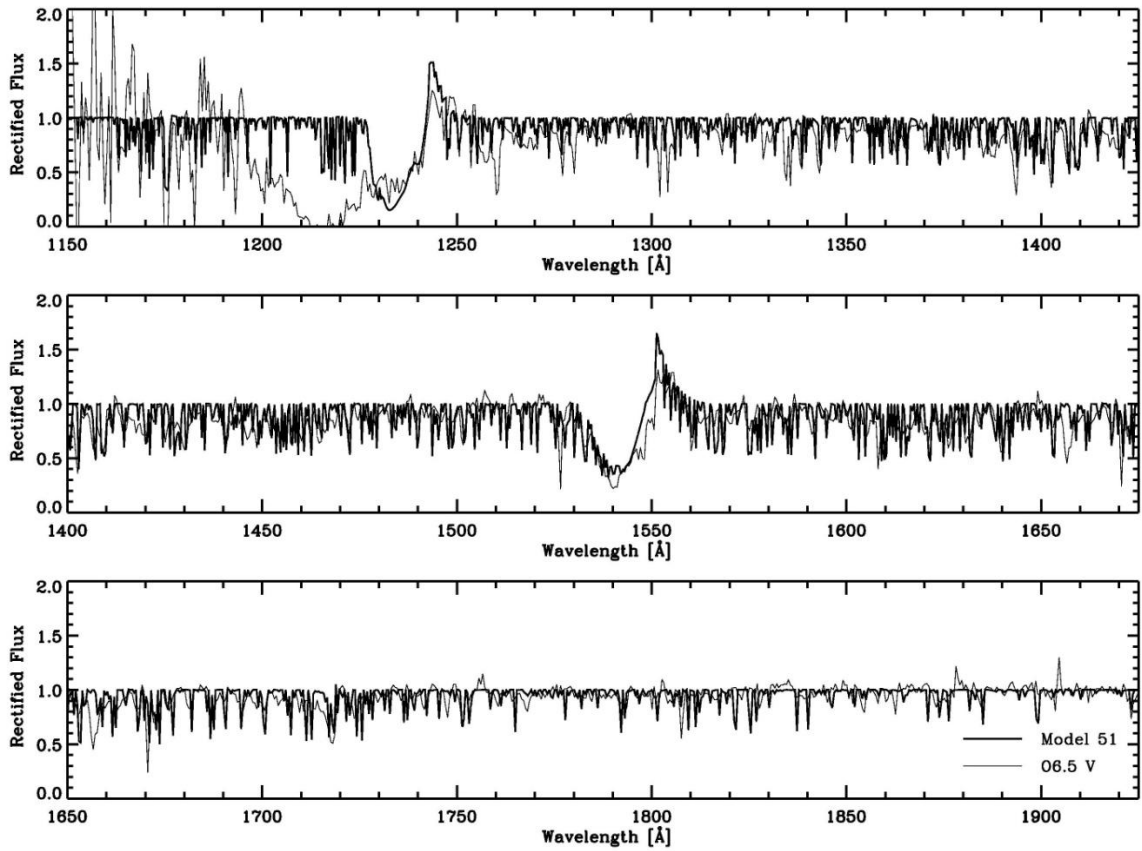

Figure 13. — Same as Figure 7, but for model 51 ($T_{\text{eff}}$ = 37,700 K; log $L$ = 5.19; $Z = Z_\odot$) and a mean O6.5 V comparison spectrum.

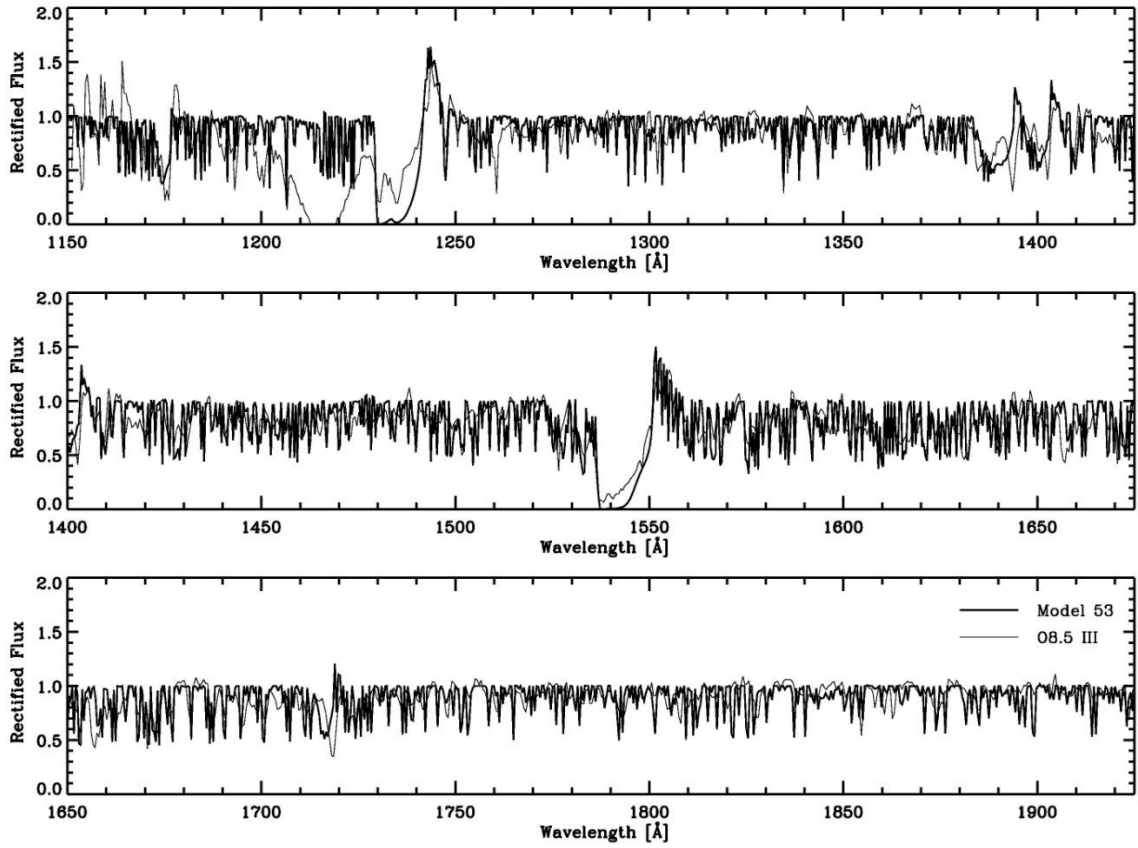

Figure 14. — Same as Figure 7, but for model 53 ($T_{\text{eff}}$ = 32,600 K; log $L$ = 5.30; $Z$ = $Z_\odot$) and a mean O8.5 III comparison spectrum.

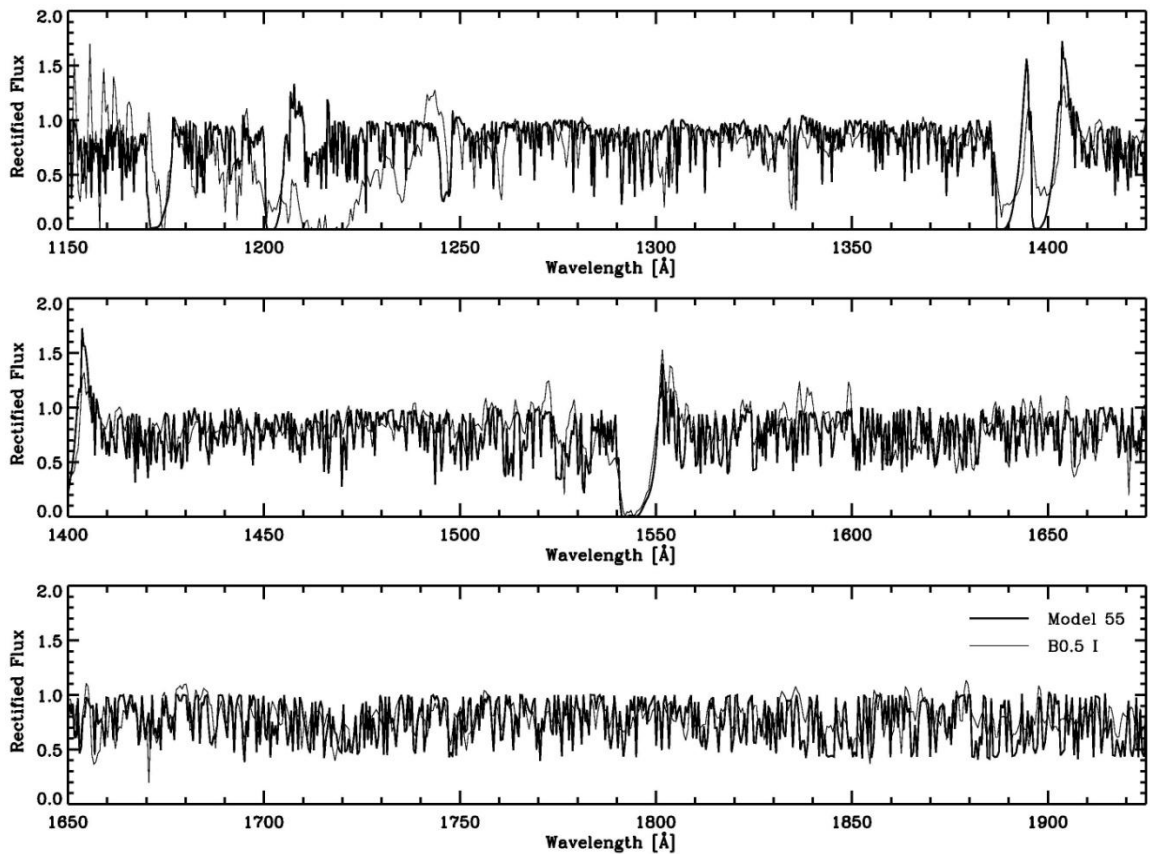

Figure 15. — Same as Figure 7, but for model 55 ($T_{\text{eff}}$ = 24,800 K; log $L$ = 5.37; $Z = Z_\odot$) and a mean B0.5 I comparison spectrum.

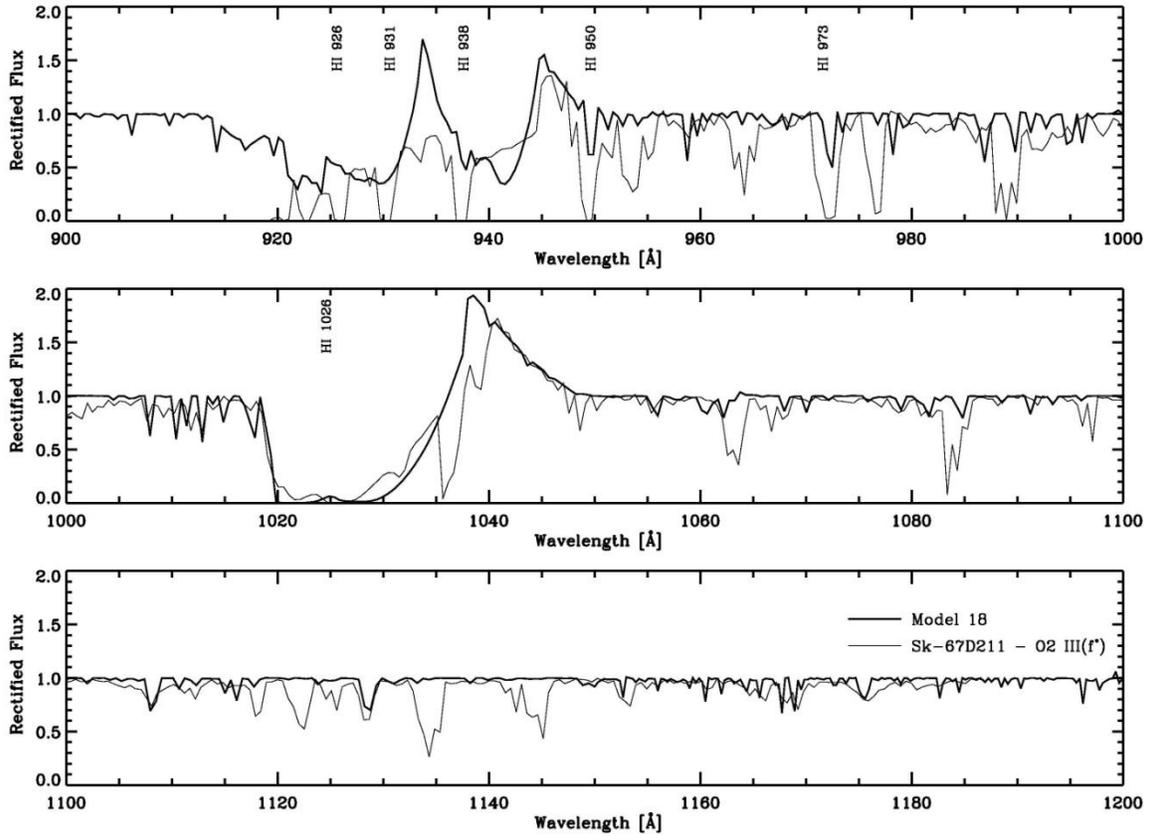

Figure 16. — Comparison of the spectrum of model 18 ($T_{\text{eff}}$ = 49,600 K; log $L$ = 5.96; $Z$ = 0.4 $Z_\odot$) with a far-UV spectrum of the LMC star Sk−67D211 observed with FUSE. The locations of the (mostly) interstellar lines of the Lyman series are indicated. See Walborn et al. (2002) for the stellar parameters of Sk −67D211 and other LMC/SMC stars with FUSE spectra.

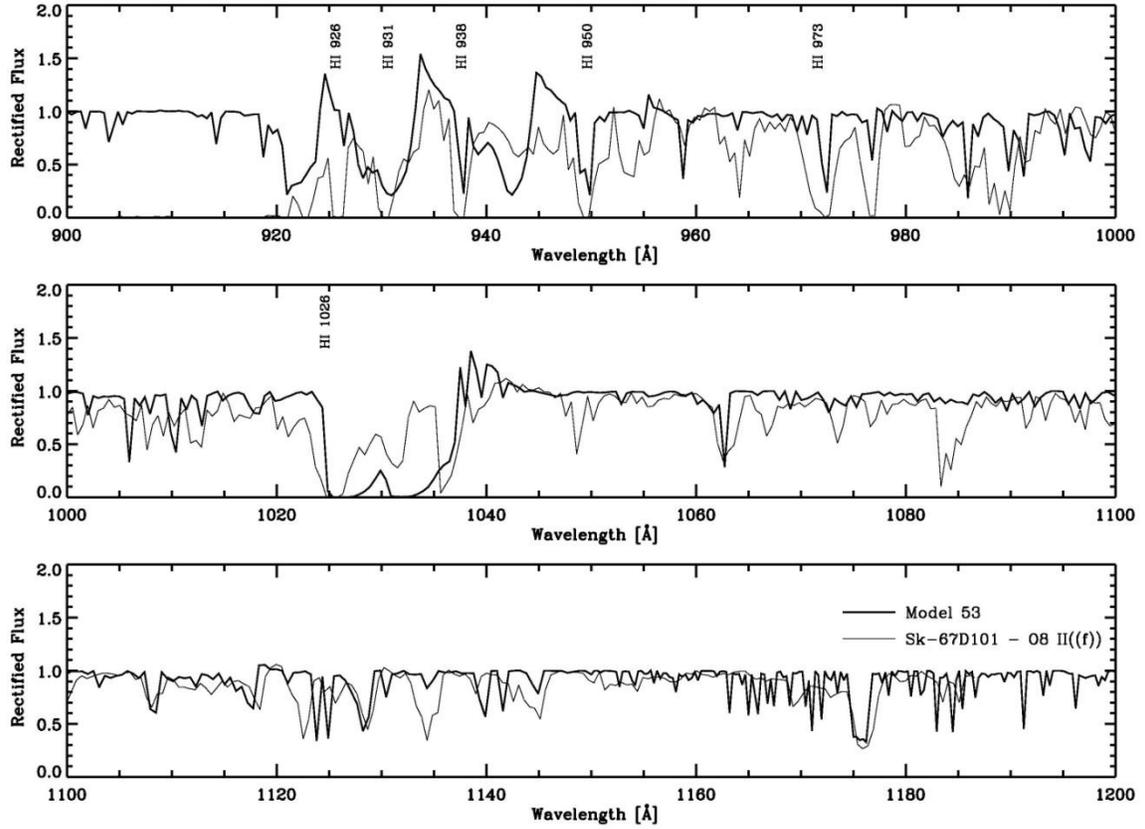

Figure 17. — Same as Figure 16 but for model 53 ($T_{\text{eff}}$ = 32,600 K; log $L$ = 5.30; $Z$ = 0.4 $Z_\odot$) and the far-UV spectrum of the LMC star Sk–67D101 observed with FUSE.

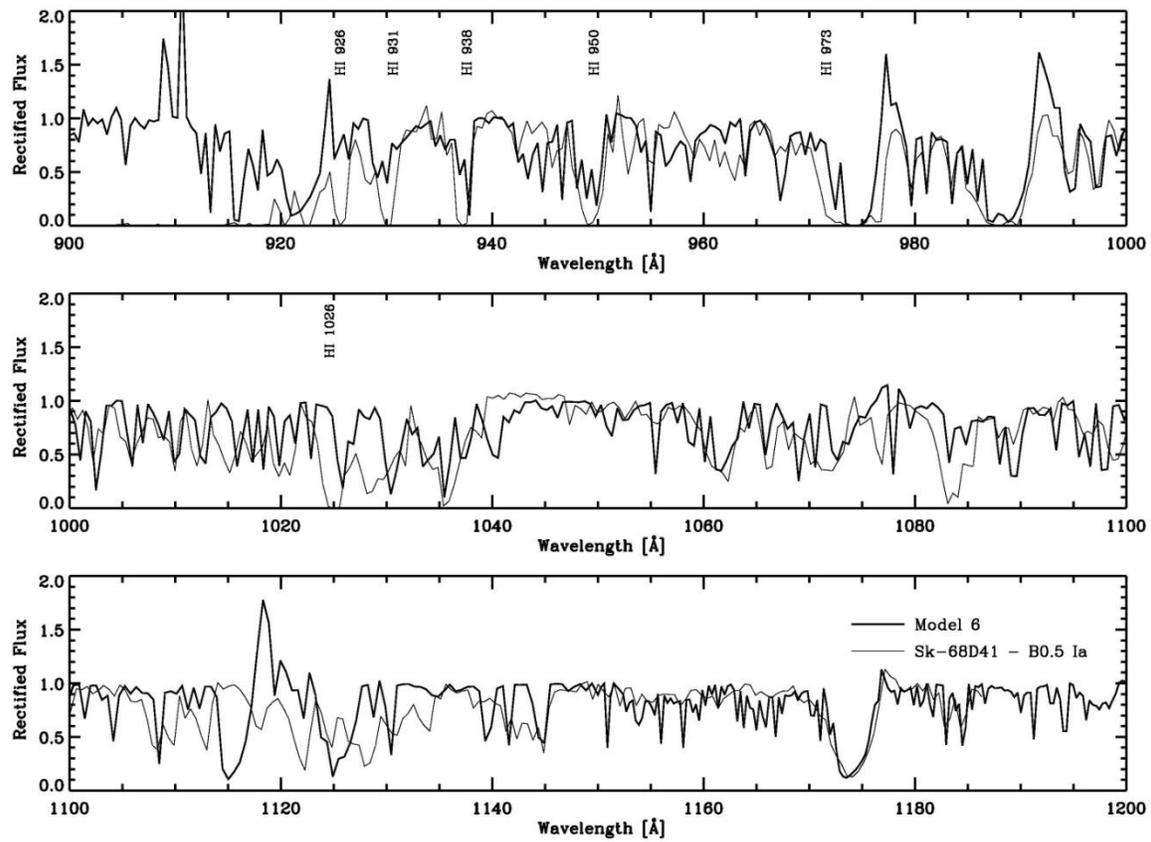

Figure 18. — Same as Figure 16 but for model 6 ($T_{\text{eff}}$ = 25,100 K; log $L$ = 6.23; $Z$ = 0.4 $Z_\odot$) and the far-UV spectrum of the LMC star Sk−66D41 observed with FUSE.

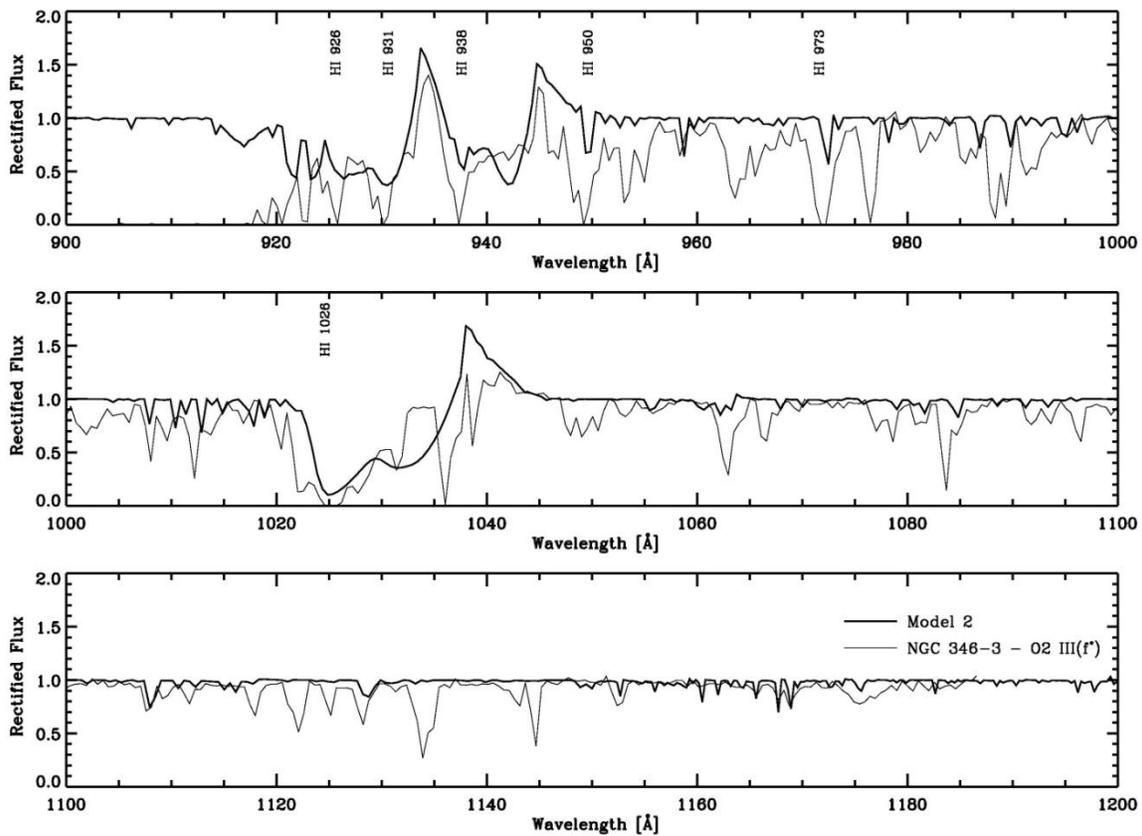

Figure 19. — Comparison of the spectrum of model 2 ($T_{\rm eff}$ = 52,100 K; log $L$ = 6.25; $Z$ = 0.2 $Z_\odot$) with a far-UV spectrum of the SMC star NGC 346-3 observed with FUSE.

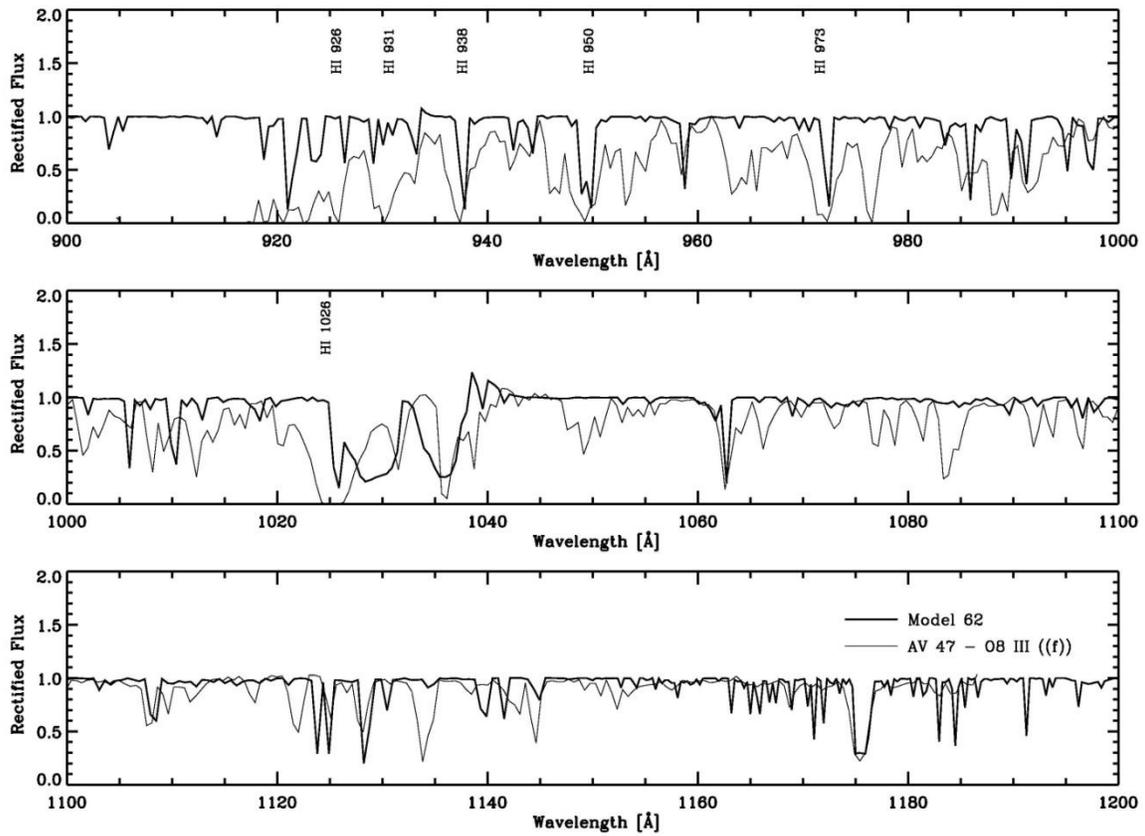

Figure 20. — Same as Figure 19 but for model 62 ($T_{\rm eff}$ = 33,000 K; log $L$ = 5.09; $Z$ = 0.2 $Z_\odot$) and the far-UV spectrum of the SMC star AV 47 observed with FUSE.

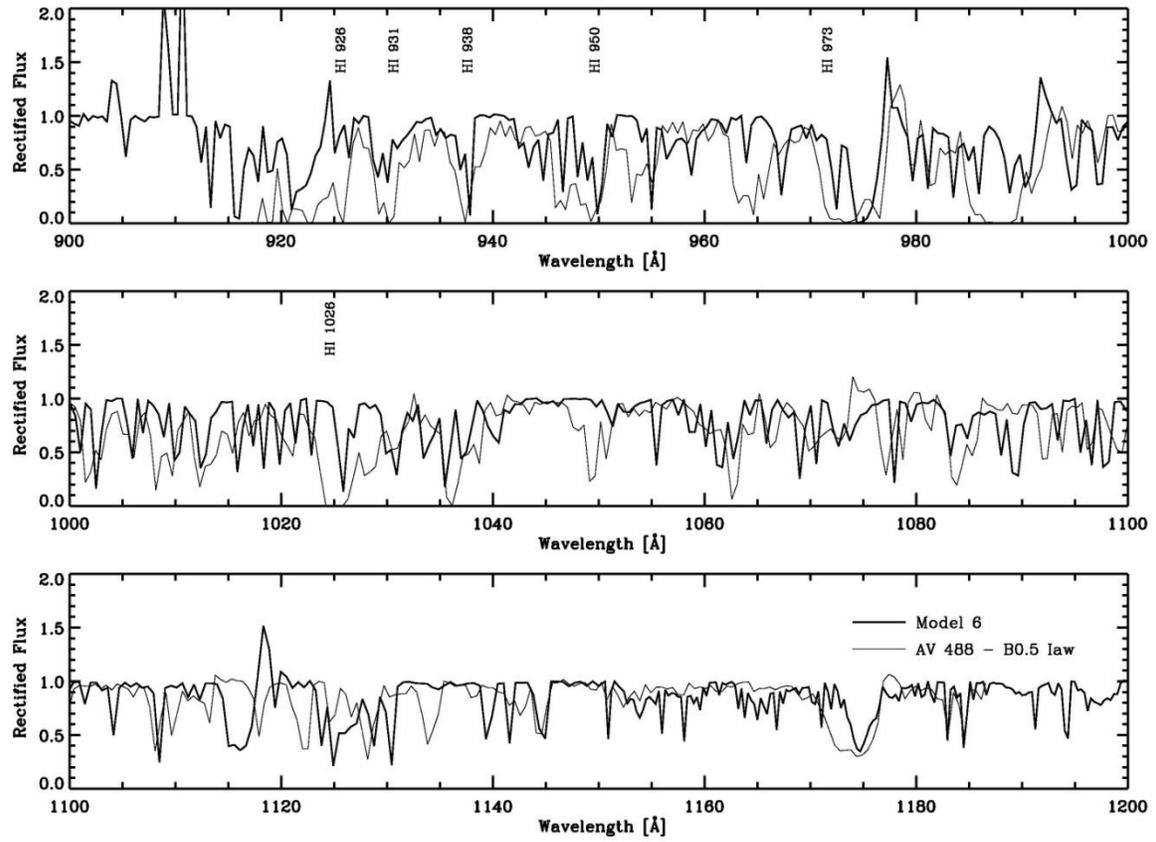

Figure 21. — Same as Figure 19 but for model 6 ($T_{\text{eff}}$ = 25,100 K; log $L$ = 6.23; $Z$ = 0.2 $Z_\odot$) and the far-UV spectrum of the SMC star AV 488 observed with FUSE.

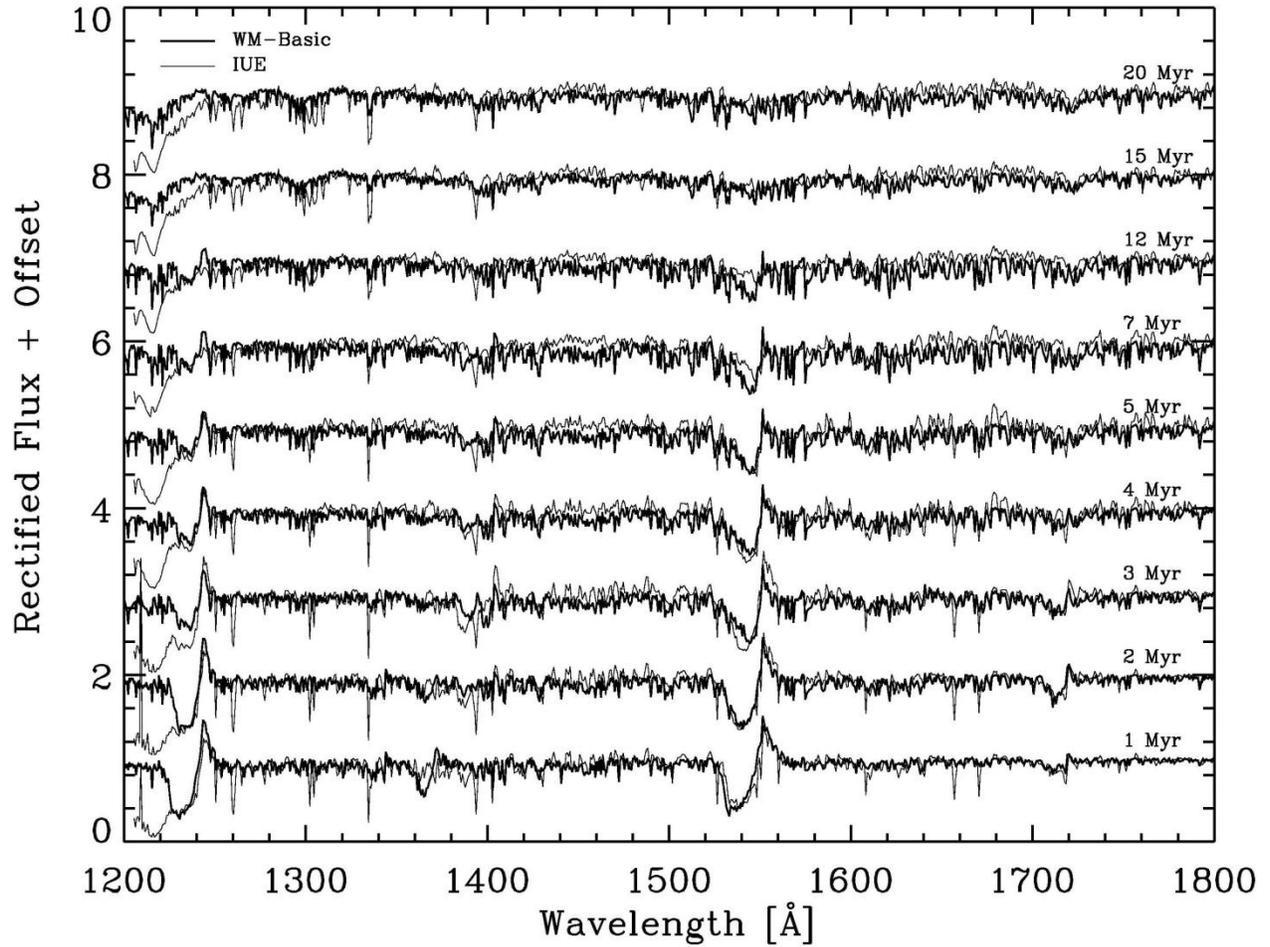

Figure 22. — Comparison of synthetic spectra of a single stellar population obtained with the WM-Basic library having $Z = Z_\odot$ (thick lines) and an empirical IUE library of Galactic stars (thin) between 1200 and 1800 Å. Each spectrum represents one age step from 1 to 20 Myr as labeled on the right. Salpeter IMF with mass limits 1 and 100 $M_\odot$.

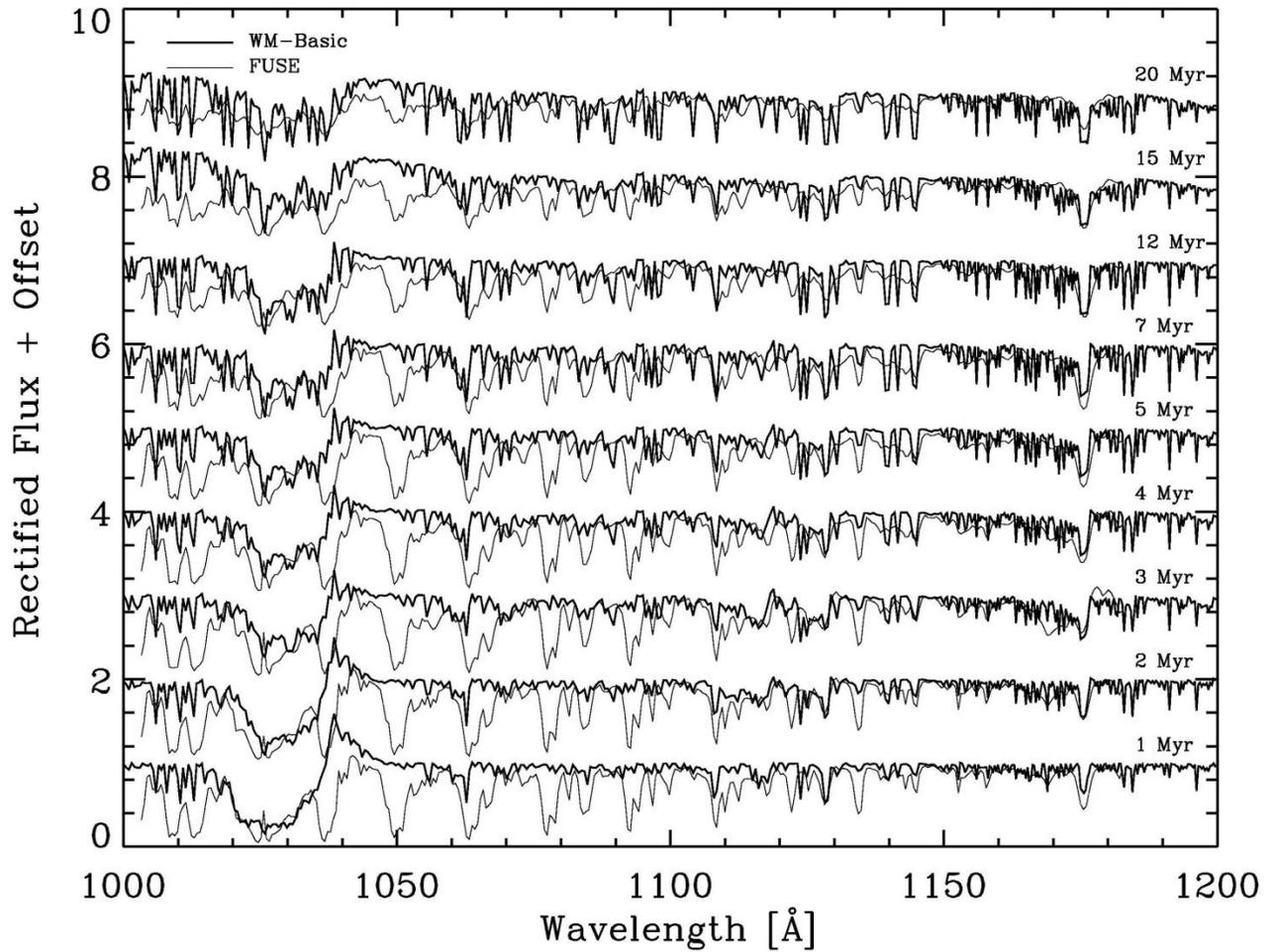

Figure 23. — Same as Figure 22, but for a comparison with an empirical library obtained with FUSE in the wavelength region 1000 to 1200 Å.

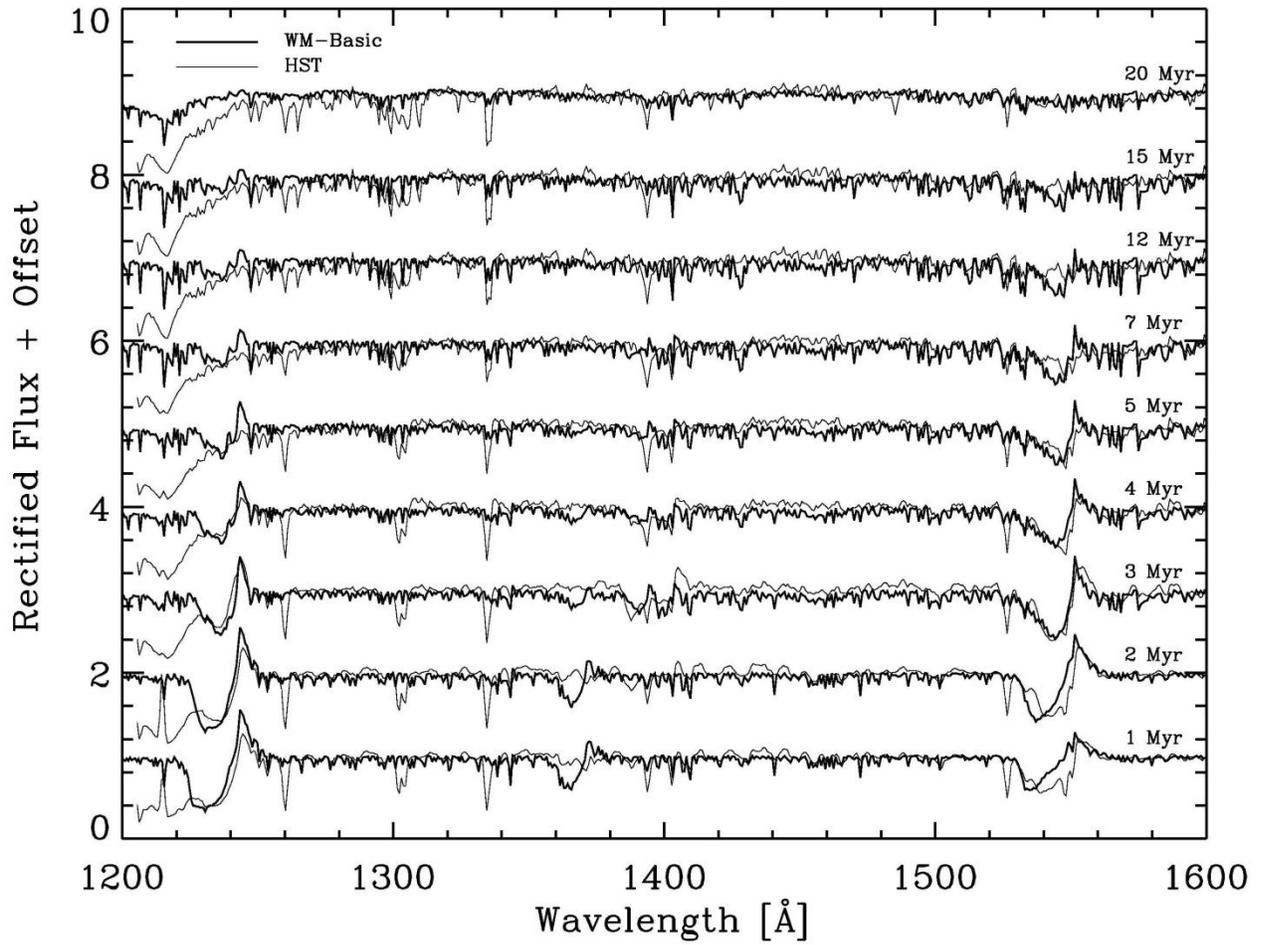

Figure 24. — Same as Figure 22, but for a comparison of the WM-Basic library having $Z = 0.4\ Z_\odot$ with an empirical library of LMC and SMC stars obtained with HST.

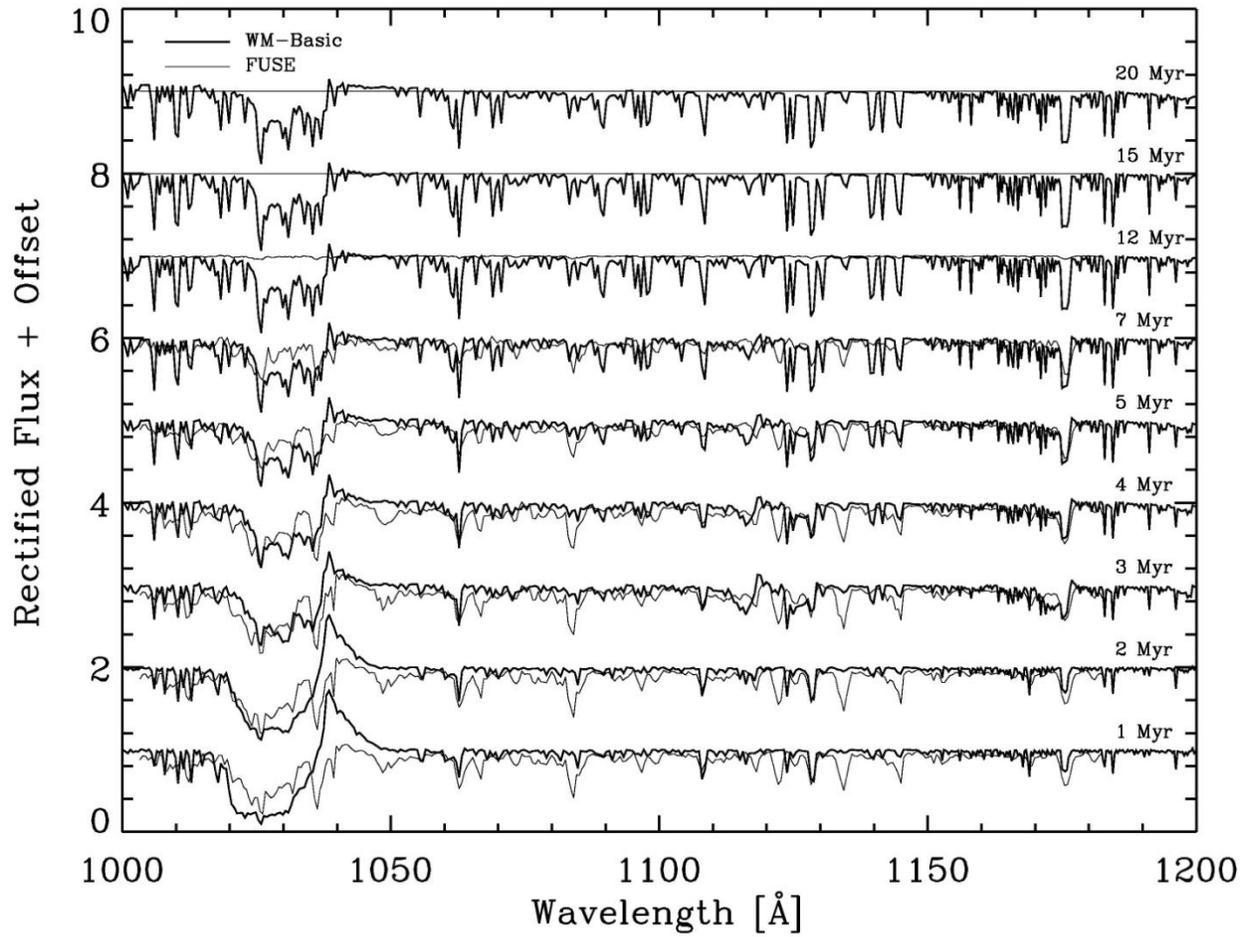

Figure 25. — Same as Figure 23, but for the WM-Basic library with $Z = 0.4\ Z_\odot$ and an empirical FUSE library of LMC and SMC stars.

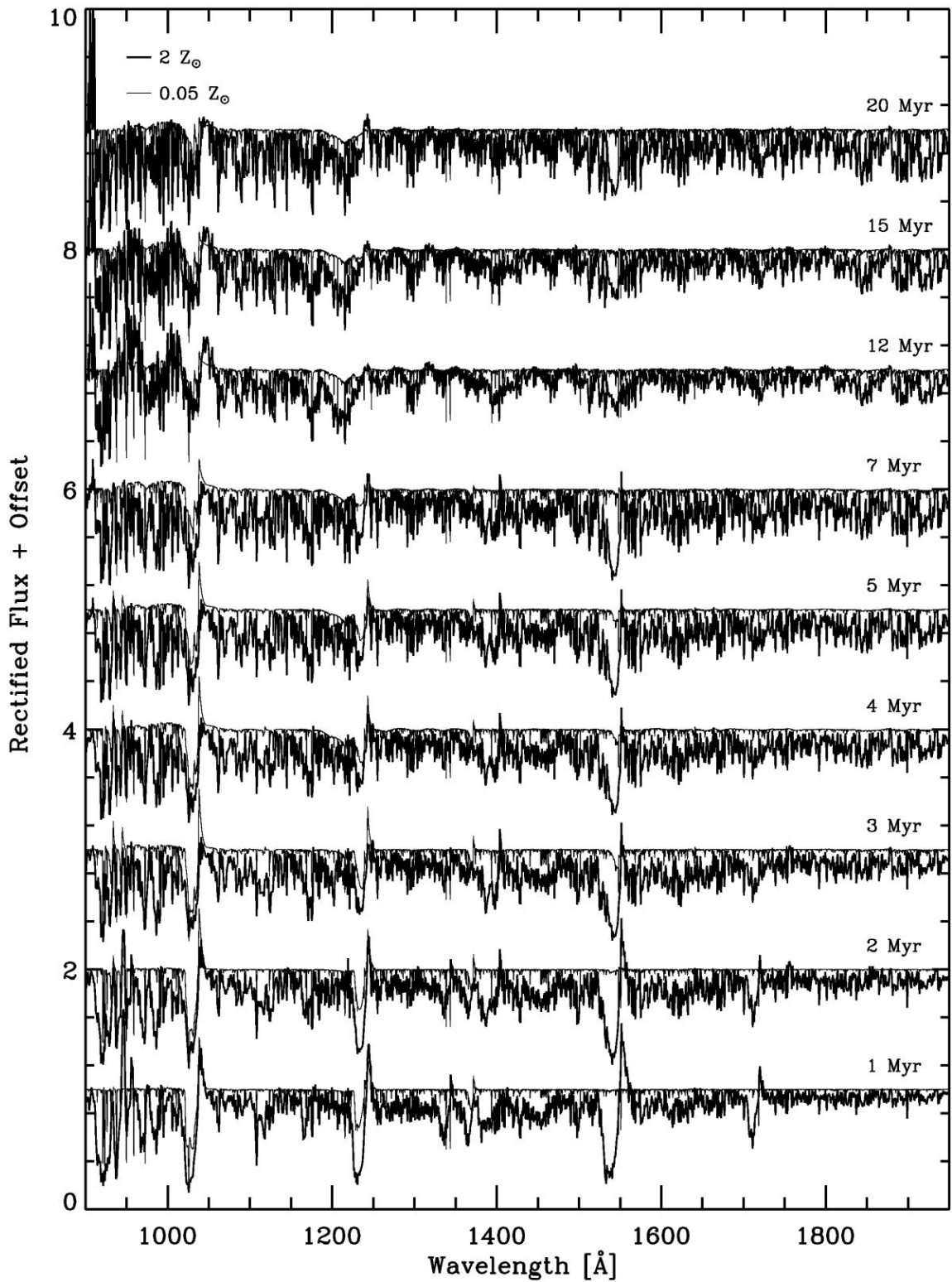

Figure 26. — Synthetic spectra between 900 and 1950 Å for instantaneous burst models. Thick: $Z = 2\,Z_\odot$; thin: $Z = 0.05\,Z_\odot$. Each spectrum represents one age step from 1 to 20 Myr as labeled on the right. Salpeter IMF with mass limits 1 and 100 $M_\odot$.

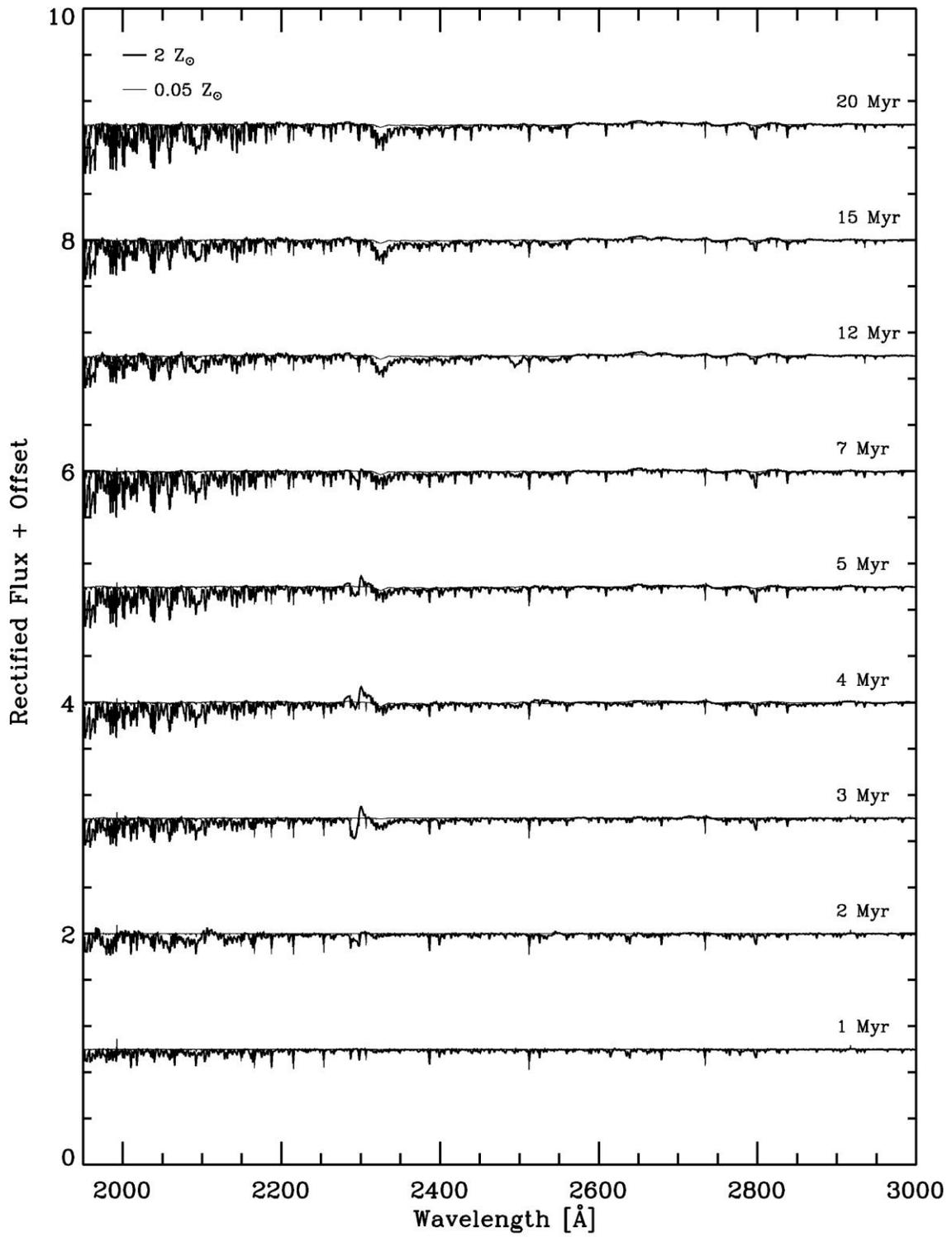

Figure 27. — Same as Figure 26, but for the wavelength range 1950 to 3000 Å.

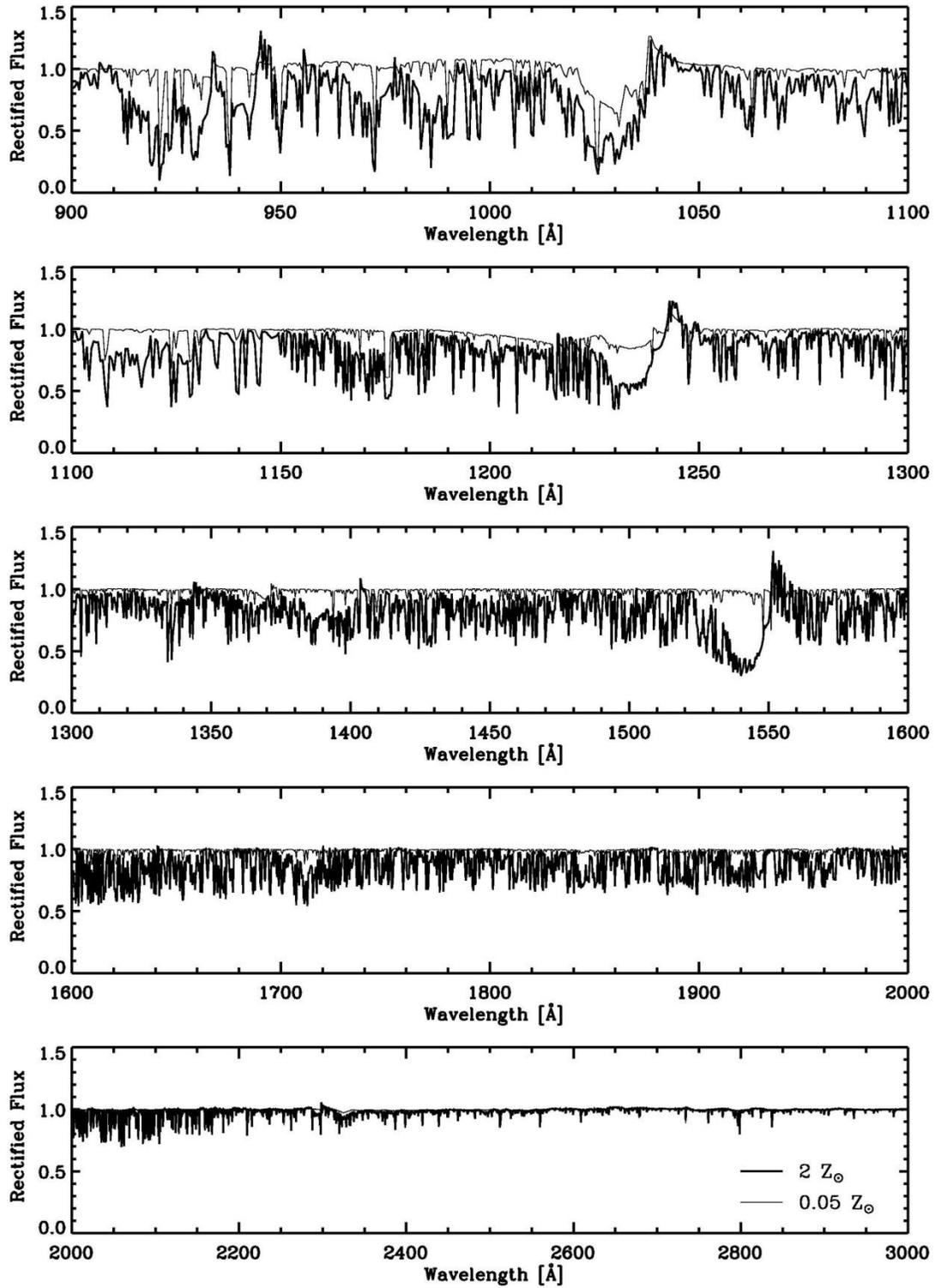

Figure 28. — Synthetic spectra between 900 and 3000 Å for models with continuous star formation. Thick: $Z = 2\ Z_\odot$; thin: $Z = 0.05\ Z_\odot$. Age: 50 Myr. Salpeter IMF with mass limits 1 and 100 $M_\odot$.

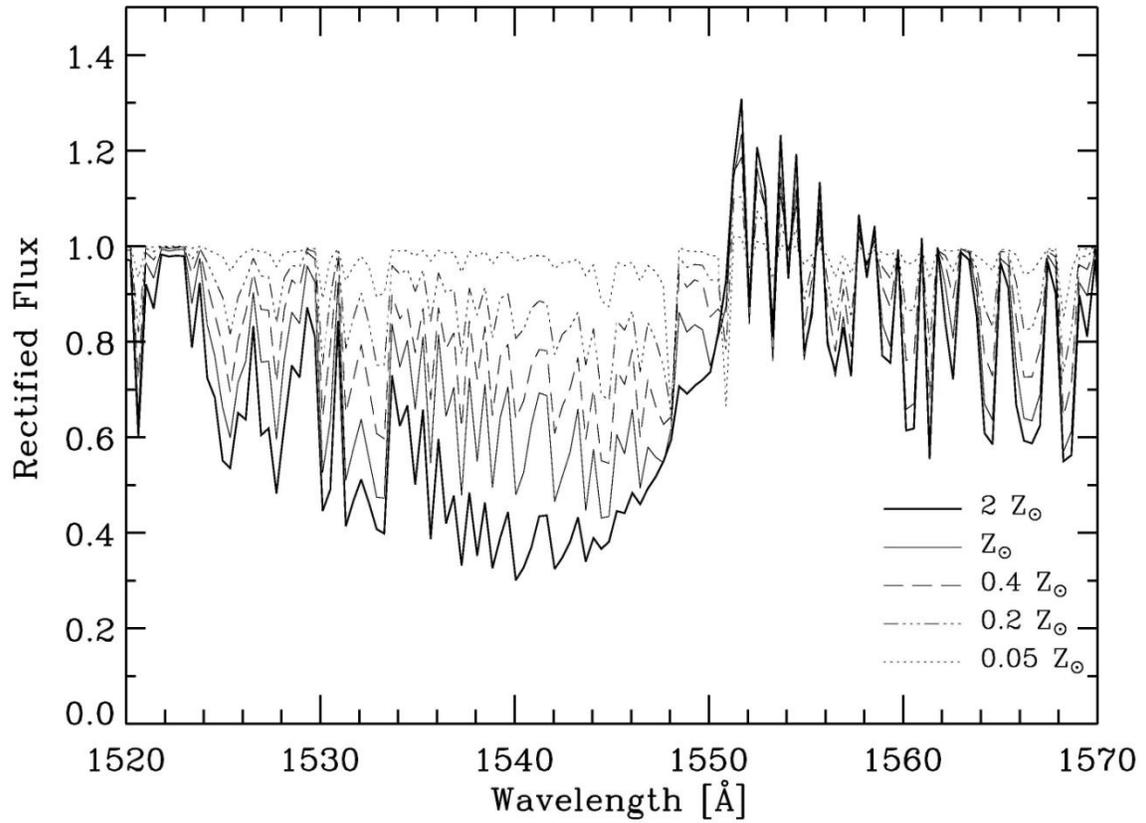

Figure 29. — C IV λ1550 for models with continuous star formation and five metallicities as labeled in the figure. Age: 50 Myr. Salpeter IMF with mass limits 1 and 100 $M_\odot$.

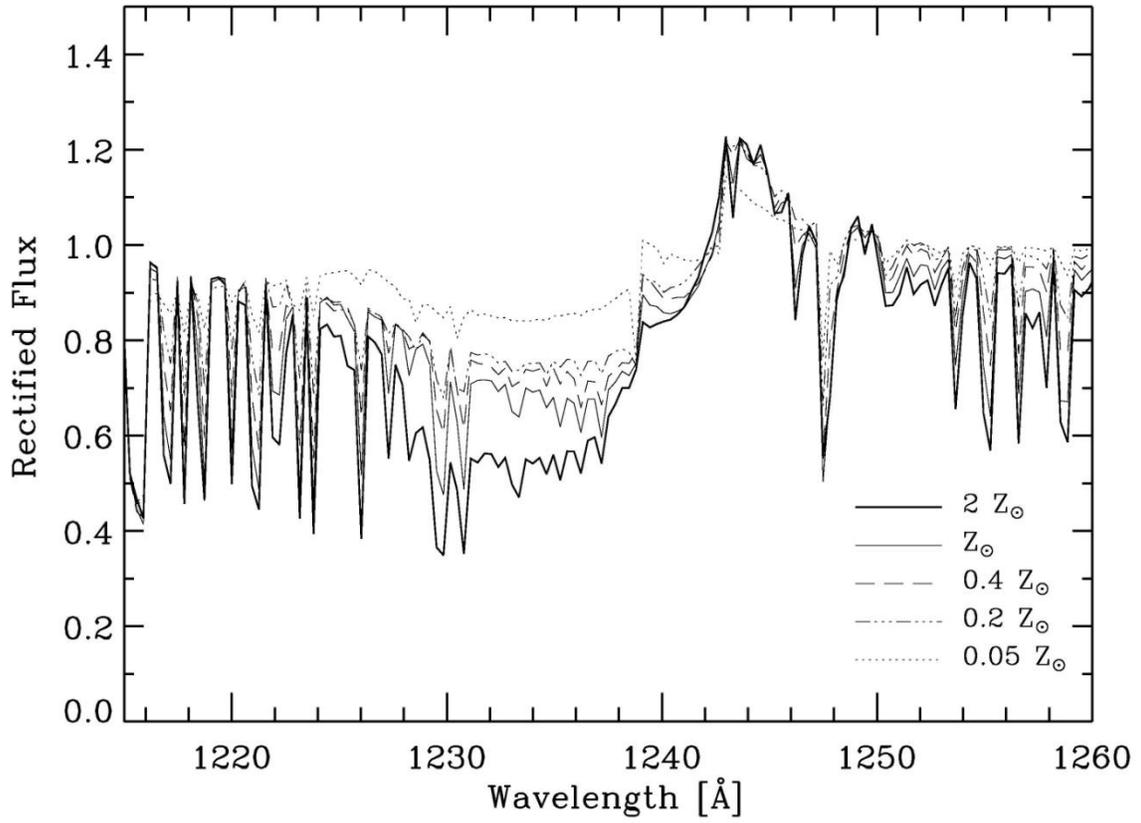

Figure 30. — Same as Figure 29, but for N V λ1241.

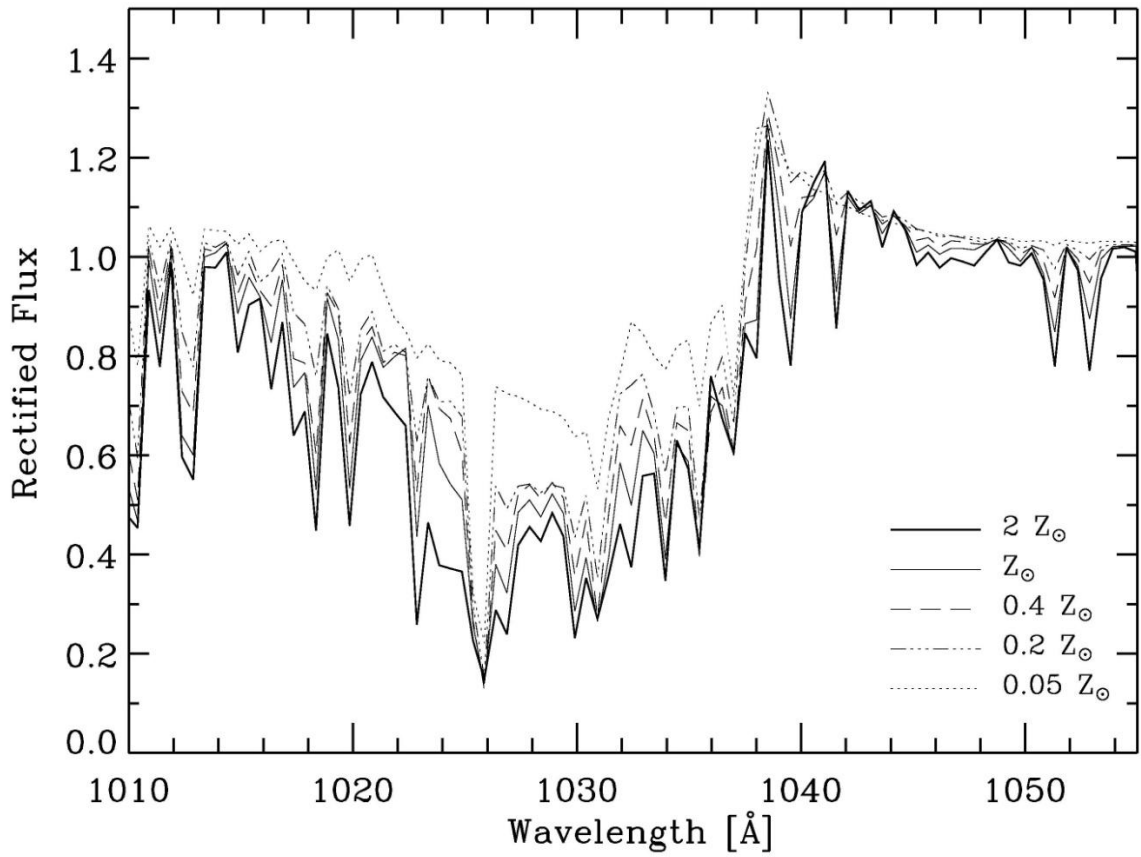

Figure 31. — Same as Figure 29, but for O VI λ1035.

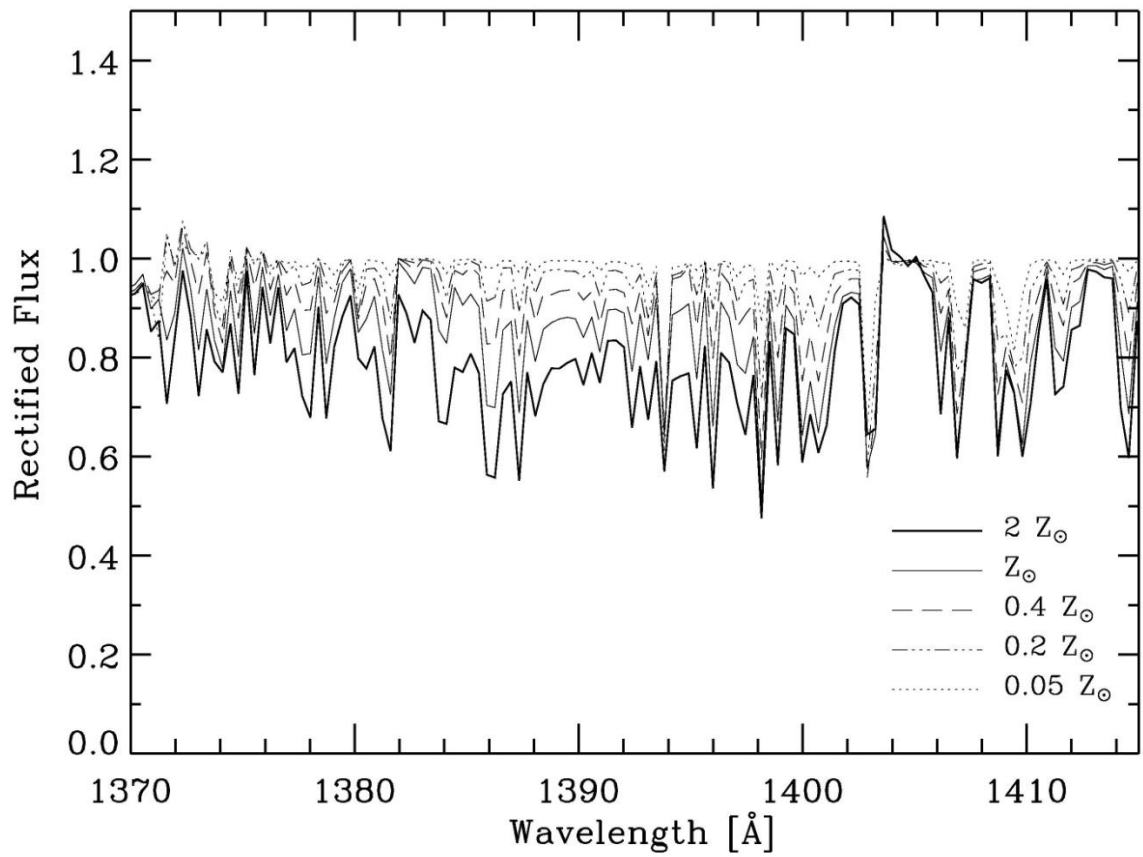

Figure 32. — Same as Figure 29, but for Si IV λ1398.

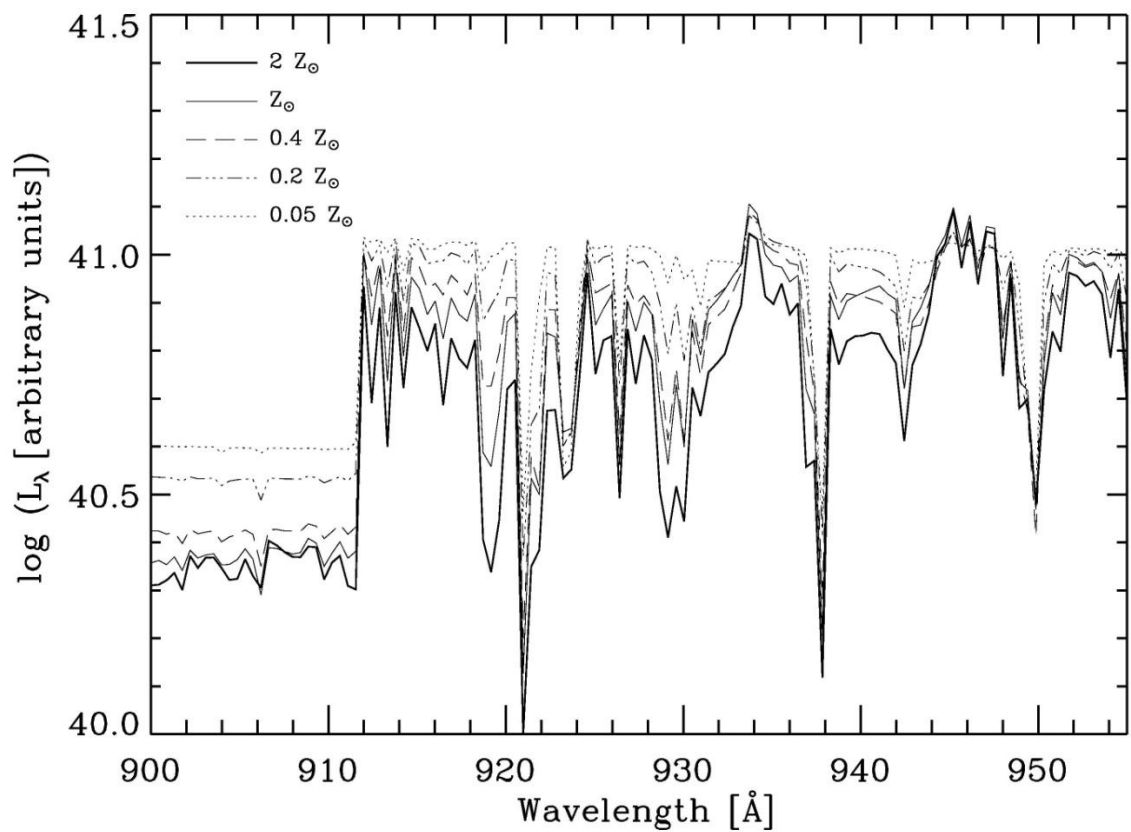

Figure 33. — Spectral region around the Lyman-break for models with continuous star formation and five metallicities as labeled in the figure. Age: 50 Myr. Salpeter IMF with mass limits 1 and 100 $M_\odot$.

# Tables

Table 1. Grid parameters related to stellar evolutionary tracks

| Model | $M$ ($M_\odot$) | $\log L$ ($L_\odot$) | $T_{\text{eff}}$ (K) | $R$ ($R_\odot$) | $v_{\text{esc}}$ (km s$^{-1}$) | He | C | N | O | Sp. Type |
|---|---|---|---|---|---|---|---|---|---|---|
| *1* | *120.0* | *6.25* | *59,870* | *12.4* | *1506* | | | | | |
| 2 | 119.6 | 6.25 | 52,061 | 16.3 | 1307 | | | | | O2 I |
| 3 | 113.2 | 6.25 | 49,139 | 18.4 | 1179 | | | | | O2 I |
| 4 | 103.4 | 6.25 | 46,967 | 20.1 | 1040 | 1.1 | 26 | 9 | 2 | O2 I |
| 5 | 91.8 | 6.25 | 46,330 | 20.6 | 916 | 1.4 | 21 | 11 | 4 | O2 I |
| 6 | 74.1 | 6.23 | 25,085 | 68.7 | 431 | 2.2 | 18 | 12 | 17 | B0.5 Ia |
| 7 | 35.8 | 5.90 | 22,339 | 59.2 | 328 | 3.0 | 18 | 12 | 31 | B1 Ia |
| 8 | 9.2 | 4.93 | 25,100 | 15.4 | 420 | 3.2 | 18 | 12 | 31 | B1 III |
| *9* | *99.6* | *6.13* | *58,567* | *11.3* | *1477* | | | | | |
| 10 | 99.8 | 6.13 | 50,928 | 14.9 | 1286 | | | | | O2 I |
| 11 | 96.5 | 6.14 | 48,499 | 16.7 | 1179 | | | | | O2 I |
| 12 | 91.2 | 6.15 | 46,122 | 18.6 | 1057 | | | | | O2 I |
| 13 | 83.5 | 6.16 | 43,501 | 21.1 | 909 | 2.0 | 22 | 11 | 20 | O3 I |
| 14 | 71.0 | 6.16 | 35,514 | 31.7 | 633 | 3.0 | 18 | 12 | 31 | O6.5 I |
| 15 | 51.8 | 6.08 | 31,527 | 36.9 | 461 | 3.0 | 18 | 12 | 31 | O9 I |
| 16 | 13.2 | 5.35 | 33,117 | 14.3 | 442 | 3.0 | 17 | 12 | 32 | O8 III |
| *17* | *80.1* | *5.96* | *57,000* | *9.8* | *1483* | | | | | |
| 18 | 79.9 | 5.96 | 49,566 | 12.9 | 1286 | | | | | O2 I |
| 19 | 78.6 | 5.98 | 47,765 | 14.2 | 1199 | | | | | O2 I |
| 20 | 76.5 | 6.00 | 45,683 | 15.9 | 1099 | | | | | O3 I |
| 21 | 72.7 | 6.01 | 42,474 | 18.8 | 970 | | | | | O3 I |
| 22 | 65.9 | 6.03 | 38,434 | 23.2 | 788 | 1.2 | 24 | 10 | 4 | O5 I |
| 23 | 57.5 | 6.05 | 36,653 | 26.2 | 641 | 1.7 | 20 | 12 | 8 | O6 I |
| 24 | 37.1 | 6.01 | 24,041 | 58.4 | 288 | 3.0 | 16 | 12 | 32 | B1.5 Ia |
| *25* | *60.1* | *5.73* | *54,635* | *8.2* | *1467* | | | | | |
| 26 | 59.9 | 5.73 | 47,509 | 10.8 | 1274 | | | | | O2 V |
| 27 | 59.4 | 5.75 | 46,277 | 11.7 | 1211 | | | | | O2 V |
| 28 | 58.7 | 5.78 | 44,963 | 12.8 | 1133 | | | | | O3 V |
| 29 | 57.5 | 5.80 | 43,138 | 14.3 | 1051 | | | | | O4 III |
| 30 | 55.7 | 5.83 | 40,319 | 16.8 | 928 | | | | | O5 III |
| 31 | 52.3 | 5.85 | 34,971 | 23.0 | 751 | | | | | O7 I |
| 32 | 46.3 | 5.88 | 25,314 | 45.1 | 488 | 1.1 | 28 | 9 | 2 | B1 Ia |
| *33* | *50.0* | *5.57* | *52,465* | *7.4* | *1444* | | | | | |
| 34 | 50.0 | 5.57 | 45,622 | 9.7 | 1255 | | | | | O2 V |
| 35 | 49.2 | 5.62 | 43,576 | 11.3 | 1137 | | | | | O3 V |
| 36 | 47.8 | 5.67 | 40,507 | 13.9 | 990 | | | | | O5 V |
| 37 | 46.6 | 5.69 | 37,807 | 16.4 | 889 | | | | | O6 III |
| 38 | 44.2 | 5.72 | 33,037 | 22.1 | 726 | | | | | O8 I |
| 39 | 39.8 | 5.75 | 24,880 | 40.4 | 499 | | | | | B1 Ia |
| 40 | 16.9 | 5.59 | 29,759 | 23.5 | 352 | | | | | O9.5 I |

| Model | $M$ | $\log L$ | $T_{\text{eff}}$ | $R$ | $v_{\text{esc}}$ | He | C | N | O | Sp. Type |
|---|---|---|---|---|---|---|---|---|---|---|
| | ($M_\odot$) | ($L_\odot$) | (K) | ($R_\odot$) | (km s$^{-1}$) | | | | | |
| *41* | *39.9* | *5.38* | *50,294* | *6.5* | *1412* | | | | | |
| 42 | 40.0 | 5.38 | 43,734 | 8.5 | 1228 | | | | | O3 V |
| 43 | 39.5 | 5.42 | 42,084 | 9.7 | 1137 | | | | | O4 V |
| 44 | 38.8 | 5.47 | 40,205 | 11.2 | 1030 | | | | | O5 V |
| 45 | 37.5 | 5.53 | 36,500 | 14.5 | 866 | | | | | O7 III |
| 46 | 36.3 | 5.56 | 32,858 | 18.5 | 742 | | | | | O8 I |
| 47 | 34.4 | 5.59 | 26,494 | 29.6 | 569 | | | | | B1 I |
| 48 | 15.7 | 5.53 | 32,953 | 17.9 | 383 | | | | | O8 I |
| *49* | *30.0* | *5.11* | *45,971* | *5.7* | *1339* | | | | | |
| 50 | 30.0 | 5.11 | 39,975 | 7.5 | 1166 | | | | | O5.5 V |
| 51 | 29.3 | 5.19 | 37,688 | 9.3 | 1023 | | | | | O6.5 V |
| 52 | 28.5 | 5.27 | 34,549 | 12.0 | 865 | | | | | O7.5 III |
| 53 | 28.0 | 5.30 | 32,589 | 14.0 | 789 | | | | | O8.5 III |
| 54 | 27.4 | 5.33 | 29,727 | 17.5 | 699 | | | | | B0 III |
| 55 | 26.6 | 5.37 | 24,838 | 26.1 | 553 | | | | | B0.5 I |
| 56 | 15.6 | 5.45 | 15,267 | 75.4 | 212 | | | | | B3 Ia |
| *57* | *25.0* | *4.91* | *43,811* | *5.0* | *1328* | | | | | |
| 58 | 25.0 | 4.91 | 38,097 | 6.5 | 1155 | | | | | O6.5 V |
| 59 | 24.6 | 4.97 | 36,525 | 7.7 | 1053 | | | | | O7 V |
| 60 | 24.4 | 5.02 | 35,524 | 8.5 | 983 | | | | | O7.5 V |
| 61 | 24.0 | 5.06 | 34,052 | 9.8 | 908 | | | | | O8.5 V |
| 62 | 23.8 | 5.09 | 32,982 | 10.7 | 856 | | | | | O8.5 III |
| 63 | 23.3 | 5.14 | 29,531 | 14.3 | 733 | | | | | B0 III |
| 64 | 22.9 | 5.18 | 26,636 | 18.2 | 634 | | | | | B0.5 III |
| 65 | 22.4 | 5.21 | 23,768 | 23.7 | 546 | | | | | B0.7 III |
| *66* | *20.0* | *4.66* | *40,422* | *4.4* | *1282* | | | | | |
| 67 | 20.0 | 4.66 | 35,150 | 5.7 | 1116 | | | | | O8 V |
| 68 | 19.9 | 4.70 | 34,298 | 6.3 | 1057 | | | | | O8 V |
| 69 | 19.6 | 4.76 | 33,160 | 7.3 | 975 | | | | | O8.5 V |
| 70 | 19.3 | 4.84 | 31,151 | 9.0 | 860 | | | | | O9.5 V |
| 71 | 19.1 | 4.89 | 29,320 | 10.8 | 780 | | | | | B0 V |
| 72 | 18.9 | 4.92 | 27,925 | 12.3 | 724 | | | | | B0.5 III |
| 73 | 18.5 | 4.98 | 24,035 | 17.9 | 590 | | | | | B1 III |
| *74* | *15.0* | *4.31* | *35,864* | *3.7* | *1221* | | | | | |
| 75 | 15.0 | 4.31 | 31,186 | 4.9 | 1062 | | | | | O9.5 V |
| 76 | 14.9 | 4.37 | 30,294 | 5.6 | 991 | | | | | B0 V |
| 77 | 14.9 | 4.40 | 30,016 | 5.9 | 963 | | | | | B0 V |
| 78 | 14.8 | 4.43 | 29,660 | 6.2 | 934 | | | | | B0 V |
| 79 | 14.8 | 4.45 | 29,253 | 6.6 | 908 | | | | | B0 V |
| 80 | 14.8 | 4.48 | 28,766 | 7.0 | 875 | | | | | B0.5 V |
| 81 | 14.6 | 4.57 | 26,393 | 9.3 | 754 | | | | | B0.7 V |
| 82 | 14.5 | 4.63 | 24,266 | 11.7 | 664 | | | | | B1 III |
| 83 | 14.4 | 4.67 | 22,909 | 13.7 | 608 | | | | | B1 III |
| 84 | 10.0 | 3.77 | 25,453 | 3.9 | 977 | | | | | B1.5 V |

| Model | $M$ ($M_\odot$) | $\log L$ ($L_\odot$) | $T_{\text{eff}}$ (K) | $R$ ($R_\odot$) | $v_{\text{esc}}$ (km s$^{-1}$) | He | C | N | O | Sp. Type |
|---|---|---|---|---|---|---|---|---|---|---|
| 85 | 10.0 | 3.88 | 24,487 | 4.8 | 880 | | | | | B1.5 V |
| 86 | 5.0 | 2.74 | 17,169 | 2.7 | 848 | | | | | B3 V |

Table 2. Stellar-wind parameters for $Z = Z_\odot$

| Model | $k$ | $\alpha$ | $\delta$ | $\log \dot{M}$ ($M_\odot$ yr$^{-1}$) | $v_\infty$ (km s$^{-1}$) |
|---|---|---|---|---|---|
| 1 | 0.234 | 0.598 | 0.050 | −4.75 | 4076 |
| 2 | 0.229 | 0.598 | 0.050 | −4.75 | 3536 |
| 3 | 0.296 | 0.598 | 0.050 | −4.51 | 3190 |
| 4 | 0.288 | 0.598 | 0.050 | −4.48 | 2817 |
| 5 | 0.278 | 0.598 | 0.050 | −4.43 | 2481 |
| 6 | 0.078 | 0.625 | 0.075 | −5.19 | 1175 |
| 7 | 0.042 | 0.607 | 0.050 | −6.06 | 830 |
| 8 | 0.116 | 0.625 | 0.075 | −6.59 | 1132 |
| 9 | 0.224 | 0.598 | 0.050 | −4.95 | 3997 |
| 10 | 0.220 | 0.598 | 0.050 | −4.95 | 3480 |
| 11 | 0.217 | 0.598 | 0.050 | −4.92 | 3190 |
| 12 | 0.282 | 0.598 | 0.050 | −4.65 | 2861 |
| 13 | 0.273 | 0.598 | 0.050 | −4.60 | 2463 |
| 14 | 0.252 | 0.598 | 0.050 | −4.53 | 1722 |
| 15 | 0.211 | 0.611 | 0.050 | −4.57 | 1247 |
| 16 | 0.146 | 0.632 | 0.050 | −5.69 | 1227 |
| 17 | 0.213 | 0.598 | 0.050 | −5.24 | 4011 |
| 18 | 0.209 | 0.598 | 0.050 | −5.24 | 3480 |
| 19 | 0.207 | 0.598 | 0.050 | −5.19 | 3246 |
| 20 | 0.204 | 0.598 | 0.050 | −5.14 | 2974 |
| 21 | 0.267 | 0.598 | 0.050 | −4.87 | 2628 |
| 22 | 0.257 | 0.597 | 0.050 | −4.80 | 2135 |
| 23 | 0.244 | 0.598 | 0.050 | −4.69 | 1747 |
| 24 | 0.056 | 0.668 | 0.075 | −5.27 | 746 |
| 25 | 0.198 | 0.598 | 0.050 | −5.63 | 3968 |
| 26 | 0.194 | 0.598 | 0.050 | −5.63 | 3446 |
| 27 | 0.193 | 0.598 | 0.050 | −5.59 | 3276 |
| 28 | 0.192 | 0.598 | 0.050 | −5.52 | 3066 |
| 29 | 0.191 | 0.598 | 0.050 | −5.47 | 2844 |
| 30 | 0.188 | 0.598 | 0.050 | −5.40 | 2514 |
| 31 | 0.245 | 0.597 | 0.050 | −5.10 | 2034 |

| Model | $k$ | $\alpha$ | $\delta$ | $\log \dot{M}$ $(M_\odot\ \text{yr}^{-1})$ | $v_\infty$ $(\text{km s}^{-1})$ |
|---|---|---|---|---|---|
| 32 | 0.102 | 0.609 | 0.075 | −5.61 | 1304 |
| 33 | 0.188 | 0.598 | 0.050 | −5.90 | 3906 |
| 34 | 0.184 | 0.598 | 0.050 | −5.90 | 3397 |
| 35 | 0.183 | 0.598 | 0.050 | −5.80 | 3078 |
| 36 | 0.181 | 0.598 | 0.050 | −5.69 | 2680 |
| 37 | 0.179 | 0.598 | 0.050 | −5.64 | 2408 |
| 38 | 0.235 | 0.597 | 0.050 | −5.32 | 1967 |
| 39 | 0.102 | 0.609 | 0.075 | −5.82 | 1333 |
| 40 | 0.073 | 0.645 | 0.075 | −5.72 | 894 |
| 41 | 0.176 | 0.598 | 0.050 | −6.22 | 3820 |
| 42 | 0.173 | 0.598 | 0.050 | −6.22 | 3324 |
| 43 | 0.172 | 0.598 | 0.050 | −6.14 | 3076 |
| 44 | 0.172 | 0.598 | 0.050 | −6.03 | 2787 |
| 45 | 0.169 | 0.598 | 0.050 | −5.90 | 2347 |
| 46 | 0.166 | 0.597 | 0.050 | −5.83 | 2011 |
| 47 | 0.116 | 0.609 | 0.075 | −5.98 | 1519 |
| 48 | 0.138 | 0.647 | 0.050 | −5.34 | 1041 |
| 49 | 0.069 | 0.598 | 0.050 | −7.34 | 3624 |
| 50 | 0.064 | 0.598 | 0.050 | −7.39 | 3154 |
| 51 | 0.069 | 0.598 | 0.050 | −7.17 | 2768 |
| 52 | 0.156 | 0.598 | 0.050 | −6.35 | 2343 |
| 53 | 0.155 | 0.598 | 0.050 | −6.29 | 2136 |
| 54 | 0.082 | 0.607 | 0.070 | −6.70 | 1869 |
| 55 | 0.125 | 0.609 | 0.075 | −6.24 | 1477 |
| 56 | 0.157 | 0.422 | 0.050 | −6.51 | 301 |
| 57 | 0.046 | 0.598 | 0.050 | −7.96 | 3592 |
| 58 | 0.045 | 0.598 | 0.050 | −7.96 | 3125 |
| 59 | 0.049 | 0.598 | 0.050 | −7.77 | 2851 |
| 60 | 0.053 | 0.598 | 0.050 | −7.62 | 2661 |
| 61 | 0.056 | 0.598 | 0.050 | −7.49 | 2458 |
| 62 | 0.058 | 0.598 | 0.050 | −7.40 | 2317 |
| 63 | 0.060 | 0.607 | 0.070 | −7.24 | 1958 |
| 64 | 0.063 | 0.607 | 0.070 | −7.11 | 1695 |
| 65 | 0.133 | 0.609 | 0.075 | −6.42 | 1458 |
| 66 | 0.030 | 0.598 | 0.050 | −8.67 | 3468 |
| 67 | 0.029 | 0.598 | 0.050 | −8.67 | 3019 |
| 68 | 0.031 | 0.598 | 0.050 | −8.55 | 2859 |
| 69 | 0.034 | 0.598 | 0.050 | −8.36 | 2640 |
| 70 | 0.038 | 0.598 | 0.050 | −8.11 | 2327 |
| 71 | 0.040 | 0.607 | 0.070 | −7.96 | 2085 |
| 72 | 0.042 | 0.607 | 0.070 | −7.87 | 1936 |

| Model | $k$ | $\alpha$ | $\delta$ | $\log \dot{M}$ ($M_\odot$ yr$^{-1}$) | $v_\infty$ (km s$^{-1}$) |
|---|---|---|---|---|---|
| 73 | 0.147 | 0.609 | 0.075 | −6.70 | 1576 |
| 74 | 0.016 | 0.599 | 0.050 | −9.66 | 3302 |
| 75 | 0.016 | 0.598 | 0.050 | −9.66 | 2874 |
| 76 | 0.017 | 0.598 | 0.050 | −9.48 | 2682 |
| 77 | 0.018 | 0.598 | 0.050 | −9.39 | 2605 |
| 78 | 0.019 | 0.607 | 0.070 | −9.30 | 2495 |
| 79 | 0.020 | 0.607 | 0.070 | −9.24 | 2427 |
| 80 | 0.021 | 0.607 | 0.070 | −9.14 | 2338 |
| 81 | 0.023 | 0.607 | 0.070 | −8.90 | 2014 |
| 82 | 0.026 | 0.607 | 0.070 | −8.69 | 1774 |
| 83 | 0.031 | 0.589 | 0.050 | −8.56 | 1599 |
| 84 | 0.015 | 0.608 | 0.070 | −10.51 | 2609 |
| 85 | 0.012 | 0.608 | 0.070 | −10.51 | 2350 |
| 86 | 2.955 | 0.403 | 0.050 | −10.08 | 1227 |

Table 3. Stellar-wind parameters for $Z = 2\, Z_\odot$

| Model | $k$ | $\alpha$ | $\delta$ | $\log \dot{M}$ ($M_\odot$ yr$^{-1}$) | $v_\infty$ (km s$^{-1}$) |
|---|---|---|---|---|---|
| 1 | 0.279 | 0.620 | 0.050 | −4.55 | 4419 |
| 2 | 0.273 | 0.620 | 0.050 | −4.55 | 3834 |
| 3 | 0.358 | 0.620 | 0.050 | −4.31 | 3458 |
| 4 | 0.349 | 0.619 | 0.050 | −4.27 | 3054 |
| 5 | 0.340 | 0.619 | 0.050 | −4.22 | 2690 |
| 6 | 0.089 | 0.639 | 0.075 | −5.02 | 1220 |
| 7 | 0.042 | 0.647 | 0.050 | −5.84 | 917 |
| 8 | 0.134 | 0.642 | 0.075 | −6.38 | 1182 |
| 9 | 0.266 | 0.620 | 0.050 | −4.74 | 4334 |
| 10 | 0.260 | 0.620 | 0.050 | −4.74 | 3773 |
| 11 | 0.257 | 0.620 | 0.050 | −4.71 | 3459 |
| 12 | 0.339 | 0.619 | 0.050 | −4.45 | 3102 |
| 13 | 0.331 | 0.619 | 0.050 | −4.39 | 2669 |
| 14 | 0.308 | 0.619 | 0.050 | −4.32 | 1866 |
| 15 | 0.248 | 0.646 | 0.050 | −4.36 | 1365 |
| 16 | 0.175 | 0.653 | 0.050 | −5.48 | 1296 |
| 17 | 0.249 | 0.620 | 0.050 | −5.03 | 4349 |
| 18 | 0.244 | 0.620 | 0.050 | −5.03 | 3773 |

| Model | $k$ | $\alpha$ | $\delta$ | $\log \dot{M}$ ($M_\odot$ yr$^{-1}$) | $v_\infty$ (km s$^{-1}$) |
|---|---|---|---|---|---|
| 19 | 0.243 | 0.620 | 0.050 | −4.99 | 3519 |
| 20 | 0.241 | 0.619 | 0.050 | −4.93 | 3224 |
| 21 | 0.319 | 0.619 | 0.050 | −4.67 | 2848 |
| 22 | 0.309 | 0.619 | 0.050 | −4.59 | 2313 |
| 23 | 0.297 | 0.620 | 0.050 | −4.49 | 1893 |
| 24 | 0.071 | 0.695 | 0.075 | −5.01 | 799 |
| 25 | 0.228 | 0.620 | 0.050 | −5.42 | 4302 |
| 26 | 0.223 | 0.620 | 0.050 | −5.42 | 3736 |
| 27 | 0.223 | 0.620 | 0.050 | −5.38 | 3552 |
| 28 | 0.223 | 0.620 | 0.050 | −5.31 | 3324 |
| 29 | 0.222 | 0.619 | 0.050 | −5.27 | 3083 |
| 30 | 0.219 | 0.619 | 0.050 | −5.19 | 2725 |
| 31 | 0.290 | 0.619 | 0.050 | −4.89 | 2205 |
| 32 | 0.120 | 0.625 | 0.075 | −5.41 | 1385 |
| 33 | 0.214 | 0.620 | 0.050 | −5.69 | 4236 |
| 34 | 0.210 | 0.620 | 0.050 | −5.69 | 3683 |
| 35 | 0.210 | 0.620 | 0.050 | −5.59 | 3337 |
| 36 | 0.209 | 0.619 | 0.050 | −5.48 | 2905 |
| 37 | 0.207 | 0.619 | 0.050 | −5.43 | 2611 |
| 38 | 0.277 | 0.619 | 0.050 | −5.11 | 2132 |
| 39 | 0.126 | 0.625 | 0.075 | −5.57 | 1416 |
| 40 | 0.081 | 0.682 | 0.075 | −5.51 | 989 |
| 41 | 0.199 | 0.620 | 0.050 | −6.01 | 4142 |
| 42 | 0.195 | 0.620 | 0.050 | −6.01 | 3604 |
| 43 | 0.195 | 0.620 | 0.050 | −5.93 | 3335 |
| 44 | 0.195 | 0.619 | 0.050 | −5.83 | 3022 |
| 45 | 0.194 | 0.619 | 0.050 | −5.70 | 2544 |
| 46 | 0.191 | 0.619 | 0.050 | −5.62 | 2179 |
| 47 | 0.136 | 0.625 | 0.075 | −5.77 | 1611 |
| 48 | 0.172 | 0.665 | 0.050 | −5.14 | 1103 |
| 49 | 0.069 | 0.620 | 0.050 | −7.19 | 3929 |
| 50 | 0.068 | 0.620 | 0.050 | −7.19 | 3420 |
| 51 | 0.077 | 0.620 | 0.050 | −6.93 | 3001 |
| 52 | 0.176 | 0.619 | 0.050 | −6.14 | 2540 |
| 53 | 0.175 | 0.619 | 0.050 | −6.08 | 2315 |
| 54 | 0.093 | 0.625 | 0.070 | −6.49 | 2002 |
| 55 | 0.146 | 0.625 | 0.075 | −6.03 | 1567 |
| 56 | 0.145 | 0.464 | 0.050 | −6.30 | 338 |
| 57 | 0.048 | 0.620 | 0.050 | −7.75 | 3895 |
| 58 | 0.047 | 0.620 | 0.050 | −7.75 | 3389 |
| 59 | 0.052 | 0.620 | 0.050 | −7.57 | 3091 |

| Model | k | α | δ | log $\dot{M}$ ($M_\odot$ yr$^{-1}$) | $v_\infty$ (km s$^{-1}$) |
|---|---|---|---|---|---|
| 60 | 0.056 | 0.619 | 0.050 | −7.41 | 2885 |
| 61 | 0.062 | 0.619 | 0.050 | −7.25 | 2665 |
| 62 | 0.063 | 0.619 | 0.050 | −7.19 | 2512 |
| 63 | 0.066 | 0.625 | 0.070 | −7.03 | 2096 |
| 64 | 0.070 | 0.625 | 0.070 | −6.91 | 1816 |
| 65 | 0.154 | 0.625 | 0.075 | −6.22 | 1546 |
| 66 | 0.030 | 0.620 | 0.050 | −8.46 | 3760 |
| 67 | 0.030 | 0.620 | 0.050 | −8.46 | 3273 |
| 68 | 0.031 | 0.620 | 0.050 | −8.34 | 3100 |
| 69 | 0.035 | 0.620 | 0.050 | −8.15 | 2862 |
| 70 | 0.039 | 0.619 | 0.050 | −7.90 | 2523 |
| 71 | 0.043 | 0.625 | 0.070 | −7.75 | 2230 |
| 72 | 0.045 | 0.625 | 0.070 | −7.66 | 2072 |
| 73 | 0.169 | 0.625 | 0.075 | −6.50 | 1670 |
| 74 | 0.016 | 0.620 | 0.050 | −9.45 | 3581 |
| 75 | 0.015 | 0.620 | 0.050 | −9.45 | 3116 |
| 76 | 0.017 | 0.620 | 0.050 | −9.27 | 2907 |
| 77 | 0.018 | 0.620 | 0.050 | −9.18 | 2824 |
| 78 | 0.020 | 0.625 | 0.070 | −9.07 | 2665 |
| 79 | 0.020 | 0.625 | 0.070 | −9.04 | 2593 |
| 80 | 0.020 | 0.625 | 0.070 | −8.98 | 2498 |
| 81 | 0.024 | 0.625 | 0.070 | −8.66 | 2154 |
| 82 | 0.027 | 0.625 | 0.070 | −8.48 | 1899 |
| 83 | 0.031 | 0.611 | 0.050 | −8.36 | 1732 |
| 84 | 0.015 | 0.625 | 0.070 | −10.30 | 2783 |
| 85 | 0.012 | 0.625 | 0.070 | −10.30 | 2509 |
| 86 | 2.415 | 0.431 | 0.050 | −9.88 | 1332 |

Table 4. Stellar-wind parameters for $Z = 0.4\, Z_\odot$

| Model | k | α | δ | log $\dot{M}$ ($M_\odot$ yr$^{-1}$) | $v_\infty$ (km s$^{-1}$) |
|---|---|---|---|---|---|
| 1 | 0.191 | 0.569 | 0.050 | −5.03 | 3675 |
| 2 | 0.187 | 0.569 | 0.050 | −5.03 | 3189 |
| 3 | 0.238 | 0.569 | 0.050 | −4.79 | 2877 |
| 4 | 0.230 | 0.569 | 0.050 | −4.75 | 2541 |
| 5 | 0.220 | 0.569 | 0.050 | −4.70 | 2239 |

| Model | $k$ | $\alpha$ | $\delta$ | $\log \dot{M}$ ($M_\odot$ yr$^{-1}$) | $v_\infty$ (km s$^{-1}$) |
|---|---|---|---|---|---|
| 6 | 0.074 | 0.578 | 0.075 | −5.46 | 1030 |
| 7 | 0.038 | 0.575 | 0.050 | −6.32 | 763 |
| 8 | 0.109 | 0.592 | 0.075 | −6.84 | 1036 |
| 9 | 0.185 | 0.569 | 0.050 | −5.23 | 3604 |
| 10 | 0.181 | 0.569 | 0.050 | −5.23 | 3138 |
| 11 | 0.178 | 0.569 | 0.050 | −5.19 | 2877 |
| 12 | 0.227 | 0.569 | 0.050 | −4.93 | 2581 |
| 13 | 0.218 | 0.569 | 0.050 | −4.87 | 2222 |
| 14 | 0.200 | 0.569 | 0.050 | −4.80 | 1554 |
| 15 | 0.173 | 0.573 | 0.050 | −4.84 | 1119 |
| 16 | 0.137 | 0.581 | 0.050 | −5.95 | 1073 |
| 17 | 0.178 | 0.569 | 0.050 | −5.52 | 3617 |
| 18 | 0.175 | 0.569 | 0.050 | −5.52 | 3138 |
| 19 | 0.173 | 0.569 | 0.050 | −5.47 | 2927 |
| 20 | 0.170 | 0.569 | 0.050 | −5.42 | 2682 |
| 21 | 0.218 | 0.569 | 0.050 | −5.15 | 2370 |
| 22 | 0.208 | 0.568 | 0.050 | −5.07 | 1926 |
| 23 | 0.194 | 0.569 | 0.050 | −4.97 | 1576 |
| 24 | 0.049 | 0.625 | 0.075 | −5.54 | 673 |
| 25 | 0.169 | 0.570 | 0.050 | −5.90 | 3578 |
| 26 | 0.166 | 0.569 | 0.050 | −5.90 | 3107 |
| 27 | 0.165 | 0.569 | 0.050 | −5.86 | 2954 |
| 28 | 0.163 | 0.569 | 0.050 | −5.80 | 2765 |
| 29 | 0.161 | 0.569 | 0.050 | −5.75 | 2565 |
| 30 | 0.158 | 0.569 | 0.050 | −5.68 | 2268 |
| 31 | 0.201 | 0.568 | 0.050 | −5.38 | 1836 |
| 32 | 0.087 | 0.580 | 0.075 | −5.89 | 1177 |
| 33 | 0.163 | 0.570 | 0.050 | −6.18 | 3522 |
| 34 | 0.160 | 0.569 | 0.050 | −6.18 | 3063 |
| 35 | 0.158 | 0.569 | 0.050 | −6.07 | 2776 |
| 36 | 0.155 | 0.569 | 0.050 | −5.96 | 2417 |
| 37 | 0.152 | 0.569 | 0.050 | −5.91 | 2173 |
| 38 | 0.196 | 0.568 | 0.050 | −5.59 | 1775 |
| 39 | 0.093 | 0.580 | 0.075 | −6.05 | 1204 |
| 40 | 0.058 | 0.625 | 0.075 | −6.00 | 854 |
| 41 | 0.155 | 0.570 | 0.050 | −6.49 | 3444 |
| 42 | 0.153 | 0.569 | 0.050 | −6.50 | 2997 |
| 43 | 0.151 | 0.569 | 0.050 | −6.41 | 2774 |
| 44 | 0.150 | 0.569 | 0.050 | −6.31 | 2514 |
| 45 | 0.146 | 0.569 | 0.050 | −6.18 | 2117 |
| 46 | 0.143 | 0.569 | 0.050 | −6.11 | 1814 |

| Model | $k$ | $\alpha$ | $\delta$ | $\log \dot{M}$ ($M_\odot$ yr$^{-1}$) | $v_\infty$ (km s$^{-1}$) |
|---|---|---|---|---|---|
| 47 | 0.102 | 0.581 | 0.075 | −6.25 | 1372 |
| 48 | 0.119 | 0.601 | 0.050 | −5.62 | 931 |
| 49 | 0.062 | 0.570 | 0.050 | −7.67 | 3267 |
| 50 | 0.061 | 0.569 | 0.050 | −7.67 | 2844 |
| 51 | 0.067 | 0.569 | 0.050 | −7.42 | 2497 |
| 52 | 0.138 | 0.569 | 0.050 | −6.62 | 2114 |
| 53 | 0.137 | 0.569 | 0.050 | −6.56 | 1927 |
| 54 | 0.075 | 0.578 | 0.070 | −6.97 | 1687 |
| 55 | 0.115 | 0.581 | 0.075 | −6.48 | 1334 |
| 56 | 0.165 | 0.380 | 0.050 | −6.79 | 268 |
| 57 | 0.045 | 0.570 | 0.050 | −8.24 | 3239 |
| 58 | 0.044 | 0.569 | 0.050 | −8.24 | 2818 |
| 59 | 0.048 | 0.569 | 0.050 | −8.05 | 2571 |
| 60 | 0.051 | 0.569 | 0.050 | −7.89 | 2400 |
| 61 | 0.053 | 0.569 | 0.050 | −7.77 | 2217 |
| 62 | 0.055 | 0.569 | 0.050 | −7.67 | 2090 |
| 63 | 0.057 | 0.578 | 0.070 | −7.52 | 1767 |
| 64 | 0.059 | 0.578 | 0.070 | −7.39 | 1530 |
| 65 | 0.115 | 0.581 | 0.075 | −6.73 | 1316 |
| 66 | 0.031 | 0.570 | 0.050 | −8.94 | 3127 |
| 67 | 0.030 | 0.569 | 0.050 | −8.94 | 2722 |
| 68 | 0.032 | 0.569 | 0.050 | −8.82 | 2578 |
| 69 | 0.034 | 0.569 | 0.050 | −8.63 | 2381 |
| 70 | 0.038 | 0.569 | 0.050 | −8.39 | 2099 |
| 71 | 0.040 | 0.578 | 0.070 | −8.23 | 1881 |
| 72 | 0.041 | 0.578 | 0.070 | −8.14 | 1747 |
| 73 | 0.134 | 0.581 | 0.075 | −6.98 | 1423 |
| 74 | 0.018 | 0.570 | 0.050 | −9.94 | 2977 |
| 75 | 0.017 | 0.569 | 0.050 | −9.94 | 2591 |
| 76 | 0.019 | 0.569 | 0.050 | −9.75 | 2418 |
| 77 | 0.020 | 0.569 | 0.050 | −9.66 | 2349 |
| 78 | 0.020 | 0.579 | 0.070 | −9.57 | 2251 |
| 79 | 0.021 | 0.579 | 0.070 | −9.51 | 2190 |
| 80 | 0.022 | 0.579 | 0.070 | −9.42 | 2109 |
| 81 | 0.024 | 0.578 | 0.070 | −9.14 | 1817 |
| 82 | 0.028 | 0.578 | 0.070 | −8.91 | 1601 |
| 83 | 0.032 | 0.560 | 0.050 | −8.84 | 1445 |
| 84 | 0.017 | 0.579 | 0.070 | −10.78 | 2353 |
| 85 | 0.014 | 0.579 | 0.070 | −10.78 | 2120 |
| 86 | 3.970 | 0.367 | 0.050 | −10.36 | 1102 |

Table 5. Stellar-wind parameters for $Z = 0.2\ Z_\odot$

| Model | $k$ | $\alpha$ | $\delta$ | $\log \dot{M}$ ($M_\odot$ yr$^{-1}$) | $v_\infty$ (km s$^{-1}$) |
|---|---|---|---|---|---|
| 1 | 0.179 | 0.535 | 0.050 | −5.24 | 3268 |
| 2 | 0.176 | 0.534 | 0.050 | −5.24 | 2835 |
| 3 | 0.219 | 0.534 | 0.050 | −5.00 | 2557 |
| 4 | 0.210 | 0.534 | 0.050 | −4.96 | 2258 |
| 5 | 0.199 | 0.533 | 0.050 | −4.91 | 1988 |
| 6 | 0.069 | 0.549 | 0.075 | −5.67 | 939 |
| 7 | 0.034 | 0.556 | 0.050 | −6.53 | 724 |
| 8 | 0.116 | 0.552 | 0.075 | −7.07 | 922 |
| 9 | 0.175 | 0.535 | 0.050 | −5.43 | 3205 |
| 10 | 0.173 | 0.534 | 0.050 | −5.43 | 2790 |
| 11 | 0.169 | 0.534 | 0.050 | −5.40 | 2557 |
| 12 | 0.210 | 0.534 | 0.050 | −5.14 | 2293 |
| 13 | 0.201 | 0.533 | 0.050 | −5.08 | 1973 |
| 14 | 0.182 | 0.532 | 0.050 | −5.01 | 1375 |
| 15 | 0.158 | 0.543 | 0.050 | −5.02 | 1028 |
| 16 | 0.124 | 0.552 | 0.050 | −6.17 | 987 |
| 17 | 0.172 | 0.535 | 0.050 | −5.72 | 3216 |
| 18 | 0.170 | 0.534 | 0.050 | −5.72 | 2790 |
| 19 | 0.167 | 0.534 | 0.050 | −5.68 | 2602 |
| 20 | 0.163 | 0.534 | 0.050 | −5.62 | 2384 |
| 21 | 0.205 | 0.534 | 0.050 | −5.36 | 2106 |
| 22 | 0.193 | 0.533 | 0.050 | −5.28 | 1710 |
| 23 | 0.179 | 0.532 | 0.050 | −5.18 | 1394 |
| 24 | 0.042 | 0.605 | 0.075 | −5.75 | 643 |
| 25 | 0.167 | 0.535 | 0.050 | −6.11 | 3181 |
| 26 | 0.165 | 0.534 | 0.050 | −6.11 | 2763 |
| 27 | 0.163 | 0.534 | 0.050 | −6.07 | 2626 |
| 28 | 0.161 | 0.534 | 0.050 | −6.00 | 2458 |
| 29 | 0.158 | 0.534 | 0.050 | −5.96 | 2280 |
| 30 | 0.154 | 0.534 | 0.050 | −5.88 | 2015 |
| 31 | 0.191 | 0.533 | 0.050 | −5.58 | 1630 |
| 32 | 0.080 | 0.558 | 0.075 | −6.10 | 1092 |
| 33 | 0.164 | 0.535 | 0.050 | −6.38 | 3132 |
| 34 | 0.162 | 0.534 | 0.050 | −6.38 | 2724 |
| 35 | 0.159 | 0.534 | 0.050 | −6.28 | 2467 |
| 36 | 0.154 | 0.534 | 0.050 | −6.17 | 2148 |
| 37 | 0.151 | 0.534 | 0.050 | −6.12 | 1930 |
| 38 | 0.190 | 0.533 | 0.050 | −5.80 | 1576 |

| Model | $k$ | $\alpha$ | $\delta$ | $\log \dot{M}$ $(M_\odot\ yr^{-1})$ | $v_\infty$ $(km\ s^{-1})$ |
|---|---|---|---|---|---|
| 39 | 0.086 | 0.558 | 0.075 | −6.26 | 1117 |
| 40 | 0.058 | 0.583 | 0.075 | −6.20 | 772 |
| 41 | 0.160 | 0.535 | 0.050 | −6.70 | 3063 |
| 42 | 0.158 | 0.534 | 0.050 | −6.70 | 2665 |
| 43 | 0.155 | 0.534 | 0.050 | −6.62 | 2466 |
| 44 | 0.152 | 0.534 | 0.050 | −6.52 | 2234 |
| 45 | 0.147 | 0.533 | 0.050 | −6.39 | 1880 |
| 46 | 0.143 | 0.533 | 0.050 | −6.31 | 1611 |
| 47 | 0.094 | 0.558 | 0.075 | −6.46 | 1272 |
| 48 | 0.107 | 0.570 | 0.050 | −5.83 | 860 |
| 49 | 0.069 | 0.535 | 0.050 | −7.88 | 2906 |
| 50 | 0.068 | 0.534 | 0.050 | −7.88 | 2529 |
| 51 | 0.074 | 0.534 | 0.050 | −7.62 | 2219 |
| 52 | 0.144 | 0.534 | 0.050 | −6.83 | 1878 |
| 53 | 0.141 | 0.533 | 0.050 | −6.77 | 1712 |
| 54 | 0.072 | 0.556 | 0.070 | −7.18 | 1565 |
| 55 | 0.105 | 0.558 | 0.075 | −6.72 | 1237 |
| 56 | 0.180 | 0.350 | 0.050 | −6.95 | 246 |
| 57 | 0.053 | 0.535 | 0.050 | −8.44 | 2880 |
| 58 | 0.052 | 0.535 | 0.050 | −8.44 | 2506 |
| 59 | 0.055 | 0.534 | 0.050 | −8.26 | 2286 |
| 60 | 0.058 | 0.534 | 0.050 | −8.10 | 2133 |
| 61 | 0.060 | 0.534 | 0.050 | −7.97 | 1970 |
| 62 | 0.062 | 0.534 | 0.050 | −7.88 | 1857 |
| 63 | 0.056 | 0.556 | 0.070 | −7.72 | 1639 |
| 64 | 0.058 | 0.556 | 0.070 | −7.60 | 1419 |
| 65 | 0.113 | 0.558 | 0.075 | −6.91 | 1221 |
| 66 | 0.038 | 0.535 | 0.050 | −9.15 | 2781 |
| 67 | 0.037 | 0.535 | 0.050 | −9.15 | 2420 |
| 68 | 0.039 | 0.534 | 0.050 | −9.03 | 2292 |
| 69 | 0.041 | 0.534 | 0.050 | −8.84 | 2116 |
| 70 | 0.045 | 0.534 | 0.050 | −8.59 | 1865 |
| 71 | 0.040 | 0.556 | 0.070 | −8.44 | 1745 |
| 72 | 0.041 | 0.556 | 0.070 | −8.35 | 1621 |
| 73 | 0.126 | 0.559 | 0.075 | −7.20 | 1320 |
| 74 | 0.023 | 0.535 | 0.050 | −10.14 | 2648 |
| 75 | 0.023 | 0.535 | 0.050 | −10.14 | 2304 |
| 76 | 0.025 | 0.534 | 0.050 | −9.96 | 2150 |
| 77 | 0.025 | 0.534 | 0.050 | −9.87 | 2088 |
| 78 | 0.022 | 0.557 | 0.070 | −9.78 | 2088 |
| 79 | 0.023 | 0.557 | 0.070 | −9.72 | 2031 |

| Model | k | α | δ | log $\dot{M}$ ($M_\odot$ yr$^{-1}$) | $v_\infty$ (km s$^{-1}$) |
|---|---|---|---|---|---|
| 80 | 0.023 | 0.556 | 0.070 | −9.63 | 1956 |
| 81 | 0.026 | 0.556 | 0.070 | −9.35 | 1686 |
| 82 | 0.028 | 0.556 | 0.070 | −9.17 | 1485 |
| 83 | 0.033 | 0.538 | 0.050 | −9.05 | 1341 |
| 84 | 0.020 | 0.557 | 0.070 | −10.99 | 2182 |
| 85 | 0.016 | 0.557 | 0.070 | −10.99 | 1966 |
| 86 | 6.048 | 0.328 | 0.050 | −10.57 | 980 |

Table 6. Stellar-wind parameters for $Z = 0.05\ Z_\odot$

| Model | k | α | δ | log $\dot{M}$ ($M_\odot$ yr$^{-1}$) | $v_\infty$ (km s$^{-1}$) |
|---|---|---|---|---|---|
| 1 | 0.155 | 0.481 | 0.050 | −5.66 | 2757 |
| 2 | 0.155 | 0.481 | 0.050 | −5.64 | 2392 |
| 3 | 0.185 | 0.480 | 0.050 | −5.41 | 2157 |
| 4 | 0.176 | 0.480 | 0.050 | −5.38 | 1905 |
| 5 | 0.164 | 0.479 | 0.050 | −5.33 | 1678 |
| 6 | 0.063 | 0.496 | 0.075 | −6.08 | 793 |
| 7 | 0.036 | 0.491 | 0.050 | −6.95 | 604 |
| 8 | 0.118 | 0.496 | 0.075 | −7.49 | 774 |
| 9 | 0.155 | 0.481 | 0.050 | −5.85 | 2704 |
| 10 | 0.155 | 0.481 | 0.050 | −5.84 | 2354 |
| 11 | 0.156 | 0.480 | 0.050 | −5.78 | 2158 |
| 12 | 0.180 | 0.480 | 0.050 | −5.55 | 1935 |
| 13 | 0.169 | 0.479 | 0.050 | −5.49 | 1665 |
| 14 | 0.150 | 0.478 | 0.050 | −5.43 | 1161 |
| 15 | 0.126 | 0.486 | 0.050 | −5.47 | 867 |
| 16 | 0.122 | 0.480 | 0.050 | −6.59 | 804 |
| 17 | 0.155 | 0.481 | 0.050 | −6.15 | 2713 |
| 18 | 0.155 | 0.481 | 0.050 | −6.14 | 2354 |
| 19 | 0.156 | 0.480 | 0.050 | −6.07 | 2195 |
| 20 | 0.148 | 0.480 | 0.050 | −6.04 | 2011 |
| 21 | 0.179 | 0.480 | 0.050 | −5.77 | 1777 |
| 22 | 0.166 | 0.479 | 0.050 | −5.69 | 1443 |
| 23 | 0.149 | 0.478 | 0.050 | −5.59 | 1177 |
| 24 | 0.041 | 0.534 | 0.075 | −6.17 | 539 |
| 25 | 0.155 | 0.481 | 0.050 | −6.55 | 2684 |

| Model | $k$ | $\alpha$ | $\delta$ | $\log \dot{M}$ $(M_\odot\ yr^{-1})$ | $v_\infty$ $(km\ s^{-1})$ |
|---|---|---|---|---|---|
| 26 | 0.155 | 0.481 | 0.050 | −6.54 | 2331 |
| 27 | 0.155 | 0.481 | 0.050 | −6.48 | 2216 |
| 28 | 0.156 | 0.480 | 0.050 | −6.39 | 2073 |
| 29 | 0.147 | 0.480 | 0.050 | −6.37 | 1923 |
| 30 | 0.142 | 0.480 | 0.050 | −6.30 | 1700 |
| 31 | 0.170 | 0.479 | 0.050 | −6.00 | 1376 |
| 32 | 0.082 | 0.494 | 0.075 | −6.51 | 891 |
| 33 | 0.155 | 0.481 | 0.050 | −6.83 | 2642 |
| 34 | 0.155 | 0.481 | 0.050 | −6.82 | 2298 |
| 35 | 0.156 | 0.480 | 0.050 | −6.68 | 2082 |
| 36 | 0.147 | 0.480 | 0.050 | −6.59 | 1812 |
| 37 | 0.143 | 0.480 | 0.050 | −6.54 | 1629 |
| 38 | 0.172 | 0.479 | 0.050 | −6.22 | 1330 |
| 39 | 0.090 | 0.494 | 0.075 | −6.67 | 911 |
| 40 | 0.054 | 0.522 | 0.075 | −6.62 | 658 |
| 41 | 0.155 | 0.481 | 0.050 | −7.15 | 2584 |
| 42 | 0.155 | 0.481 | 0.050 | −7.14 | 2248 |
| 43 | 0.155 | 0.481 | 0.050 | −7.04 | 2080 |
| 44 | 0.156 | 0.480 | 0.050 | −6.89 | 1885 |
| 45 | 0.143 | 0.480 | 0.050 | −6.80 | 1587 |
| 46 | 0.137 | 0.479 | 0.050 | −6.73 | 1360 |
| 47 | 0.101 | 0.495 | 0.075 | −6.88 | 1038 |
| 48 | 0.099 | 0.494 | 0.050 | −6.24 | 701 |
| 49 | 0.080 | 0.481 | 0.050 | −8.29 | 2451 |
| 50 | 0.079 | 0.481 | 0.050 | −8.29 | 2133 |
| 51 | 0.083 | 0.480 | 0.050 | −8.04 | 1872 |
| 52 | 0.146 | 0.480 | 0.050 | −7.25 | 1585 |
| 53 | 0.142 | 0.479 | 0.050 | −7.18 | 1444 |
| 54 | 0.085 | 0.492 | 0.070 | −7.59 | 1277 |
| 55 | 0.115 | 0.495 | 0.075 | −7.13 | 1010 |
| 56 | 0.208 | 0.293 | 0.050 | −7.41 | 207 |
| 57 | 0.065 | 0.481 | 0.050 | −8.86 | 2429 |
| 58 | 0.064 | 0.481 | 0.050 | −8.86 | 2114 |
| 59 | 0.067 | 0.480 | 0.050 | −8.67 | 1928 |
| 60 | 0.069 | 0.480 | 0.050 | −8.52 | 1799 |
| 61 | 0.070 | 0.480 | 0.050 | −8.39 | 1663 |
| 62 | 0.071 | 0.480 | 0.050 | −8.30 | 1567 |
| 63 | 0.071 | 0.492 | 0.070 | −8.14 | 1337 |
| 64 | 0.072 | 0.492 | 0.070 | −8.01 | 1158 |
| 65 | 0.127 | 0.495 | 0.075 | −7.32 | 997 |
| 66 | 0.050 | 0.481 | 0.050 | −9.57 | 2346 |

| Model | $k$ | $\alpha$ | $\delta$ | log $\dot{M}$ ($M_\odot$ yr$^{-1}$) | $v_\infty$ (km s$^{-1}$) |
|---|---|---|---|---|---|
| 67 | 0.049 | 0.481 | 0.050 | −9.57 | 2042 |
| 68 | 0.051 | 0.481 | 0.050 | −9.44 | 1934 |
| 69 | 0.053 | 0.480 | 0.050 | −9.26 | 1785 |
| 70 | 0.055 | 0.480 | 0.050 | −9.01 | 1574 |
| 71 | 0.056 | 0.493 | 0.070 | −8.86 | 1424 |
| 72 | 0.057 | 0.492 | 0.070 | −8.76 | 1322 |
| 73 | 0.148 | 0.495 | 0.075 | −7.60 | 1077 |
| 74 | 0.035 | 0.481 | 0.050 | −10.56 | 2233 |
| 75 | 0.034 | 0.481 | 0.050 | −10.56 | 1944 |
| 76 | 0.036 | 0.481 | 0.050 | −10.38 | 1814 |
| 77 | 0.037 | 0.481 | 0.050 | −10.29 | 1762 |
| 78 | 0.037 | 0.493 | 0.070 | −10.19 | 1704 |
| 79 | 0.037 | 0.493 | 0.070 | −10.13 | 1657 |
| 80 | 0.038 | 0.493 | 0.070 | −10.04 | 1596 |
| 81 | 0.041 | 0.493 | 0.070 | −9.77 | 1375 |
| 82 | 0.042 | 0.492 | 0.070 | −9.58 | 1211 |
| 83 | 0.047 | 0.477 | 0.050 | −9.46 | 1108 |
| 84 | 0.038 | 0.494 | 0.070 | −11.41 | 1781 |
| 85 | 0.030 | 0.494 | 0.070 | −11.40 | 1605 |
| 86 | 0.030 | 0.274 | 0.050 | −10.98 | 825 |

Table 7. Parameters of the comparison stars used in Figure 7 through Figure 15

| Sp. Type | Star 1 | Star 2 | Star 3 | log $L$ ($L_\odot$) | $T_{eff}$ ( K) |
|---|---|---|---|---|---|
| O3 V | HD 93250 | HD 164794 | HDE 303308 | 5.78 | 42,900 |
| O3 I | HD 93129A | HD151932 | HD190429A | 5.93 | 42,600 |
| O5 I | HD 14947 | HD 15558 | HD 66811 | 6.03 | 39,800 |
| O6 III | HD15558 | HD 93130 | HD 190864 | 5.63 | 38,600 |
| O6.5 V | HD 5005A | HD 12993 | HD 42088 | 5.23 | 37,900 |
| O8.5 III | HD 36861 | HD 186980 | HD 193443 | 5.33 | 33,600 |
| B0.5 I | HD 115842 | HD 152234 | HD 213087 | 5.21 | 26,000 |
| B1 Ia | HD 91316 | HD 148688 | HD 193237 | 5.66 | 21,500 |
| B1.5 Ia | HD 13841 | HD 152236 | HD 190603 | 5.83 | 20,500 |